%% file: paper.tex
\documentclass[draftclsnofoot,onecolumn]{IEEEtran}
\usepackage{cite}
\usepackage{amsmath,amssymb,amsfonts}
\usepackage{graphicx,color}
\usepackage{textcomp}
\def\BibTeX{{\rm B\kern-.05em{\sc i\kern-.025em b}\kern-.08em
    T\kern-.1667em\lower.7ex\hbox{E}\kern-.125emX}}

\usepackage[ruled]{algorithm2e}
\usepackage{array}
\usepackage[caption=false,font=footnotesize]{subfig}
\usepackage{textcomp}
\usepackage{stfloats}
\usepackage{url}
\usepackage{verbatim}

\usepackage{pgfplots}
\usepackage{amsthm}
\usepackage{multirow}
\usepackage{cleveref}
\usepackage{fontawesome}
\usepackage[nomain,acronym,toc,symbols]{glossaries-extra}
\usepackage{mathtools}
\usepackage{siunitx}

\DeclareMathOperator*{\argmin}{arg\,min}
\DeclarePairedDelimiter{\abs}{\lvert}{\rvert}
\DeclarePairedDelimiter{\floor}{\lfloor}{\rfloor}
\DeclarePairedDelimiter{\ceil}{\lceil}{\rceil}
\DeclarePairedDelimiterXPP{\CN}[1]{\mathcal{CN}}{(}{)}{}{#1}
\DeclarePairedDelimiterXPP{\E}[1]{\mathbb{E}}{[}{]}{}{#1}

\setabbreviationstyle[acronym]{long-short}
\newacronym{snr}{SNR}{signal-to-noise ratio}
\newacronym{ris}{RIS}{reconfigurable intelligent surface}
\newacronym{los}{LoS}{line-of-sight}
\newacronym{bs}{BS}{base station}
\newacronym[longplural=angles of arrival]{aoa}{AoA}{angle of arrival}
\newacronym[longplural=angles of departure]{aod}{AoD}{angle of departure}
\newacronym{mmwave}{mmWave}{millimeter wave}
\newacronym{csi}{CSI}{channel state information}
\newacronym{siso}{SISO}{single-input single-output}
\newacronym{fs}{FS}{full search}
\newacronym{ts}{TS}{tracking-based search}
\newacronym{UnitCellSplit}{UCS}{unit cell split}
\newacronym{uav}{UAV}{unmanned aerial vehicle}
\newacronym{eh}{EH}{energy harvesting}
\newacronym{rhs}{RHS}{right-hand side}
\newacronym{PowerSplit}{PS}{power splitting}
\newacronym{ElementSplit}{ES}{element splitting}
\newacronym{TimeSplit}{TS}{time splitting}
\newacronym{rf}{RF}{radio-frequency}
\newacronym{dc}{DC}{direct-current}
\newacronym{iot}{IoT}{Internet of Things}

\crefname{equation}{}{}

\pgfplotsset{compat=1.16}
\usetikzlibrary{plotmarks}
\definecolor{color1}{HTML}{3a5cbc}
\definecolor{color3}{HTML}{94c4cd}
\definecolor{color2}{HTML}{f36a79}
\definecolor{color4}{HTML}{b42f54}
\pgfplotscreateplotcyclelist{colorblindfriendlysolidmarkers4}{
  {color1,solid,mark=o},
  {color2,solid,mark=square},
  {color3,solid,mark=triangle},
  {color4,solid,mark=diamond}
}
\pgfplotsset{
  line plot style/.style={
    font=\footnotesize,
    cycle list name=colorblindfriendlysolidmarkers4,
    xmajorgrids=false,
    ymajorgrids=true,
    xminorgrids=false,
    yminorgrids=false,
    enlarge x limits=false,
    line width=1pt,
    axis line style={line width=0.5pt},
    grid style={line width=0.5pt},
    legend style={draw=none,line width=1pt},
    legend cell align={left},
    minor tick num=1,
    tick style={line width=0.5pt},
    major tick length=3pt,
    minor tick length=1.5pt,
    mark options={solid},
    width=8cm,
    height=5.3cm,
  }
}

\usepgfplotslibrary{external}
\usepgfplotslibrary{fillbetween}
\usetikzlibrary{patterns}
\usetikzlibrary{shapes.geometric}
\usetikzlibrary{shapes.misc}
\usetikzlibrary{arrows.meta}
\usetikzlibrary{positioning}
\usetikzlibrary{spy}
\usetikzlibrary{decorations.pathmorphing}
\usetikzlibrary{calc}
\tikzset{external/figure name={laue}}

\definecolor{TechMetallic}{RGB}{119, 159, 181}
\definecolor{TechDarkMetallic}{RGB}{65, 116, 141}
\colorlet{BaseColor}{TechMetallic}
\colorlet{BaseDarkColor}{TechDarkMetallic}
\colorlet{SeparationLineColor}{TechDarkMetallic}
\colorlet{BaseColorA}{BaseColor!62.5}
\colorlet{BaseColorB}{BaseColor!37.5}
\colorlet{BaseColorC}{BaseColor!25.0}
\colorlet{BaseColorD}{BaseColor!12.5}
\colorlet{BaseDarkColorA}{BaseDarkColor!62.5}
\colorlet{BaseDarkColorB}{BaseDarkColor!37.5}
\colorlet{BaseDarkColorC}{BaseDarkColor!25.0}
\colorlet{BaseDarkColorD}{BaseDarkColor!12.5}

\tikzset{
    pics/newbeam/.style n args={5}{
    code = {%
      \begin{scope}
        \def\beamlength{#2}
        \def\controlwidth{#3}
        \def\tipwidth{#4}
        \def\controllength{0.8*\beamlength}
        \def\controlwidthhalf{0.5*\controlwidth}
        \def\tipwidthhalf{0.5*\tipwidth}

        \coordinate (origin) at #1;
        \coordinate (tip) at ([xshift=\beamlength]origin);
        \coordinate (controllower) at ([xshift=\controllength,yshift=-\controlwidthhalf]origin);
        \coordinate (tiplower) at ([xshift=\beamlength,yshift=-\tipwidthhalf]origin);
        \coordinate (controlupper) at ([xshift=\controllength,yshift=\controlwidthhalf]origin);
        \coordinate (tipupper) at ([xshift=\beamlength,yshift=\tipwidthhalf]origin);
        \draw [fill=#5] (origin) .. controls (controllower) and (tiplower) .. (tip) .. controls (tipupper) and (controlupper) .. (origin);
      \end{scope}
    }
  }
}

\loadglsentries{glskeys.tex}
\loadglsentries{symbols.tex}

\input{macros.tex}

\begin{document}

\title{Codebook-Based Self-Sustainable RIS: Optimal Splitting Schemes and Power Allocation}

\author{FRIEDEMANN~LAUE\IEEEauthorrefmark{1} \IEEEmembership{(Graduate~Student~Member,~IEEE)}, SEBASTIAN~LOTTER\IEEEauthorrefmark{1} \IEEEmembership{(Member,~IEEE)}, NIKITA~SHANIN\IEEEauthorrefmark{1} \IEEEmembership{(Member,~IEEE)}, AND ROBERT~SCHOBER\IEEEauthorrefmark{1} \IEEEmembership{(Fellow,~IEEE)}%
\thanks{\IEEEauthorrefmark{1}Institute for Digital Communications (IDC), Friedrich-Alexander-Universität Erlangen-Nürnberg (FAU), 91058 Erlangen, Germany}
\thanks{CORRESPONDING AUTHOR: Friedemann Laue (e-mail: friedemann.laue@fau.de).}
\thanks{The work of Robert Schober was supported in part by the Federal Ministry for Research, Technology and Space (BMFTR) in Germany in the Program of “Souverän. Digital. Vernetzt.” joint Project 6G-RIC under Project 16KISK023; in part by the Deutsche Forschungsgemeinschaft (DFG, German Research Foundation) through Project SFB 1483 under Project 442419336 (EmpkinS) and through Project SCHO 831/15-1; and in part by Horizon Europe Marie Skodowska-Curie Actions (MSCA)-UNITE under Project 101129618.}
\thanks{This work was presented in part at the IEEE Global Communications Conference Workshop, 2023 [DOI: 10.1109/GCWkshps58843.2023.10464537].}
}

\maketitle

\begin{abstract}
    This paper studies the codebook-based configuration of a \gls{ris} that extends the coverage of a \gls{bs} while utilizing energy harvesting to facilitate self-sustainable operation.
    For a given coverage area, we design a \gls{ris} codebook and propose a mathematical framework for analyzing the efficiency of three common energy harvesting schemes: \gls{PowerSplit}, \gls{ElementSplit}, and \gls{TimeSplit}.
    Thereby, we use a tile-based architecture at the \gls{ris} to exploit the advantages of both \gls{rf} combining and \gls{dc} combining.
    Moreover, we account for deterministic and random transmit signals for beam training and data transmission, respectively, and show their impact on the \gls{rf}-\gls{dc} conversion efficiencies at the rectifiers.
    Our main objective is to minimize the average transmit power at the \gls{bs} by jointly optimizing the splitting ratio for the incident signal at the \gls{ris} and the power allocated to each \gls{ris} codeword.
    While the optimal power allocation is derived analytically, we show that the optimal splitting ratio can be determined by performing a grid search over a single optimization variable.
    Our performance evaluation reveals that the efficiency of the optimized splitting schemes depends on the adopted power consumption model and the number of tiles at the \gls{ris}.
    In particular, our results show that depending on the system parameters a different splitting scheme will achieve the lowest transmit power at the \gls{bs}.
\end{abstract}


\glsresetall

\section{INTRODUCTION}
\label{sec:introduction}
\input{sections/introduction.tex}

\section{SYSTEM MODEL}
\label{sec:system}
\input{sections/system_model.tex}

\section{ENERGY HARVESTING AT RIS}
\label{sec:harvesting}
\input{sections/energy_harvesting.tex}

\section{OPTIMAL SPLITTING SCHEMES AND POWER ALLOCATION}%
\label{sec:optimization}
\input{sections/optimization_problem.tex}

\section{PERFORMANCE EVALUATION}
\label{sec:evaluation}
\input{sections/performance_evaluation.tex}

\section{CONCLUSION}
\label{sec:conclusion}
\input{sections/conclusion.tex}

\section*{APPENDIX}
\input{sections/appendix.tex}

\bibliographystyle{IEEEtran}
\bibliography{IEEEabrv,bibliography}

\end{document}

%% file: glskeys.tex
\glsaddkey*%
  {Squared}%
  {\ensuremath{\glsentrytext{\glslabel}^2}}%
  {\glsentrySquared}%
  {\GlsentrySquared}%
  {\glsSquared}%
  {\GlsSquared}%
  {\GLSSquared}

\glsaddkey*%
  {Hermitian}%
  {\ensuremath{\glsentrytext{\glslabel}^H}}%
  {\glsentryHermitian}%
  {\GlsentryHermitian}%
  {\glsHermitian}%
  {\GlsHermitian}%
  {\GLSHermitian}

\glsaddkey*%
  {FullSearch}%
  {\ensuremath{\glsentrytext{\glslabel}^\text{FS}}}%
  {\glsentryFullSearch}%
  {\GlsentryFullSearch}%
  {\glsFullSearch}%
  {\GlsFullSearch}%
  {\GLSFullSearch}

\glsaddkey*%
  {TrackingSearch}%
  {\ensuremath{\glsentrytext{\glslabel}^\text{TS}}}%
  {\glsentryTrackingSearch}%
  {\GlsentryTrackingSearch}%
  {\glsTrackingSearch}%
  {\GlsTrackingSearch}%
  {\GLSTrackingSearch}

\glsaddkey*%
  {DiodeLinear}%
  {\ensuremath{\glsentrytext{\glslabel}^\text{lin}}}%
  {\glsentryDiodeLinear}%
  {\GlsentryDiodeLinear}%
  {\glsDiodeLinear}%
  {\GlsDiodeLinear}%
  {\GLSDiodeLinear}

\glsaddkey*%
  {SaturationNonLinear}%
  {\ensuremath{\glsentrytext{\glslabel}^\mathsf{sig}}}%
  {\glsentrySaturationNonLinear}%
  {\GlsentrySaturationNonLinear}%
  {\glsSaturationNonLinear}%
  {\GlsSaturationNonLinear}%
  {\GLSSaturationNonLinear}

\glsaddkey*%
  {Instantaneous}%
  {\ensuremath{\glsentrytext{\glslabel}^\mathsf{sin}}}%
  {\glsentryInstantaneous}%
  {\GlsentryInstantaneous}%
  {\glsInstantaneous}%
  {\GlsInstantaneous}%
  {\GLSInstantaneous}

\glsaddkey*%
  {Optimal}%
  {\ensuremath{\glsentrytext{\glslabel}^*}}%
  {\glsentryOptimal}%
  {\GlsentryOptimal}%
  {\glsOptimal}%
  {\GlsOptimal}%
  {\GLSOptimal}

\glsaddkey*%
  {SupAverage}%
  {\ensuremath{\glsentrytext{\glslabel}^\mathrm{avg}}}%
  {\glsentrySupAverage}%
  {\GlsentrySupAverage}%
  {\glsSupAverage}%
  {\GlsSupAverage}%
  {\GLSSupAverage}

\glsaddkey*%
  {SupMax}%
  {\ensuremath{\glsentrytext{\glslabel}^\mathrm{max}}}%
  {\glsentrySupMax}%
  {\GlsentrySupMax}%
  {\glsSupMax}%
  {\GlsSupMax}%
  {\GLSSupMax}

\glsaddkey*%
  {SubMax}%
  {\ensuremath{\glsentrytext{\glslabel}_\mathrm{max}}}%
  {\glsentrySubMax}%
  {\GlsentrySubMax}%
  {\glsSubMax}%
  {\GlsSubMax}%
  {\GLSSubMax}

\glsaddkey*%
  {SupMin}%
  {\ensuremath{\glsentrytext{\glslabel}^\mathrm{min}}}%
  {\glsentrySupMin}%
  {\GlsentrySupMin}%
  {\glsSupMin}%
  {\GlsSupMin}%
  {\GLSSupMin}

\glsaddkey*%
  {DcComb}%
  {\ensuremath{\glsentrytext{\glslabel}^\text{DC}}}%
  {\glsentryDcComb}%
  {\GlsentryDcComb}%
  {\glsDcComb}%
  {\GlsDcComb}%
  {\GLSDcComb}

\glsaddkey*%
  {SubDc}%
  {\ensuremath{\glsentrytext{\glslabel}_\mathrm{DC}}}%
  {\glsentrySubDc}%
  {\GlsentrySubDc}%
  {\glsSubDc}%
  {\GlsSubDc}%
  {\GLSSubDc}

\glsaddkey*%
  {RfComb}%
  {\ensuremath{\glsentrytext{\glslabel}^\text{RF}}}%
  {\glsentryRfComb}%
  {\GlsentryRfComb}%
  {\glsRfComb}%
  {\GlsRfComb}%
  {\GLSRfComb}

\glsaddkey*%
  {SubRf}%
  {\ensuremath{\glsentrytext{\glslabel}_\mathrm{RF}}}%
  {\glsentrySubRf}%
  {\GlsentrySubRf}%
  {\glsSubRf}%
  {\GlsSubRf}%
  {\GLSSubRf}

\glsaddkey*%
  {SubRx}%
  {\ensuremath{\glsentrytext{\glslabel}_\mathrm{rx}}}%
  {\glsentrySubRx}%
  {\GlsentrySubRx}%
  {\glsSubRx}%
  {\GlsSubRx}%
  {\GLSSubRx}

\glsaddkey*%
  {SubTx}%
  {\ensuremath{\glsentrytext{\glslabel}_\mathrm{tx}}}%
  {\glsentrySubTx}%
  {\GlsentrySubTx}%
  {\glsSubTx}%
  {\GlsSubTx}%
  {\GLSSubTx}

\glsaddkey*%
  {SubStageSupRis}%
  {\ensuremath{\glsentrytext{\glslabel}_\mathrm{q}^\mathrm{ris}}}%
  {\glsentrySubStageSupRis}%
  {\GlsentrySubStageSupRis}%
  {\glsSubStageSupRis}%
  {\GlsSubStageSupRis}%
  {\GLSSubStageSupRis}

\glsaddkey*%
  {SubStageWordSupRis}%
  {\ensuremath{\glsentrytext{\glslabel}_\mathrm{q,b}^\mathrm{ris}}}%
  {\glsentrySubStageWordSupRis}%
  {\GlsentrySubStageWordSupRis}%
  {\glsSubStageWordSupRis}%
  {\GlsSubStageWordSupRis}%
  {\GLSSubStageWordSupRis}

\glsaddkey*%
  {LastIteration}%
  {\ensuremath{\glsentrytext{\glslabel}^{(t)}}}%
  {\glsentryLastIteration}%
  {\GlsentryLastIteration}%
  {\glsLastIteration}%
  {\GlsLastIteration}%
  {\GLSLastIteration}

\glsaddkey*%
  {FirstIteration}%
  {\ensuremath{\glsentrytext{\glslabel}^{(0)}}}%
  {\glsentryFirstIteration}%
  {\GlsentryFirstIteration}%
  {\glsFirstIteration}%
  {\GlsFirstIteration}%
  {\GLSFirstIteration}

\glsaddkey*%
  {MonomialExponent}%
  {\ensuremath{\glsentrytext{\glslabel}^{\prime}}}%
  {\glsentryMonomialExponent}%
  {\GlsentryMonomialExponent}%
  {\glsMonomialExponent}%
  {\GlsMonomialExponent}%
  {\GLSMonomialExponent}

\glsaddkey*{SubTraining}%
  {\ensuremath{\glsentrytext{\glslabel}_{\mathrm{bt}}}}%
  {\glsentrySubTraining}%
  {\GlsentrySubTraining}%
  {\glsSubTraining}%
  {\GlsSubTraining}%
  {\GLSSubTraining}

\glsaddkey*{SubEstimation}%
  {\ensuremath{\glsentrytext{\glslabel}_{\mathrm{ce}}}}%
  {\glsentrySubEstimation}%
  {\GlsentrySubEstimation}%
  {\glsSubEstimation}%
  {\GlsSubEstimation}%
  {\GLSSubEstimation}

\glsaddkey*{SubHarvesting}%
  {\ensuremath{\glsentrytext{\glslabel}_{\mathrm{eh}}}}%
  {\glsentrySubHarvesting}%
  {\GlsentrySubHarvesting}%
  {\glsSubHarvesting}%
  {\GlsSubHarvesting}%
  {\GLSSubHarvesting}

\glsaddkey*{SubHarvested}%
  {\ensuremath{\glsentrytext{\glslabel}_{\mathrm{h}}}}%
  {\glsentrySubHarvested}%
  {\GlsentrySubHarvested}%
  {\glsSubHarvested}%
  {\GlsSubHarvested}%
  {\GLSSubHarvested}

\glsaddkey*{SubConsumed}%
  {\ensuremath{\glsentrytext{\glslabel}_{\mathrm{c}}}}%
  {\glsentrySubConsumed}%
  {\GlsentrySubConsumed}%
  {\glsSubConsumed}%
  {\GlsSubConsumed}%
  {\GLSSubConsumed}

\glsaddkey*{SubReflection}%
  {\ensuremath{\glsentrytext{\glslabel}_{\mathrm{SR}}}}%
  {\glsentrySubReflection}%
  {\GlsentrySubReflection}%
  {\glsSubReflection}%
  {\GlsSubReflection}%
  {\GLSSubReflection}

\glsaddkey*{SubFrame}%
  {\ensuremath{\glsentrytext{\glslabel}_{\mathrm{fr}}}}%
  {\glsentrySubFrame}%
  {\GlsentrySubFrame}%
  {\glsSubFrame}%
  {\GlsSubFrame}%
  {\GLSSubFrame}

\glsaddkey*{SubSymbol}%
  {\ensuremath{\glsentrytext{\glslabel}_{\mathrm{s}}}}%
  {\glsentrySubSymbol}%
  {\GlsentrySubSymbol}%
  {\glsSubSymbol}%
  {\GlsSubSymbol}%
  {\GLSSubSymbol}

\glsaddkey*{SubDelay}%
  {\ensuremath{\glsentrytext{\glslabel}_{\mathrm{d}}}}%
  {\glsentrySubDelay}%
  {\GlsentrySubDelay}%
  {\glsSubDelay}%
  {\GlsSubDelay}%
  {\GLSSubDelay}

\glsaddkey*{SubResponse}%
  {\ensuremath{\glsentrytext{\glslabel}_{\mathrm{r}}}}%
  {\glsentrySubResponse}%
  {\GlsentrySubResponse}%
  {\glsSubResponse}%
  {\GlsSubResponse}%
  {\GLSSubResponse}

\glsaddkey*{SubSubframe}%
  {\ensuremath{\glsentrytext{\glslabel}_{\mathrm{sf}}}}%
  {\glsentrySubSubframe}%
  {\GlsentrySubSubframe}%
  {\glsSubSubframe}%
  {\GlsSubSubframe}%
  {\GLSSubSubframe}

\glsaddkey*{SubOverhead}%
  {\ensuremath{\glsentrytext{\glslabel}_{\mathrm{OH}}}}%
  {\glsentrySubOverhead}%
  {\GlsentrySubOverhead}%
  {\glsSubOverhead}%
  {\GlsSubOverhead}%
  {\GLSSubOverhead}

\glsaddkey*{SubData}%
  {\ensuremath{\glsentrytext{\glslabel}_{\mathrm{dt}}}}%
  {\glsentrySubData}%
  {\GlsentrySubData}%
  {\glsSubData}%
  {\GlsSubData}%
  {\GLSSubData}

\glsaddkey*{SubDataSupPrime}%
  {\ensuremath{\glsentrytext{\glslabel}_{\mathrm{dt}}^{\prime}}}%
  {\glsentrySubDataSupPrime}%
  {\GlsentrySubDataSupPrime}%
  {\glsSubDataSupPrime}%
  {\GlsSubDataSupPrime}%
  {\GLSSubDataSupPrime}

\glsaddkey*{SubVar}%
  {\ensuremath{\glsentrytext{\glslabel}_{\mathrm{q}}}}%
  {\glsentrySubVar}%
  {\GlsentrySubVar}%
  {\glsSubVar}%
  {\GlsSubVar}%
  {\GLSSubVar}

\glsaddkey*{SubVarP}%
  {\ensuremath{\glsentrytext{\glslabel}_{\mathrm{p}}}}%
  {\glsentrySubVarP}%
  {\GlsentrySubVarP}%
  {\glsSubVarP}%
  {\GlsSubVarP}%
  {\GLSSubVarP}

\glsaddkey*{SubIncident}%
  {\ensuremath{\glsentrytext{\glslabel}_\mathrm{inc}}}%
  {\glsentrySubIncident}%
  {\GlsentrySubIncident}%
  {\glsSubIncident}%
  {\GlsSubIncident}%
  {\GLSSubIncident}

\glsaddkey*{SubReflected}%
  {\ensuremath{\glsentrytext{\glslabel}_\mathrm{ref}}}%
  {\glsentrySubReflected}%
  {\GlsentrySubReflected}%
  {\glsSubReflected}%
  {\GlsSubReflected}%
  {\GLSSubReflected}

\glsaddkey*{SubSaturation}%
  {\ensuremath{\glsentrytext{\glslabel}_{\mathrm{sat}}}}%
  {\glsentrySubSaturation}%
  {\GlsentrySubSaturation}%
  {\glsSubSaturation}%
  {\GlsSubSaturation}%
  {\GLSSubSaturation}

\glsaddkey*{SubTrainingSupMin}%
  {\ensuremath{\glsentrytext{\glslabel}_{\mathrm{bt}}^{\mathrm{min}}}}%
  {\glsentrySubTrainingSupMin}%
  {\GlsentrySubTrainingSupMin}%
  {\glsSubTrainingSupMin}%
  {\GlsSubTrainingSupMin}%
  {\GLSSubTrainingSupMin}

\glsaddkey*{SubTrainingSupMax}%
  {\ensuremath{\glsentrytext{\glslabel}_{\mathrm{bt}}^{\mathrm{max}}}}%
  {\glsentrySubTrainingSupMax}%
  {\GlsentrySubTrainingSupMax}%
  {\glsSubTrainingSupMax}%
  {\GlsSubTrainingSupMax}%
  {\GLSSubTrainingSupMax}

\glsaddkey*{SubDataSupMin}%
  {\ensuremath{\glsentrytext{\glslabel}_{\mathrm{dt}}^{\mathrm{min}}}}%
  {\glsentrySubDataSupMin}%
  {\GlsentrySubDataSupMin}%
  {\glsSubDataSupMin}%
  {\GlsSubDataSupMin}%
  {\GLSSubDataSupMin}

\glsaddkey*{SubHarvestingSupSin}%
  {\ensuremath{\glsentrytext{\glslabel}_{\mathrm{EH}}^{\mathrm{sin}}}}%
  {\glsentrySubHarvestingSupSin}%
  {\GlsentrySubHarvestingSupSin}%
  {\glsSubHarvestingSupSin}%
  {\GlsSubHarvestingSupSin}%
  {\GLSSubHarvestingSupSin}

\glsaddkey*{SubHarvestingSupSig}%
  {\ensuremath{\glsentrytext{\glslabel}_{\mathrm{EH}}^{\mathrm{sig}}}}%
  {\glsentrySubHarvestingSupSig}%
  {\GlsentrySubHarvestingSupSig}%
  {\glsSubHarvestingSupSig}%
  {\GlsSubHarvestingSupSig}%
  {\GLSSubHarvestingSupSig}

\glsaddkey*{SubVarSupSplitPower}%
  {\ensuremath{\glsentrytext{\glslabel}_{\mathrm{p}}^{\mathrm{PS}}}}%
  {\glsentrySubVarSupSplitPower}%
  {\GlsentrySubVarSupSplitPower}%
  {\glsSubVarSupSplitPower}%
  {\GlsSubVarSupSplitPower}%
  {\GLSSubVarSupSplitPower}

\glsaddkey*{SubVarSupSplitElement}%
  {\ensuremath{\glsentrytext{\glslabel}_{\mathrm{X}}^{\mathrm{ES}}}}%
  {\glsentrySubVarSupSplitElement}%
  {\GlsentrySubVarSupSplitElement}%
  {\glsSubVarSupSplitElement}%
  {\GlsSubVarSupSplitElement}%
  {\GLSSubVarSupSplitElement}

\glsaddkey*{SubVarSupSplitTime}%
  {\ensuremath{\glsentrytext{\glslabel}_{\mathrm{X}}^{\mathrm{TS}}}}%
  {\glsentrySubVarSupSplitTime}%
  {\GlsentrySubVarSupSplitTime}%
  {\glsSubVarSupSplitTime}%
  {\GlsSubVarSupSplitTime}%
  {\GLSSubVarSupSplitTime}

\glsaddkey*{SubVarSupMax}%
  {\ensuremath{\glsentrytext{\glslabel}_{\mathrm{q}}^{\mathrm{max}}}}%
  {\glsentrySubVarSupMax}%
  {\GlsentrySubVarSupMax}%
  {\glsSubVarSupMax}%
  {\GlsSubVarSupMax}%
  {\GLSSubVarSupMax}

\glsaddkey*{SubDataTransmissionSupSplitPower}%
  {\ensuremath{\glsentrytext{\glslabel}_{\mathrm{DT}}^{\mathrm{PS}}}}%
  {\glsentrySubDataTransmissionSupSplitPower}%
  {\GlsentrySubDataTransmissionSupSplitPower}%
  {\glsSubDataTransmissionSupSplitPower}%
  {\GlsSubDataTransmissionSupSplitPower}%
  {\GLSSubDataTransmissionSupSplitPower}

\glsaddkey*{SubDataTransmissionSupSplitElement}%
  {\ensuremath{\glsentrytext{\glslabel}_{\mathrm{DT}}^{\mathrm{ES}}}}%
  {\glsentrySubDataTransmissionSupSplitElement}%
  {\GlsentrySubDataTransmissionSupSplitElement}%
  {\glsSubDataTransmissionSupSplitElement}%
  {\GlsSubDataTransmissionSupSplitElement}%
  {\GLSSubDataTransmissionSupSplitElement}

\glsaddkey*{SubDataTransmissionSupSplitTime}%
  {\ensuremath{\glsentrytext{\glslabel}_{\mathrm{DT}}^{\mathrm{TS}}}}%
  {\glsentrySubDataTransmissionSupSplitTime}%
  {\GlsentrySubDataTransmissionSupSplitTime}%
  {\glsSubDataTransmissionSupSplitTime}%
  {\GlsSubDataTransmissionSupSplitTime}%
  {\GLSSubDataTransmissionSupSplitTime}

\glsaddkey*{SubTrainingSupSplitPower}%
  {\ensuremath{\glsentrytext{\glslabel}_{\mathrm{BT}}^{\mathrm{PS}}}}%
  {\glsentrySubTrainingSupSplitPower}%
  {\GlsentrySubTrainingSupSplitPower}%
  {\glsSubTrainingSupSplitPower}%
  {\GlsSubTrainingSupSplitPower}%
  {\GLSSubTrainingSupSplitPower}

\glsaddkey*{SubTrainingSupSplitElement}%
  {\ensuremath{\glsentrytext{\glslabel}_{\mathrm{BT}}^{\mathrm{ES}}}}%
  {\glsentrySubTrainingSupSplitElement}%
  {\GlsentrySubTrainingSupSplitElement}%
  {\glsSubTrainingSupSplitElement}%
  {\GlsSubTrainingSupSplitElement}%
  {\GLSSubTrainingSupSplitElement}

\glsaddkey*{SubTrainingSupSplitTime}%
  {\ensuremath{\glsentrytext{\glslabel}_{\mathrm{BT}}^{\mathrm{TS}}}}%
  {\glsentrySubTrainingSupSplitTime}%
  {\GlsentrySubTrainingSupSplitTime}%
  {\glsSubTrainingSupSplitTime}%
  {\GlsSubTrainingSupSplitTime}%
  {\GLSSubTrainingSupSplitTime}

\glsaddkey*{SubHarvestedSupSplitPower}%
  {\ensuremath{\glsentrytext{\glslabel}_{\mathrm{h}}^{\mathrm{PS}}}}%
  {\glsentrySubHarvestedSupSplitPower}%
  {\GlsentrySubHarvestedSupSplitPower}%
  {\glsSubHarvestedSupSplitPower}%
  {\GlsSubHarvestedSupSplitPower}%
  {\GLSSubHarvestedSupSplitPower}

\glsaddkey*{SubHarvestedSupSplitElement}%
  {\ensuremath{\glsentrytext{\glslabel}_{\mathrm{h}}^{\mathrm{ES}}}}%
  {\glsentrySubHarvestedSupSplitElement}%
  {\GlsentrySubHarvestedSupSplitElement}%
  {\glsSubHarvestedSupSplitElement}%
  {\GlsSubHarvestedSupSplitElement}%
  {\GLSSubHarvestedSupSplitElement}

\glsaddkey*{SubHarvestedSupSplitTime}%
  {\ensuremath{\glsentrytext{\glslabel}_{\mathrm{h}}^{\mathrm{TS}}}}%
  {\glsentrySubHarvestedSupSplitTime}%
  {\GlsentrySubHarvestedSupSplitTime}%
  {\glsSubHarvestedSupSplitTime}%
  {\GlsSubHarvestedSupSplitTime}%
  {\GLSSubHarvestedSupSplitTime}

\glsaddkey*{SubHarvestingSupAverage}%
  {\ensuremath{\glsentrytext{\glslabel}_{\mathrm{EH}}^{\mathrm{avg}}}}%
  {\glsentrySubHarvestingSupAverage}%
  {\GlsentrySubHarvestingSupAverage}%
  {\glsSubHarvestingSupAverage}%
  {\GlsSubHarvestingSupAverage}%
  {\GLSSubHarvestingSupAverage}

\glsaddkey*{SubHarvestingSupInstantaneous}%
  {\ensuremath{\glsentrytext{\glslabel}_{\mathrm{EH}}^{\mathrm{ins}}}}%
  {\glsentrySubHarvestingSupInstantaneous}%
  {\GlsentrySubHarvestingSupInstantaneous}%
  {\glsSubHarvestingSupInstantaneous}%
  {\GlsSubHarvestingSupInstantaneous}%
  {\GLSSubHarvestingSupInstantaneous}

\glsaddkey*{SubDcSupAverage}%
  {\ensuremath{\glsentrytext{\glslabel}_{\mathrm{DC}}^{\mathrm{avg}}}}%
  {\glsentrySubDcSupAverage}%
  {\GlsentrySubDcSupAverage}%
  {\glsSubDcSupAverage}%
  {\GlsSubDcSupAverage}%
  {\GLSSubDcSupAverage}

\glsaddkey*{SubDcSupInstantaneous}%
  {\ensuremath{\glsentrytext{\glslabel}_{\mathrm{DC}}^{\mathrm{ins}}}}%
  {\glsentrySubDcSupInstantaneous}%
  {\GlsentrySubDcSupInstantaneous}%
  {\glsSubDcSupInstantaneous}%
  {\GlsSubDcSupInstantaneous}%
  {\GLSSubDcSupInstantaneous}

\glsaddkey*{SubRfSupAverage}%
  {\ensuremath{\glsentrytext{\glslabel}_{\mathrm{RF}}^{\mathrm{avg}}}}%
  {\glsentrySubRfSupAverage}%
  {\GlsentrySubRfSupAverage}%
  {\glsSubRfSupAverage}%
  {\GlsSubRfSupAverage}%
  {\GLSSubRfSupAverage}

\glsaddkey*{SubRfSupInstantaneous}%
  {\ensuremath{\glsentrytext{\glslabel}_{\mathrm{RF}}^{\mathrm{ins}}}}%
  {\glsentrySubRfSupInstantaneous}%
  {\GlsentrySubRfSupInstantaneous}%
  {\glsSubRfSupInstantaneous}%
  {\GlsSubRfSupInstantaneous}%
  {\GLSSubRfSupInstantaneous}

%% file: symbols.tex
\glsxtrnewsymbol[
  description={Gain}
]{Gain}{\ensuremath{G}}

\glsxtrnewsymbol[
  description={Set of tiles}
]{SetTiles}{\ensuremath{\mathcal{T}}}
\glsxtrnewsymbol[
  description={Set of unit cells}
]{SetCells}{\ensuremath{\mathcal{U}}}
\glsxtrnewsymbol[
  description={Set of subareas}
]{SetLocations}{\ensuremath{\mathcal{L}}}

\glsxtrnewsymbol[
  description={Transmit Symbol}
]{Symbol}{\ensuremath{x}}

\glsxtrnewsymbol[
  description={Accuracy of grid search}
]{Accuracy}{\ensuremath{\alpha}}

\glsxtrnewsymbol[
  description={Exponent in DC loss}
]{LossExponent}{\ensuremath{\beta}}

\glsxtrnewsymbol[
  description={Codebook}
]{Codebook}{\ensuremath{\mathcal{B}}}
\glsxtrnewsymbol[
  description={Codebook size}
]{NumberCodewords}{\ensuremath{M}}
\glsxtrnewsymbol[
  description={Codebook size in Y direction}
]{NumberCodewordsY}{\ensuremath{M_y}}
\glsxtrnewsymbol[
  description={Codebook size in Z direction}
]{NumberCodewordsZ}{\ensuremath{M_z}}
\glsxtrnewsymbol[
  description={Number of codewords per training}
]{NumberCodewordsTraining}{\ensuremath{M_t}}

\glsxtrnewsymbol[
  description={Number of items}
]{Number}{\ensuremath{N}}

\glsxtrnewsymbol[
  description={n-th phase of the incident channel.}
]{PhaseIncident}{\ensuremath{\phi_{\mathrm{inc},n}}}
\glsxtrnewsymbol[
  description={n-th phase of the reflected channel.}
]{PhaseReflected}{\ensuremath{\phi_{\mathrm{ref},n}}}
\glsxtrnewsymbol[
  description={n-th phase shift at RIS.}
]{ShiftUnitCell}{\ensuremath{\theta_n}}
\glsxtrnewsymbol[
  description={n-th phase shift error at RIS for b-th codeword.}
]{ShiftErrorUnitCell}{\ensuremath{\tilde{\theta}_{n,b}}}
\glsxtrnewsymbol[
  description={n-th phase shift at RF combining.}
]{ShiftCombining}{\ensuremath{\varphi_n}}
\glsxtrnewsymbol[
  description={n-th phase shift error at RF combining.}
]{ShiftErrorCombining}{\ensuremath{\tilde{\varphi}_n}}
\glsxtrnewsymbol[
  description={Number of unit cells}
]{NumberUnitCells}{\ensuremath{N_\mathrm{UC}}}
\glsxtrnewsymbol[
  description={Number of unit cells in reflection mode}
]{NumberUnitCellsReflection}{\ensuremath{N_\mathrm{UC,r}}}
\glsxtrnewsymbol[
  description={Number of unit cells in harvesting mode}
]{NumberUnitCellsHarvesting}{\ensuremath{N_\mathrm{UC,h}}}
\glsxtrnewsymbol[
  description={Number of groups}
]{NumberGroups}{\ensuremath{N_\mathrm{TL}}}
\glsxtrnewsymbol[
  description={Gain of the RIS}
]{GainRis}{\ensuremath{G_\mathrm{RIS}}}
\glsxtrnewsymbol[
  description={Gain of a unit cell}
]{GainUnitCell}{\ensuremath{G_\mathrm{UC}}}
\glsxtrnewsymbol[
  description={Set of elements in tile.}
]{SetElementsTile}{\ensuremath{\mathcal{S}^m}}

\glsxtrnewsymbol[
  description={Ratio of unit cells in reflection mode}
]{RatioUnitCellsReflection}{\ensuremath{N_r/N_\mathrm{UC}}}
\glsxtrnewsymbol[
  description={Split ratio of the considered split policy.}
]{Split}{\ensuremath{\rho}}
\glsxtrnewsymbol[
  description={Ratio of unit cells in reflection mode}
]{RatioUnitCellsReflectionGrid}{\ensuremath{\rho_n}}

\glsxtrnewsymbol[
  description={Coverage area size}
]{SizeCoverageArea}{\ensuremath{A}}
\glsxtrnewsymbol[
  description={Coverage area size along \(y\) axis}
]{SizeCoverageAreaY}{\ensuremath{A_y}}
\glsxtrnewsymbol[
  description={Coverage area size along \(z\) axis}
]{SizeCoverageAreaZ}{\ensuremath{A_z}}
\glsxtrnewsymbol[
  description={Side length of subareas}
]{LengthSideSubarea}{\ensuremath{L}}

\glsxtrnewsymbol[
  description={Insertion loss of a phase shifter}
]{LossInsertion}{\ensuremath{\delta}}
\glsxtrnewsymbol[
  description={Quantization error of phase shifter}
]{ErrorPhase}{\ensuremath{\epsilon}}

\glsxtrnewsymbol[
  description={Time duration}
]{TimeDuration}{\ensuremath{T}}
\glsxtrnewsymbol[
  description={Frame model parameter}
]{ParameterFrame}{\ensuremath{\nu}}
\glsxtrnewsymbol[
  description={Number of subframes}
]{NumberSubframes}{\ensuremath{N_\mathrm{SF}}}

\glsxtrnewsymbol[
  description={Training overhead for FS}
]{OverheadTraining}{\ensuremath{\tau}}

\glsxtrnewsymbol[
  description={Power of a symbol.}
]{Power}{\ensuremath{P}}
\glsxtrnewsymbol[
  description={Power consumed.}
]{PowerConsumed}{\ensuremath{Q}}
\glsxtrnewsymbol[
  description={Average power of symbol alphabet.}
]{PowerAverage}{\ensuremath{\bar{P}}}
\glsxtrnewsymbol[
  description={Energy}
]{Energy}{\ensuremath{E}}
\glsxtrnewsymbol[
  description={Model function}
]{FunctionModel}{\ensuremath{h}}
\glsxtrnewsymbol[
  description={PDF}
]{FunctionPdf}{\ensuremath{f}}
\glsxtrnewsymbol[
  description={Harvested energy}
]{EnergyHarvested}{\ensuremath{E_h}}
\glsxtrnewsymbol[
  description={Consumed energy}
]{EnergyConsumed}{\ensuremath{E_c}}

\glsxtrnewsymbol[
  description={EH model function.}
]{ModelEnergyHarvesting}{\ensuremath{f_\text{EH}}}
\glsxtrnewsymbol[
  description={Lower bound of the EH model function.}
]{ModelEnergyHarvestingLowerBound}{\ensuremath{\hat{f}_\text{EH}}}
\glsxtrnewsymbol[
  description={Transmit power}
]{PowerTransmitterX}{\ensuremath{P_p}}
\glsxtrnewsymbol[
  description={Transmit power for phase one, training.}
]{PowerTransmitterTraining}{\ensuremath{P_1}}
\glsxtrnewsymbol[
  description={Transmit power for phase two, data.}
]{PowerTransmitterData}{\ensuremath{P_2}}
\glsxtrnewsymbol[
  description={Noise power at receiver}
]{PowerNoise}{\ensuremath{\sigma^2}}
\glsxtrnewsymbol[
  description={Normalized harvested energy}
]{EnergyNormalizedHarvested}{\ensuremath{\bar{E}_h}}
\glsxtrnewsymbol[
  description={Approximated harvested energy}
]{EnergyNormalizedHarvestedApproximation}{\ensuremath{\hat{E}_h}}
\glsxtrnewsymbol[
  description={Factor for approximated harvested energy}
]{FactorEnergyNormalizedHarvestedApproximation}{\ensuremath{\tilde{E}_h}}
\glsxtrnewsymbol[
  description={Normalized consumed energy}
]{EnergyNormalizedConsumed}{\ensuremath{\bar{E}_c}}
\glsxtrnewsymbol[
  description={Normalized consumed energy at the transmitter}
]{EnergyNormalizedTransmitter}{\ensuremath{\bar{E}_t}}
\glsxtrnewsymbol[
  description={Consumed energy per unit cell reconfiguration}
]{EnergyReconfiguration}{\ensuremath{E_r}}
\glsxtrnewsymbol[
  description={Consumed power per controller}
]{PowerController}{\ensuremath{P_c}}
\glsxtrnewsymbol[
  description={RF input power for harvesting with RF combining}
]{PowerInCombRf}{\ensuremath{P_\text{in,RF}}}
\glsxtrnewsymbol[
  description={RF input power for harvesting with DC combining}
]{PowerInCombDc}{\ensuremath{P_\text{in,DC}}}
\glsxtrnewsymbol[
  description={RF input power for harvesting with X combining}
]{PowerInCombX}{\ensuremath{P_\text{in,q}}}
\glsxtrnewsymbol[
  description={RF input power for harvesting}
]{PowerIn}{\ensuremath{P_\text{in}}}
\glsxtrnewsymbol[
  description={RF input power for harvesting per element}
]{PowerRfElement}{\ensuremath{P_{\text{in},n}}}
\glsxtrnewsymbol[
  description={DC power from RF harvesting with RF combining}
]{PowerOutCombRf}{\ensuremath{P_\text{out,RF}}}
\glsxtrnewsymbol[
  description={DC power from RF harvesting with DC combining}
]{PowerOutCombDc}{\ensuremath{P_\text{out,DC}}}
\glsxtrnewsymbol[
  description={DC power from RF harvesting with X combining}
]{PowerOutCombX}{\ensuremath{P_\text{out,c}}}
\glsxtrnewsymbol[
  description={DC output power from harvesting}
]{PowerOut}{\ensuremath{P_\text{out}}}
\glsxtrnewsymbol[
  description={Bound of DC power from RF harvesting with X combining}
]{PowerOutBoundCombX}{\ensuremath{\bar{P}_\text{out,c}}}

\glsxtrnewsymbol[
  description={SNR}
]{Snr}{\ensuremath{\gamma}}
\glsxtrnewsymbol[
  description={Effective communication rate}
]{RateEffective}{\ensuremath{R}}

\glsxtrnewsymbol[
  description={Normalized achievable SNR for data symbol}
]{SnrFactorData}{\ensuremath{\tilde{\gamma}}}

\glsxtrnewsymbol[
  description={Pathloss}
]{Pathloss}{\ensuremath{L}}
\glsxtrnewsymbol[
  description={Average pathloss}
]{PathlossAverage}{\ensuremath{\bar{L}}}

\glsxtrnewsymbol[
  description={Gain of transmit antenna}
]{GainTransmitter}{\ensuremath{G_\text{TX}}}
\glsxtrnewsymbol[
  description={Gain of receive antenna}
]{GainReceiver}{\ensuremath{G_\text{RX}}}
\glsxtrnewsymbol[
  description={Common pathloss of incident channel}
]{PathlossCommonIncident}{\ensuremath{\beta_{i}}}
\glsxtrnewsymbol[
  description={Pathloss of incident channel}
]{PathlossIncident}{\ensuremath{\beta_{i,m}}}
\glsxtrnewsymbol[
  description={Pathloss of reflected channel}
]{PathlossReflected}{\ensuremath{\beta_{r,m}}}

\glsxtrnewsymbol[
  description={Distance between RIS and user}
]{DistanceRisUser}{\ensuremath{d_r}}
\glsxtrnewsymbol[
  description={Distance between RIS and AP}
]{DistanceRisAp}{\ensuremath{d_i}}
\glsxtrnewsymbol[
  description={Velocity of mobile user}
]{Velocity}{\ensuremath{v}}

\glsxtrnewsymbol[
  description={Loss in effective communication rate}
]{RateEffectiveLoss}{\ensuremath{\tilde{R}}}
\glsxtrnewsymbol[
  description={Maximum effective communication rate}
]{RateEffectiveOptimal}{\ensuremath{R^{\star}}}
\glsxtrnewsymbol[
  description={Efficient of energy harvesting}
]{EfficiencyHarvesting}{\ensuremath{\eta}}

\glsxtrnewsymbol[
  description={Average beam width}
]{WidthBeam}{\ensuremath{w}}
\glsxtrnewsymbol[
  description={Average beam width}
]{WidthBeamAverage}{\ensuremath{\bar{w}}}

\glsxtrnewsymbol[
  description={Max side length of subareas}
]{LengthSideSubareaMax}{\ensuremath{L^{\text{max}}}}
\glsxtrnewsymbol[
  description={Distance between RIS and center of coverage area}
]{DistanceRisAreaCenter}{\ensuremath{d_A}}
\glsxtrnewsymbol[
  description={Distance between RIS and center of the \(m\)th subarea}
]{DistanceRisSubarea}{\ensuremath{d_m}}
\glsxtrnewsymbol[
  description={Wavelength}
]{Wavelength}{\ensuremath{\lambda}}
\glsxtrnewsymbol[
  description={Number of unit cell reconfigurations per frame}
]{NumberReconfigurations}{\ensuremath{N_c}}

\glsxtrnewsymbol[
  description={Sigmoid function for non-linear energy harvesting.}
]{SigmoidHarvesting}{\ensuremath{\Psi}}
\glsxtrnewsymbol[
  description={Compensation function for non-linear energy harvesting.}
]{SigmoidCompensationHarvesting}{\ensuremath{\Omega}}
\glsxtrnewsymbol[
  description={Power saturation for non-linear energy harvesting.}
]{SaturationHarvesting}{\ensuremath{P^\text{sat}}}
\glsxtrnewsymbol[
  description={Steepness of sigmoid function.}
]{SigmoidSteepness}{\ensuremath{a}}
\glsxtrnewsymbol[
  description={Turn-on of sigmoid function.}
]{SigmoidTurnOn}{\ensuremath{b}}
\glsxtrnewsymbol[
  description={K-Factor of Rician channel.}
]{FactorK}{\ensuremath{K}}
\glsxtrnewsymbol[
  description={Number pilots estimation.}
]{NumberPilotsEstimation}{\ensuremath{N_e}}
\glsxtrnewsymbol[
  description={Number pilots training.}
]{NumberPilotsTraining}{\ensuremath{N_t}}
\glsxtrnewsymbol[
  description={Phase variation for incident channel.}
]{PhaseVariationIncident}{\ensuremath{A_{q_x,q_y}^{i,m}}}
\glsxtrnewsymbol[
  description={Phase variation for reflected channel.}
]{PhaseVariationReflected}{\ensuremath{A_{q_x,q_y}^{r,m}}}
\glsxtrnewsymbol[
  description={Phase variation for general channel.}
]{PhaseVariationGeneral}{\ensuremath{A_{q_x,q_y}^{s,m}}}
\glsxtrnewsymbol[
  description={Elevation angle.}
]{AngleElevation}{\ensuremath{\theta_{s,m}}}
\glsxtrnewsymbol[
  description={Elevation angle.}
]{AngleElevationReflected}{\ensuremath{\theta_{r,m}}}
\glsxtrnewsymbol[
  description={Elevation angle.}
]{AngleElevationIncident}{\ensuremath{\theta_{i,m}}}
\glsxtrnewsymbol[
  description={Azimuth angle.}
]{AngleAzimuth}{\ensuremath{\phi_{s,m}}}
\glsxtrnewsymbol[
  description={Azimuth angle.}
]{AngleAzimuthReflected}{\ensuremath{\phi_{r,m}}}
\glsxtrnewsymbol[
  description={Azimuth angle.}
]{AngleAzimuthIncident}{\ensuremath{\phi_{i,m}}}
\glsxtrnewsymbol[
  description={Phase shift of \(m\)th codeword.}
]{PhaseShiftCodeword}{\ensuremath{\varphi_{q_x,q_y}^{m}}}
\glsxtrnewsymbol[
  description={Center location of subarea \(m\).}
]{CenterSubarea}{\ensuremath{c_m}}
\glsxtrnewsymbol[
  description={Side length of unit cell.}
]{LengthSideUnitCell}{\ensuremath{\delta}}

\glsxtrnewsymbol[
  description={Element of argument vector of harvested power}
]{ArgumentPowerHarvested}{\ensuremath{p}}
\glsxtrnewsymbol[
  description={Auxiliary optimization variable}
]{AuxiliaryVariable}{\ensuremath{z}}

%% file: macros.tex
\newcommand{\stage}{q}
\newcommand{\stageEh}{\mathrm{eh}}
\newcommand{\stageBt}{\mathrm{bt}}
\newcommand{\stageDt}{\mathrm{dt}}

\newcommand{\policy}{p}
\newcommand{\policyPs}{\mathrm{PS}}
\newcommand{\policyEs}{\mathrm{ES}}
\newcommand{\policyTs}{\mathrm{TS}}

\newcommand{\ris}{\mathrm{ris}}
\newcommand{\tile}{\mathrm{tl}}

\newcommand{\harvesting}{\mathrm{h}}
\newcommand{\reflecting}{\mathrm{r}}
\newcommand{\consuming}{\mathrm{c}}
\newcommand{\sym}{\mathrm{s}}

\newcommand{\dc}{\mathrm{dc}}
\newcommand{\rf}{\mathrm{rf}}
\newcommand{\shifter}{\mathrm{sh}}
\newcommand{\cell}{\mathrm{uc}}
\newcommand{\fr}{\mathrm{fr}}
\newcommand{\bs}{\mathrm{bs}}
\newcommand{\diode}{\mathrm{on}}
\newcommand{\bits}{\mathrm{B}}

\newcommand{\threshold}{\mathrm{thr}}
\newcommand{\maximum}{\mathrm{max}}
\newcommand{\minimum}{\mathrm{min}}
\newcommand{\static}{\mathrm{sta}}

\newcommand{\incident}{\mathrm{inc}}
\newcommand{\reflected}{\mathrm{ref}}

\newcommand{\TileGainMin}{\gls{Gain}_{\tile,\minimum}}
\newcommand{\TileGainMax}{\gls{Gain}_{\tile,\maximum}}

\newcommand{\Ucr}{_{\mathrm{uc},\reflecting}}
\newcommand{\RfVarVar}{_{\rf,\stage,m}^\policy}
\newcommand{\DcVarVar}{_{\dc,\stage,m}^\policy}
\newcommand{\PdfDt}{\gls{FunctionPdf}_{\gls{Power}_\stageDt^\policy(b)}(\gls{Power}_\stageDt^\policy(b); \gls{PowerAverage}_\stageDt^\policy(b))}
\newcommand{\PdfRfDt}{\gls{FunctionPdf}_{\gls{Power}_{\rf,\stageDt,m}^\policy(b)}(\gls{Power}_{\rf,\stageDt,m}^\policy(b); \gls{PowerAverage}_{\rf,\stageDt,m}^\policy(b))}
\newcommand{\PdfX}{\gls{FunctionPdf}_x(x; \bar{x})}

%% file: sections/introduction.tex
\noindent
\IEEEPARstart{I}{n} recent years, \glspl{ris} have gained significant attention in both academia and industry as they have the ability to improve the system performance of future wireless communication networks.
In particular, the link quality between two communication nodes can be enhanced by appropriate configuration of the \gls{ris} unit cells.
For example, each unit cell can be configured to impose a specific phase shift on the impinging electromagnetic wave, which results in signal reflection in a desired direction.
The key challenge is to determine the optimal phase-shift configuration with low computational overhead.
Recent studies have addressed this issue by, e.g., adopting multi-agent deep reinforcement learning to optimize the beamforming codebooks for a \gls{ris}-assisted communication network in the \gls{mmwave} band~\cite{abdallah2024multiagentdeep}.
Moreover, the optimal configuration of both active and passive \gls{ris} unit cells was investigated for batteryless \gls{iot} networks~\cite{ahmed2024revolutionizingbatterylessiot} and a multi-cell scenario with simultaneously transmitting and reflecting \glspl{ris}~\cite{ahmed2025optimizingsmallcell}.
The configuration of a beyond-diagonal \gls{ris} was recently analyzed in~\cite{sousadesena2025diagonalrismulti}, where the authors optimized the scattering matrix considering the frequency-dependent characteristics of the \gls{ris}.

While having great potential for improving the performance of wireless communication networks, one of the key advantages of \glspl{ris} is their low power consumption.
In fact, the power consumption of a \gls{ris} is much lower than that of a \gls{bs} or an active relay because a \gls{ris} does not require \gls{rf} chains, baseband processing, or power amplifiers.
Nevertheless, a power supply must still be available for \gls{ris} operation.
However, since not all potential locations for \gls{ris} deployment provide access to the power grid, batteries or solar panels may need to be attached to the \gls{ris}, leading to higher cost for installation and maintenance.

As an alternative, \glspl{ris} with energy harvesting capabilities, also known as \emph{autonomous \gls{ris}}, \emph{self-sustainable \gls{ris}}, \emph{zero-energy \gls{ris}}, or \emph{perpetual \gls{ris}}, have recently been proposed~\cite{wu2022intelligentreflectingsurface,hu2021robustsecuresum,zou2022robustbeamformingoptimization,zheng2023zeroenergydevice,ma2023beamformingoptimizationreconfigurable,ma2023wirelesspoweredintelligent,xu2022selfsustainablewireless,ma2022reconfigurableintelligentsurface,liu2022jointtrajectorydesign,peng2023energyharvestingreconfigurable,xie2023sumratemaximization,wang2024multifunctionalreconfigurable,magbool2025multifunctionalris,ntontin2022millimeterwavevs,ntontin2022autonomousreconfigurableintelligent,ntontin2023timeunitcell,ntontin2023wirelessenergyharvesting,tyrovolas2023zeroenergyreconfigurablea,youn2023liquidcrystaldriven,albanese2024aresautonomousris}.
The idea is to split the impinging signal at the \gls{ris} and only reflect part of the signal to the receiver, while the other part is fed to a rectifier circuit for energy harvesting~\cite{wu2022intelligentreflectingsurface}.
Since the transmitter-\gls{ris} channel is often characterized by a strong \gls{los}, such application of \gls{rf} energy harvesting is particularly suitable for \gls{ris}-assisted networks~\cite{clerckx2022foundationswirelessinformation}.
In this way, the \gls{ris} becomes independent of external power supply.

In the context of self-sustainable \gls{ris}, most works in the literature study the joint optimization of active beamforming at the transmitter and passive beamforming at the \gls{ris} for various use cases.
For example, beamforming that is robust to imperfect \gls{csi} was investigated in~\cite{hu2021robustsecuresum,zou2022robustbeamformingoptimization,zheng2023zeroenergydevice}.
Moreover, the authors of~\cite{ma2023beamformingoptimizationreconfigurable,ma2023wirelesspoweredintelligent} focused on broadcast channels, whereas self-sustainable \glspl{ris} for backscatter communication were studied in~\cite{xu2022selfsustainablewireless,ma2022reconfigurableintelligentsurface}.
Furthermore, the authors of~\cite{liu2022jointtrajectorydesign,peng2023energyharvestingreconfigurable,xie2023sumratemaximization} used self-sustainable \glspl{ris} to optimize \gls{uav}-assisted networks, and the combination of energy harvesting and multi-functional \glspl{ris} was analyzed in~\cite{wang2024multifunctionalreconfigurable,magbool2025multifunctionalris}.

These existing works typically assume that the incident signal is split in the time domain, element domain, or power domain.
Moreover, the signals intended for energy harvesting are combined in either the \gls{rf} domain or the \gls{dc} domain, and the rectifier is usually modeled by a linear or non-linear function.
Furthermore, the model adopted for the \gls{ris} power consumption in most works scales linearly with the number of unit cells.

The authors of \cite{ntontin2022millimeterwavevs,ntontin2022autonomousreconfigurableintelligent,ntontin2023timeunitcell,ntontin2023wirelessenergyharvesting} proposed a dynamic power consumption model for the \gls{ris}, which includes not only the static power consumption resulting from the \gls{ris} control circuits, but also the additional energy required to adapt the \gls{ris} phase shifts to the instantaneous channels.
In this context, the optimal configuration and placement of the \gls{ris} was studied in~\cite{ntontin2022autonomousreconfigurableintelligent}, and the energy harvesting efficiencies at \gls{mmwave} and THz frequencies were compared in~\cite{ntontin2022millimeterwavevs}.

Furthermore, the performances for \gls{TimeSplit}, \gls{ElementSplit}, and \gls{PowerSplit} were investigated in~\cite{ntontin2023timeunitcell}, which revealed that \gls{ElementSplit} and \gls{PowerSplit} provide higher data rates for the considered range of \gls{ris} power consumption.
A similar comparison based on the joint energy-data rate outage probability was performed in~\cite{tyrovolas2023zeroenergyreconfigurablea}, where the authors concluded that \gls{PowerSplit} provides the best energy efficiency, but the optimal splitting scheme generally depends on the \gls{ris} placement or target data rate at the user.

Under the constraint of self-sustainable operation, the above works mainly focused on the optimal configuration of the individual \gls{ris} phase shifts, resulting in an overhead that scales with the number of unit cells.
Alternatively, a \gls{ris} can be configured based on a predefined codebook by means of beam training.
Especially for large \glspl{ris}, beam training is an efficient approach to \gls{ris} configuration because the overhead scales only with the number of codewords, which is typically much smaller than the number of unit cells~\cite{an2022codebookbasedsolutions}.

In the context of beam training and codebook-based configuration, only a few works have studied self-sustainable \glspl{ris}~\cite{youn2023liquidcrystaldriven,albanese2024aresautonomousris}, where the focus was mainly on self-configuration.
For example, a \gls{PowerSplit} scheme was considered in~\cite{albanese2024aresautonomousris}, where \gls{dc} power measurements and a \gls{ris} codebook are used to scan the three-dimensional space for active network devices.
Based on these power measurements, the \glspl{aoa} can be estimated at the \gls{ris}, allowing the configuration of the \gls{ris} unit cells without external control signaling.
A similar concept was experimentally validated in~\cite{youn2023liquidcrystaldriven}.
However, as these works focus on self-configuration, they do not explore the full potential of codebook-based configuration for self-sustainable \glspl{ris}.
In particular, the concept of self-configuration is different from conventional beam training in terms of, e.g., frame structure and received signal strength.
Moreover, the results in \cite{albanese2024aresautonomousris} are based on a \gls{PowerSplit} scheme with a fixed splitting ratio, and the authors did not attempt to optimize the splitting ratio and did not consider other splitting schemes.
Similarly, a combining network in the \gls{rf} domain was assumed for energy harvesting, but the impact of \gls{dc} combining or hybrid combining was not investigated.
In addition, although \gls{ris} deployment seems most beneficial for high operating frequencies~\cite{yang2024limitations5gris}, a performance analysis considering rectifiers explicitly designed for energy harvesting in the \gls{mmwave} band is missing.

Motivated by the above considerations, this paper studies energy harvesting at \glspl{ris} that are configured based on beam training.
More specifically, we focus on a self-sustainable \gls{ris} that extends the coverage of a \gls{bs} operating in the \gls{mmwave} band.
To evaluate the energy harvesting capabilities of the \gls{ris}, we examine the three common splitting schemes, namely \gls{PowerSplit}, \gls{ElementSplit}, and \gls{TimeSplit}.
Moreover, we adopt a tile-based architecture for the \gls{ris} in order to highlight the trade-off between \gls{rf} combining and \gls{dc} combining.
Our goal is to determine the optimal splitting ratio as well as the optimal power allocation for each codeword that minimizes the average transmit power at the \gls{bs}.
The main contributions of this paper can be summarized as follows:
\begin{itemize}
    \item We propose a mathematical framework for studying self-sustainable \glspl{ris} that are configured based on beam training, which can be applied to \gls{PowerSplit}, \gls{ElementSplit}, and \gls{TimeSplit} for energy harvesting. The considered codebook of \gls{ris} phase shifts is designed to fully cover a given area of interest.
    \item We provide a comprehensive model for energy harvesting at the \gls{ris} including a tile-based architecture that implements hybrid combining. The model takes into account the gains and losses induced by both \gls{rf} and \gls{dc} combining as well as the power consumption for both signal reflection and energy harvesting at the \gls{ris}.
    \item Based on a rectifier model that is specifically adapted to \gls{mmwave} applications, we show that the different transmit strategies used for beam training and data transmission lead to different rectifier characteristics, which has an impact on the efficiency of the \gls{rf}-\gls{dc} conversion during energy harvesting.
    \item We formulate an optimization problem to minimize the average transmit power at the \gls{bs}. We show that the optimization problem can efficiently be solved by a grid search over one optimization variable, which yields the optimal power allocation at the \gls{bs} for each codeword as well as the optimal splitting ratio for each considered splitting scheme.
    \item We evaluate the optimized splitting schemes under different assumptions on the \gls{ris} power consumption, revealing the characteristics of each considered splitting scheme.
    In particular, the \gls{TimeSplit} scheme typically benefits from efficient \gls{rf}-\gls{dc} conversion and the \gls{PowerSplit} scheme performs well for static power consumption. Due to its restricted feasible set, the \gls{ElementSplit} scheme may have limited performance but is robust to changes in the power model.
    Finally, our analysis shows that the number of tiles at the \gls{ris} has a strong impact on the transmit power required at the \gls{bs}.
\end{itemize}

Different from its conference version~\cite{laue2023beamtrainingself}, this paper adopts a tile-based architecture for the \gls{ris} and analyzes energy harvesting at the \gls{ris} based on hybrid combining.
Thereby, the gains and losses for both \gls{dc} and \gls{rf} combining are taken into account.
Moreover, our analysis includes the effect of the transmit strategies employed during beam training and data transmission, respectively, leading to different efficiencies of the \gls{rf}-\gls{dc} conversion.
Furthermore, in contrast to~\cite{laue2023beamtrainingself}, here we provide optimal solutions for both the power allocation at the \gls{bs} and the signal split at the \gls{ris}, considering the three common splitting schemes, i.e., \gls{TimeSplit}, \gls{ElementSplit}, and \gls{PowerSplit}.

The remainder of this paper is organized as follows. In Section~\ref{sec:system}, we describe the considered communication system, present the beam training model, introduce the considered splitting schemes, and formulate the optimization problem for optimal system design.
Section~\ref{sec:harvesting} provides the details of the energy harvesting at the \gls{ris}, including models for the rectifier and power consumption.
These models are adopted in Section~\ref{sec:optimization} to solve the optimization problem for each considered splitting scheme.
Finally, we evaluate the performance of the optimized system designs in Section~\ref{sec:evaluation}.
Conclusions are drawn in Section~\ref{sec:conclusion}.

\emph{Notations}: Variables and constants are denoted by small or capital letters, sets are denoted by calligraphic letters, and \(\abs*{\mathcal{X}}\) represents the cardinality of a set \(\mathcal{X}\).
The complex Gaussian distribution with mean value \(m\) and variance \(C\) is denoted by \(\CN{m, C}\).
Moreover, \(\ln(x)\) and \(f^{-1}(x)\) denote the natural logarithm of \(x\) and the functional inverse of function \(f(x)\), respectively.
The sets of positive integers and real numbers are denoted by \(\mathbb{N}\) and \(\mathbb{R}\), respectively.
The operator \(x \gtrapprox y\) indicates that the inequality holds and \(x\) and \(y\) are approximately equal.
Furthermore, \(\ceil*{x}\) denotes the ceiling function of \(x\), giving the smallest integer greater than or equal to \(x\). Similarly, \(\floor*{x}\) denotes the floor function of \(x\), returning the largest integer less than or equal to \(x\).

%% file: sections/system_model.tex
\begin{figure}[!t]
    \centering
    \begin{tikzpicture}[font=\footnotesize]
        \def\risposx{0cm}
        \def\risposy{0cm}
        \coordinate (rispos) at (\risposx,\risposy);
        \coordinate (bspos) at ([xshift=-1cm,yshift=-2.5cm]\risposx,\risposy);
        \coordinate (areacenter) at (3cm,-3cm);
        \def\numtilessqrt{3}
        \def\numcellssqrt{2}
        \def\areawidth{3.5cm}
        \def\arealength{2.5cm}
        \def\beamwidth{5.6mm}
        \def\footprintwidth{1.2cm}
        \def\footprintheight{1.6cm}
        \def\subareawidth{\areawidth / 3}
        \def\subareaheight{\arealength / 3}
        \def\standwidth{2mm}
        \def\standheight{8mm}
        \def\bswidth{0.7cm}
        \def\antennaheight{1cm}
        \def\cellsize{2mm}
        \def\cellspacing{0.5mm}
        \def\tilesize{\numcellssqrt*\cellsize + (1+\numcellssqrt)*\cellspacing}
        \def\tilespacing{0mm}
        \def\rissize{\numtilessqrt*(\tilesize) + (1+\numtilessqrt)*\tilespacing}

        \node [draw,rectangle,minimum size=\rissize,fill=BaseDarkColorA] (ris) at (rispos) {};
        \foreach \m in {1,2,...,\numtilessqrt} {
            \foreach \n in {1,2,...,\numtilessqrt} {
                \node [draw,rectangle,minimum size=\tilesize,fill=BaseDarkColorB] (tile\m\n) at ([xshift=(\m - 0.5 - \numtilessqrt / 2)*(\tilesize + \tilespacing),yshift=(\n - 0.5 - \numtilessqrt / 2)*(\tilesize + \tilespacing)]ris) {};

                \foreach \k in {1,2,...,\numcellssqrt} {
                    \foreach \l in {1,2,...,\numcellssqrt} {
                        \node [draw,rectangle,minimum size=\cellsize,inner sep=0cm,fill=BaseDarkColorD] (cell\k\l) at ([xshift=(\k - 0.5 - \numcellssqrt / 2)*(\cellsize + \cellspacing),yshift=(\l - 0.5 - \numcellssqrt / 2)*(\cellsize + \cellspacing)]tile\m\n) {};
                    }
                }
            }
        }
        \node [draw,rectangle,minimum width=0.8*\standwidth,minimum height=\standheight,fill=BaseDarkColor,inner sep=0mm,anchor=north] (risstand) at (ris.south) {};
        \node [draw,rectangle,minimum width=\standwidth,minimum height=0.7*\standheight,fill=BaseDarkColor,inner sep=0mm,anchor=south] at (risstand.south) {};
        \node [anchor=north] at (risstand.south) {\glsxtrshort{ris}};

        \node [draw,fill=BaseColorD] (battery) at ([xshift=0.9*\rissize]ris.center) {\faBatteryThreeQuarters};
        \draw [-Latex] ([yshift=-0.3*\rissize]ris.east) -| (battery.south);
        \draw [-Latex] (battery.north) |- ([yshift=0.3*\rissize]ris.east);        

        \node [draw,rectangle,fill=BaseColorB,xslant=-0.5,minimum height=\arealength,minimum width=\areawidth] (area) at (areacenter) {};
        \draw [dotted,xslant=-0.5] ([xshift=-\areawidth / 6]area.north) -- ++(0cm,-\arealength) ([xshift=\areawidth / 6]area.north) -- ++(0cm,-\arealength);
        \draw [dotted,xslant=-0.5] ([yshift=-\arealength / 6]area.west) -- ++(\areawidth,0cm) ([yshift=\arealength / 6]area.west) -- ++(\areawidth,0cm);
        \draw [shorten <=-0.5mm,{Circle[length=1mm]}-] ([xshift=-0.25*\areawidth]area.north east) -- +(2mm,6mm) node [above,inner sep=1mm] {coverage area};
        \draw [|<->|,xslant=-0.5] ([yshift=1mm]area.north west) -- ([yshift=1mm]area.north east) node [midway,above] {\(\gls{SizeCoverageAreaY}\)};
        \draw [|<->|,xslant=-0.5] ([xshift=-1mm]area.north west) -- ([xshift=-2mm]area.south west) node [midway,left] {\(\gls{SizeCoverageAreaZ}\)};
        \path (areacenter) node [draw,cross out,minimum size=1mm,xslant=-0.5,inner sep=0mm] {} ([xshift=3mm,yshift=-2mm]areacenter) node [xslant=-0.5,font=\footnotesize] {\((x_c,y_c,z_c)\)};

        \node [minimum width=\subareawidth,minimum height=\subareaheight,draw=none,fill=none,xslant=-0.5,anchor=south east] (subareabottomright) at (area.south east) {};
        \draw [|<->|] ([yshift=-1mm]subareabottomright.south west) -- ([yshift=-1mm]subareabottomright.south east) node [midway,below] {\(\gls{LengthSideSubarea}_{y,\policy}\)};
        \draw [|<->|] ([xshift=1mm]subareabottomright.north east) -- ([xshift=1mm]subareabottomright.south east) node [midway,right] {\(\gls{LengthSideSubarea}_{z,\policy}\)};

        \draw[dash dot,{Rays[]}-{Triangle[]},thick,xslant=-0.5] ([yshift=0.4*\arealength]area.center) .. controls ([yshift=-0.15*\arealength]area.east) .. ([xshift=-0.1*\areawidth,yshift=0.25*\arealength]area.south) coordinate [pos=0.4] (postrajectory);
        \draw [shorten <=-0.5mm,{Circle[length=1mm]}-] (postrajectory) -- +(2mm,15mm) node [above,inner sep=1mm] {user trajectory};

        \node [minimum width=\subareawidth,minimum height=\subareaheight,fill=none,draw=none,anchor=south west,xslant=-0.5] (subarea) at (area.south west) {};
        \node [ellipse,minimum width=\footprintwidth,minimum height=\footprintheight,rotate around={33:(subarea.center)},fill=BaseColorD,opacity=0.8] (footprint) at (subarea) {};
        \draw [{Circle[length=1mm]}-,shorten <=-0.5mm] ([xshift=-1mm,yshift=-3mm]footprint.center) -- +(-5mm, -4mm) node [left,inner sep=1mm] {beam footprint};
        
        \path pic [rotate=-115] {
            newbeam={(\risposx,\risposy)}{20mm}{5mm}{5mm}{BaseColor}
        };
        \path pic [rotate=-59.5] {
            newbeam={(\risposx,\risposy)}{21mm}{5mm}{5mm}{BaseColor}
        };
        \draw [dashed] (ris) -- (footprint.west);
        \draw [dashed] (ris) -- (footprint.east);
        \draw [|-|] ([shift=(-59.5+\beamwidth/2:29mm)]ris) arc [start angle=-59.5+\beamwidth/2,end angle=-59.5-\beamwidth/2,radius=29mm] node [pos=0.6,above] {\(\Delta \theta\)};
    
        \node [draw,minimum width=\bswidth,minimum height=0.7*\antennaheight] (bs) at (bspos) {};
        \node [draw,rectangle,minimum width=0.8*\standwidth,minimum height=1.5*\standheight,fill=BaseDarkColor,inner sep=0mm,anchor=north] (bsstand) at (bs.south) {};
        \node [draw,rectangle,minimum width=\standwidth,minimum height=0.7*1.5*\standheight,fill=BaseDarkColor,inner sep=0mm,anchor=south] at (bsstand.south) {};
        \node [draw,fill=BaseColorC,minimum width=0.4*\bswidth,minimum height=\antennaheight] at (bs.west) {};
        \node [draw,fill=BaseColorC,minimum width=0.4*\bswidth,minimum height=\antennaheight] at (bs.center) {};
        \node [draw,fill=BaseColorC,minimum width=0.4*\bswidth,minimum height=\antennaheight] at (bs.east) {};
        \node [anchor=north] at (bsstand.south) {\glsxtrshort{bs}};

        \draw[->] (4.5cm,1mm) -- +(-1cm,0mm) node[below] {$y$};
        \draw[->] (4.5cm,1mm) -- +(0mm,1cm) node[right] {$x$};
        \draw[->] (4.5cm,1mm) -- +(2.5mm,-5mm) node[right] {$z$};
    \end{tikzpicture}
    \caption{\Glsxtrshort{ris}-assisted communication system with a partitioned coverage area and a mobile user.}
    \label{fig:system}
\end{figure}
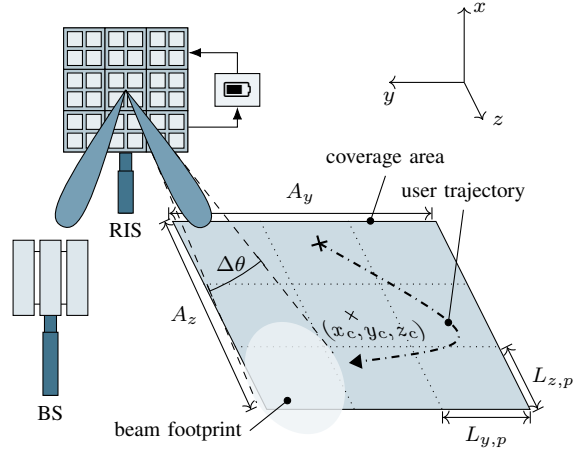
\noindent
As illustrated in Fig.~\ref{fig:system}, we consider a downlink communication system comprising a multiple-antenna \gls{bs}, a passive square-shaped \gls{ris}, and a mobile user, where the \gls{ris} is employed to overcome severe blockage of the direct \gls{bs}-user link.
In particular, the \gls{ris} reflects the incident wave from the \gls{bs} to a given coverage area of interest, where a mobile user travels on an unknown trajectory.
Since the \gls{bs} directs the transmitted signal to the \gls{ris} using beamforming, the end-to-end transmission can be modeled as an equivalent \gls{siso} link with a corresponding beamforming gain at the transmitter.

We assume a rectangular coverage area of size \(\gls{SizeCoverageArea} = \gls{SizeCoverageAreaY} \gls{SizeCoverageAreaZ}\) that is located in the \(y\)-\(z\) plane and centered at coordinates \((x_c, y_c, z_c)\), considering a Cartesian coordinate system whose origin is located at the center of the \gls{ris}.
The set of all locations within the coverage area is denoted by \(\gls{SetLocations}\).

Moreover, the \gls{ris} is composed of a set of tiles, denoted by \(\gls{SetTiles}\), and each tile comprises one or multiple unit cells, cf. Fig.~\ref{fig:system}. The indices of the unit cells forming the \(m\)th tile are collected in set \(\gls{SetCells}_m\), \(m \in \gls{SetTiles}\), such that the total number of unit cells is given by \(\gls{Number}_\cell = \sum_{m \in \gls{SetTiles}} \abs*{\gls{SetCells}_m}\). Since it is convenient from a manufacturing point of view, we assume the same number of unit cells for each tile.
In addition, we restrict our analysis to square-shaped tiles for simplicity.
Hence, as the \gls{ris} itself is square-shaped, the feasible set for the total number of tiles is given by
\(
\abs*{\gls{SetTiles}} = \gls{Number}_\tile \in \{n^2 \mid n \in \mathbb{N} \wedge \sqrt{\gls{Number}_\cell} / n \in \mathbb{N}\}
\), i.e., the set of squared integer divisors of \(\sqrt{\gls{Number}_\cell}\).

Self-sustainable operation of the \gls{ris} is achieved by splitting the incident signal into two parts, where one part is absorbed and used for energy harvesting at the \gls{ris}, and the other part is reflected towards the user.
In this work, we analyze the signal split in the time, element, and power domain, leading to the \gls{TimeSplit}, \gls{ElementSplit}, and \gls{PowerSplit} scheme, respectively.
Throughout this paper, we refer to a particular scheme by \(\policy \in \{\policyPs, \policyEs, \policyTs\}\).

\subsection{RIS CODEBOOK DESIGN}
\noindent
The phase shifts for signal reflection at the \gls{ris} unit cells are chosen from a predefined \gls{ris} codebook, denoted by \(\gls{Codebook}_\policy\), where each codeword is designed to illuminate a particular subarea of the coverage area.
The side lengths of the subareas are denoted by \(\gls{LengthSideSubarea}_{y,\policy}\) and \(\gls{LengthSideSubarea}_{z,\policy}\), cf. Fig.~\ref{fig:system}.
For codebook design, we assume narrow beams, i.e., each codeword is given by a linear phase-shift profile that results from a desired \gls{aoa} and \gls{aod} at the \gls{ris}~\cite{najafi2020physicsbasedmodeling}.
In this work, the \gls{aoa} for each codeword is aligned with the \gls{los} to the \gls{bs}, and the \glspl{aod} for the codewords are given by the directions to the centers of the subareas.
The resulting codebook comprises
\begin{equation}\label{eq:codebook-size}
    \abs*{\gls{Codebook}_\policy}
    = \ceil*{
        \pi
        \frac{\sqrt{0.5 \gls{Number}\Ucr^\policy} \gls{SizeCoverageAreaZ}}{2.782\ d_{\reflected,\minimum}}
    }
    \ceil*{
        \pi
        \frac{\sqrt{0.5 \gls{Number}\Ucr^\policy} \gls{SizeCoverageAreaY}}{2.782\ d_{\reflected,\minimum}}
    }
\end{equation}
codewords, see Appendix~\ref{sec:codebook-size}, where \(\gls{Number}\Ucr^\policy\) and \(d_{\reflected,\minimum}\) denote the number of \gls{ris} unit cells contributing to signal reflection and the minimum distance between the \gls{ris} and the coverage area, respectively.

\subsection{BEAM TRAINING}
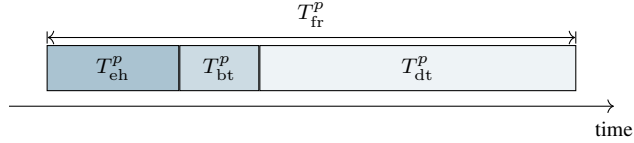
\begin{figure}[!t]
    \centering
    \begin{tikzpicture}[font=\footnotesize]
        \def\framewidth{7cm}
        \def\blockheight{0.6cm}
        \def\horizontalspace{1cm}

        \tikzset{every node/.style={rectangle,draw,minimum height=\blockheight}}

        \node [name=eh,fill=BaseColorA,minimum width=0.25*\framewidth] at (0,0) {\(\gls{TimeDuration}_\stageEh^\policy\)};
        \node [name=bt,right=0cm of eh.east,anchor=west,fill=BaseColorB,minimum width=0.15*\framewidth] {\(\gls{TimeDuration}_\stageBt^\policy\)};
        \node [name=dt,fill=BaseColorD,right=0cm of bt.east,anchor=west,minimum width=0.60*\framewidth] {\(\gls{TimeDuration}_\stageDt^\policy\)};

        \draw [|<->|] ([yshift=0.1cm]eh.north west) -- ([yshift=0.1cm]dt.north east) node [midway,above,draw=none] {\(\gls{TimeDuration}_\fr^\policy\)};

        \draw [->] ([xshift=-0.5cm,yshift=-0.2cm]eh.south west) -- ([xshift=0.5cm,yshift=-0.2cm]dt.south east) node [draw=none,below] {time};
    \end{tikzpicture}
    \caption{Transmission frame comprising stages for energy harvesting, beam training, and data transmission.}
    \label{fig:frame}
\end{figure}
\noindent
In order to keep the \gls{ris} reflection beam aligned with the current location of the mobile user, the employed \gls{ris} codeword is periodically updated based on beam training.
To this end, we consider a frame-based transmission protocol, where each frame is partitioned into three stages, see Fig.~\ref{fig:frame}.
In the following, we use \(\stage \in \{\stageEh, \stageBt, \stageDt\}\) to refer to a particular stage.

Stage \(\stage = \stageEh\) is dedicated to energy harvesting and is used only by the \gls{TimeSplit} scheme, because the \gls{ElementSplit} and \gls{PowerSplit} schemes harvest energy during beam training and data transmission.
Stage \(\stage = \stageBt\) is used for beam training, where pilot symbols are transmitted by the \gls{bs} while the \gls{ris} changes the phase-shift configuration based on codebook \(\gls{Codebook}_\policy\).
The index of the codeword providing the highest received power at the user is fed back to the \gls{bs} and applied for the subsequent data transmission.
Assuming exhaustive search\footnote{It has been shown that beam training based on exhaustive search can still lead to small beam training overhead~\cite{laue2025performancetradeoffoverhead}. More elaborate beam training strategies may be considered in future extensions of this work.} of the codebook and one pilot symbol per codeword, the duration of the beam training stage is given by
\begin{equation}
    \gls{TimeDuration}_\stageBt^\policy = (\glsSubSymbol{TimeDuration} + \glsSubResponse{TimeDuration}) \abs*{\gls{Codebook}_\policy} + \glsSubDelay{TimeDuration} + \glsSubResponse{TimeDuration},
\end{equation}
where \(\glsSubSymbol{TimeDuration}\), \(\glsSubResponse{TimeDuration}\), and \(\glsSubDelay{TimeDuration}\) denote the symbol duration, the \gls{ris} response time\footnote{The \gls{ris} response time is the time required to change from one stable phase-shift configuration to another~\cite{pan202110240element,liu2022simulationfieldtrial}.}, and the training feedback delay\footnote{Note that \(\glsSubDelay{TimeDuration} = 0\) can be realized using self-configuration of the \gls{ris}~\cite{tajin2023incidentpowerrelative,tavana2023energyharvestingmaximizationa,youn2023liquidcrystaldriven,albanese2024aresautonomousris}.}.

Using periodic beam training, continuous coverage of the mobile user's location can be achieved if the total frame duration \(\gls{TimeDuration}_\fr^\policy = \gls{TimeDuration}_\stageEh^\policy + \gls{TimeDuration}_\stageBt^\policy + \gls{TimeDuration}_\stageDt^\policy\) corresponds to the beam coherence time, i.e., the time duration for which the user is covered by a particular \gls{ris} beam.
However, the beam coherence time cannot be determined exactly if the user's trajectory is unknown.
Therefore, we adopt a parametric model for the frame duration using the diagonal of the subareas as a reference.
Then, the frame duration is given by~\cite{laue2025performancetradeoffoverhead}
\begin{equation}\label{eq:duration-frame}
    \gls{TimeDuration}_\fr^\policy = \gls{TimeDuration}_\stageEh^\policy + \gls{TimeDuration}_\stageBt^\policy + \gls{TimeDuration}_\stageDt^\policy = \frac{\gls{ParameterFrame}}{\gls{Velocity}} \sqrt{\gls{LengthSideSubarea}_{y,\policy}^2 + \gls{LengthSideSubarea}_{z,\policy}^2},
\end{equation}
where \(\gls{Velocity}\) denotes the user velocity and \(\gls{ParameterFrame} > 0\) is a design parameter.
The latter can be used to strike a balance between the frequency of beam training and the probability of beam misalignment.

Based on \eqref{eq:duration-frame}, the time duration for data transmission is given by
\begin{equation}\label{eq:duration_data}
    \gls{TimeDuration}_\stageDt^\policy = \gls{TimeDuration}_\fr^\policy - \gls{TimeDuration}_\stageEh^\policy - \gls{TimeDuration}_\stageBt^\policy.
\end{equation}
In order to perform coherent demodulation of the received symbols, \(\gls{TimeDuration}_\stageDt^\policy\) is further divided into subframes, where each subframe comprises a channel estimation stage and a data transmission stage.
We adopt a subframe duration based on the channel coherence time for user velocity \(\gls{Velocity}\), given by \(\glsSubSubframe{TimeDuration} = \frac{3\gls{Wavelength}}{4\sqrt{\pi} \gls{Velocity}}\)~\cite[Eq. (4.40.c)]{rappaport2001wirelesscommunications}, where \(\gls{Wavelength}\) denotes the wavelength of the considered carrier frequency.
Since there are \(\gls{Number}_\mathrm{sf}^\policy = \frac{\gls{TimeDuration}_\stageDt^\policy}{\glsSubSubframe{TimeDuration}}\) subframes per data transmission stage and we assume \(\glsSubEstimation{Number}\) pilot symbols per channel estimation, the effective duration for data transmission, \(\bar{\gls{TimeDuration}}_\stageDt^\policy\), is given by
\begin{equation}\label{eq:duration_data_no_estimation}
    \begin{split}
        \bar{\gls{TimeDuration}}_\stageDt^\policy &= \gls{TimeDuration}_\stageDt^\policy - \glsSubEstimation{Number} \glsSubSymbol{TimeDuration} \gls{Number}_\mathrm{sf}^\policy\\
        &= \gls{TimeDuration}_\stageDt^\policy \left(
            1 - \frac{\glsSubEstimation{Number} \glsSubSymbol{TimeDuration}}{\glsSubSubframe{TimeDuration}}
        \right).
    \end{split}
\end{equation}

\subsection{SIGNAL MODEL}
\noindent
Our considered codebook design is based on the \gls{los} paths between \gls{bs}, \gls{ris}, and user, which is a reasonable approach when both the \gls{bs}-\gls{ris} channel and the \gls{ris}-user channel have dominant \gls{los} components.
Following this assumption, we restrict our theoretical analysis to deterministic channels.
Then, the received \gls{snr} at user location \(l \in \gls{SetLocations}\) for selected codeword \(b_l\) is given by
\begin{equation}\label{eq:snr}
    \gls{Snr}_\stage^\policy (l) = \frac{
        \glsSubRx{Gain}
        \gls{Gain}_\reflected(l)
        \gls{Gain}_\ris^\policy(l)
        \glsSubIncident{Gain}
        \glsSubTx{Gain}
        \gls{PowerAverage}_\stage^\policy(b_l)
    }{
        \gls{PowerNoise}
    }
\end{equation}
where \(\glsSubRx{Gain}\), \(\gls{Gain}_\reflected(l)\), \(\gls{Gain}_\ris^\policy(l)\), \(\glsSubIncident{Gain}\), \(\glsSubTx{Gain}\), \(\gls{PowerAverage}_\stage^\policy(b_l)\), and \(\gls{PowerNoise}\) denote the antenna gain at the receiver, the pathloss of the \gls{ris}-user channel, the \gls{ris} gain, the pathloss of the \gls{bs}-\gls{ris} channel, the antenna gain at the transmitter, the average transmit power at the \gls{bs} for codeword \(b_l\), and the noise power at the receiver, respectively.
For \(\gls{Gain}_\reflected(l)\) and \(\glsSubIncident{Gain}\), we adopt the free-space pathloss model, leading to \(\glsSubIncident{Gain} = \frac{\glsSquared{Wavelength}}{4^2 \pi^2 d_\incident^2}\) and \(\gls{Gain}_\reflected(l) = \frac{\glsSquared{Wavelength}}{4^2 \pi^2 d_\reflected^2(l)}\) for \gls{bs}-\gls{ris} distance \(d_\incident\) and \gls{ris}-user distance \(d_\reflected(l)\).
The latter is given by \(d_{\reflected,l} = \sqrt{x_l^2 + y_l^2 + z_l^2}\), where \((x_l, y_l, z_l)\) denote the coordinates of user location \(l \in \gls{SetLocations}\).
Moreover, we note that \(\gls{PowerAverage}_\stage^\policy(b_l)\) is a function of the codeword employed by the \gls{ris}.
Here, codeword-dependent power allocation is used to compensate for the distance-dependent pathloss of the \gls{ris}-user link.
For example, the codewords that illuminate the most distant subareas require a higher \gls{bs} transmit power to achieve a required minimum \gls{snr} at the user.

For further specification of \(\gls{PowerAverage}_\stage^\policy(b_l)\) in \eqref{eq:snr}, let \(x_\stage^\policy(b_l)\) denote a symbol in complex baseband representation transmitted when the \gls{ris} employs codeword \(b_l\).
The corresponding \gls{rf} signal in time domain is given by \(x_{\rf,\stage}^\policy(b_l, t) = \sqrt{2} \real{x_\stage^\policy(b_l) e^{j2\pi f t}}\).
Then, the instantaneous transmit power is given by
\begin{equation}
    \gls{Power}_\stage^\policy(b_l) = \frac{1}{\glsSubSymbol{TimeDuration}}
    \int_{\glsSubSymbol{TimeDuration}} [x_{\rf,\stage}^\policy(b_l, t)]^2 dt = \abs*{x_\stage^\policy(b_l)}^2.
\end{equation}
Consequently, the average transmit power for codeword \(b_l\) is given by \(\gls{PowerAverage}_\stage^\policy(b_l) = \E{\gls{Power}_\stage^\policy(b_l)}\), where the expectation is over all symbols transmitted for codeword \(b_l\) in stage \(\stage\) of all considered frames.
For data transmission, we assume Gaussian symbols in this work, i.e., \(\gls{PowerAverage}_\stageDt^\policy(b_l)\) is given by the variance of \(x_\stageDt^\policy(b_l) \sim \CN{0, \gls{PowerAverage}_\stageDt^\policy(b_l)}\).
In contrast, during energy harvesting and beam training, no information is conveyed from the \gls{bs} to the user.
Thus, it is sufficient to transmit unmodulated sinusoidal signals during these stages of the transmission frame, i.e., one can assume that \(\gls{Power}_\stageEh^\policy(b_l)\) and \(\gls{Power}_\stageBt^\policy(b_l)\) are deterministic such that \(\gls{PowerAverage}_\stageEh^\policy(b_l) = \gls{Power}_\stageEh^\policy(b_l)\) and \(\gls{PowerAverage}_\stageBt^\policy(b_l) = \gls{Power}_\stageBt^\policy(b_l)\).

Furthermore, we note that the average transmit power at the \gls{bs} is limited by a maximum value, denoted by \(\gls{PowerAverage}_{\stage,\maximum}^\policy\), which is specified in Section~\ref{sec:power_constraints}.

Finally, based on \eqref{eq:duration_data_no_estimation} and \eqref{eq:snr}, the effective achievable rate is given by
\begin{equation}\label{eq:rate}
    \gls{RateEffective}_\policy(l)
    = \frac{\bar{\gls{TimeDuration}}_\stageDt^\policy}{\gls{TimeDuration}_\fr^\policy}
    \log_2\left(1 + \gls{Snr}_\stageDt^\policy(l)\right).
\end{equation}

\subsection{SPLITTING SCHEMES}\label{sec:splitting-schemes}
\noindent
Each considered splitting scheme can be characterized by subsets \(\gls{SetTiles}_{\reflecting,\stage}^\policy \subseteq \gls{SetTiles}\) and \(\gls{SetTiles}_{\harvesting,\stage}^\policy \subseteq \gls{SetTiles}\), which denote the tiles used for signal reflection and energy harvesting, respectively, in stage \(\stage\) of the transmission frame.
As shown in the following, these subsets result in a specific definition of the splitting ratio \(\rho_\policy\) for splitting scheme \(\policy\), which balances the powers available for energy harvesting at the \gls{ris} and for signal reflection to the user, respectively.
Hence, the splitting ratio controls the trade-off between the harvested power and the achievable data rate.

\subsubsection{TIME SPLITTING}\label{sec:time-split}
The \gls{TimeSplit} scheme uses all tiles in stage \(\stage = \stageEh\) for energy harvesting, and all tiles for signal reflection in stages \(\stage \in \{\stageBt, \stageDt\}\).
Thus, we have \(\gls{SetTiles}_{\harvesting,\stageEh}^\policyTs = \gls{SetTiles}_{\reflecting,\stageBt}^\policyTs = \gls{SetTiles}_{\reflecting,\stageDt}^\policyTs = \gls{SetTiles}\) and \(\gls{SetTiles}_{\reflecting,\stageEh}^\policyTs = \gls{SetTiles}_{\harvesting,\stageBt}^\policyTs = \gls{SetTiles}_{\harvesting,\stageDt}^\policyTs = \emptyset\), as well as \(\gls{Number}\Ucr^\policyTs = \gls{Number}_\cell\).
Here, the splitting ratio is defined as the fraction of the transmission frame that is actually used for signal reflection, given by \(\rho_\policyTs = 1 - \frac{\gls{TimeDuration}_\stageEh^\policyTs}{\gls{TimeDuration}_\fr^\policyTs} \in [0, 1]\).
A special feature of the \gls{TimeSplit} scheme is the separation of energy harvesting and signal reflection in the time domain.
Therefore, the transmit strategy and transmit power at the \gls{bs} can be designed individually for these purposes.

\subsubsection{ELEMENT SPLITTING}\label{sec:element-split}
\begin{figure}[!t]
    \centering
    \begin{tikzpicture}[font=\footnotesize]
        \newcommand{\risposx}{0cm}
        \newcommand{\risposy}{0cm}
        \coordinate (rispos) at (\risposx,\risposy);
        \newcommand{\numtilessqrt}{4}
        \newcommand{\numcellssqrt}{2}
        \newcommand{\numtileouterlow}{1}
        \newcommand{\numtileouterhigh}{4}
        \newcommand{\cellsize}{2mm}
        \newcommand{\cellspacing}{0.5mm}
        \newcommand{\tilesize}{\numcellssqrt*\cellsize + (1+\numcellssqrt)*\cellspacing}
        \newcommand{\tilespacing}{0.5mm}
        \newcommand{\rissize}{\numtilessqrt*(\tilesize) + (1+\numtilessqrt)*\tilespacing}
    
        \node [draw,rectangle,minimum size=\rissize,fill=BaseDarkColorA] (ris) at (rispos) {};
        \def\m{1}
        \foreach \n in {1,...,\numtilessqrt} {
            \node [draw,rectangle,minimum size=\tilesize,fill=BaseDarkColorB] (tile\m\n) at ([xshift=(\m - 0.5 - \numtilessqrt / 2)*(\tilesize + \tilespacing),yshift=(\n - 0.5 - \numtilessqrt / 2)*(\tilesize + \tilespacing)]ris) {};

            \foreach \k in {1,2,...,\numcellssqrt} {
                \foreach \l in {1,2,...,\numcellssqrt} {
                    \node [draw,rectangle,minimum size=\cellsize,inner sep=0cm,fill=BaseDarkColorD] (cell\k\l) at ([xshift=(\k - 0.5 - \numcellssqrt / 2)*(\cellsize + \cellspacing),yshift=(\l - 0.5 - \numcellssqrt / 2)*(\cellsize + \cellspacing)]tile\m\n) {};
                }
            }
        }
        \foreach \m in {2,3} {
            \foreach \n in {1,\numtilessqrt} {
                \node [draw,rectangle,minimum size=\tilesize,fill=BaseDarkColorB] (tile\m\n) at ([xshift=(\m - 0.5 - \numtilessqrt / 2)*(\tilesize + \tilespacing),yshift=(\n - 0.5 - \numtilessqrt / 2)*(\tilesize + \tilespacing)]ris) {};

                \foreach \k in {1,2,...,\numcellssqrt} {
                    \foreach \l in {1,2,...,\numcellssqrt} {
                        \node [draw,rectangle,minimum size=\cellsize,inner sep=0cm,fill=BaseDarkColorD] (cell\k\l) at ([xshift=(\k - 0.5 - \numcellssqrt / 2)*(\cellsize + \cellspacing),yshift=(\l - 0.5 - \numcellssqrt / 2)*(\cellsize + \cellspacing)]tile\m\n) {};
                    }
                }
            }
        }
        \def\m{4}
        \foreach \n in {1,...,\numtilessqrt} {
            \node [draw,rectangle,minimum size=\tilesize,fill=BaseDarkColorB] (tile\m\n) at ([xshift=(\m - 0.5 - \numtilessqrt / 2)*(\tilesize + \tilespacing),yshift=(\n - 0.5 - \numtilessqrt / 2)*(\tilesize + \tilespacing)]ris) {};

            \foreach \k in {1,2,...,\numcellssqrt} {
                \foreach \l in {1,2,...,\numcellssqrt} {
                    \node [draw,rectangle,minimum size=\cellsize,inner sep=0cm,fill=BaseDarkColorD] (cell\k\l) at ([xshift=(\k - 0.5 - \numcellssqrt / 2)*(\cellsize + \cellspacing),yshift=(\l - 0.5 - \numcellssqrt / 2)*(\cellsize + \cellspacing)]tile\m\n) {};
                }
            }
        }
        \foreach \m in {2,3} {
            \foreach \n in {2,3} {
                \node [draw,rectangle,minimum size=\tilesize,fill=BaseDarkColor] (tile\m\n) at ([xshift=(\m - 0.5 - \numtilessqrt / 2)*(\tilesize + \tilespacing),yshift=(\n - 0.5 - \numtilessqrt / 2)*(\tilesize + \tilespacing)]ris) {};

                \foreach \k in {1,2,...,\numcellssqrt} {
                    \foreach \l in {1,2,...,\numcellssqrt} {
                        \node [draw,rectangle,minimum size=\cellsize,inner sep=0cm,fill=BaseDarkColorA] (cell\k\l) at ([xshift=(\k - 0.5 - \numcellssqrt / 2)*(\cellsize + \cellspacing),yshift=(\l - 0.5 - \numcellssqrt / 2)*(\cellsize + \cellspacing)]tile\m\n) {};
                    }
                }
            }
        }

        \draw [semithick] (tile23.center) node[circle,fill=black,inner sep=0pt,minimum size=2*\cellspacing] {} -- ([xshift=-(\tilesize),yshift=(\tilesize)]ris.west) node [left] {tile in reflection mode};

        \draw [semithick] (tile12.center) node[circle,fill=black,inner sep=0pt,minimum size=2*\cellspacing] {} -- ([xshift=-(\tilesize),yshift=-(\tilesize)]ris.west) node [left] {tile in harvesting mode};

        \draw [semithick] ([xshift=-0.25*(\tilesize),yshift=0.25*(\tilesize)]tile11.center) node[circle,fill=black,inner sep=0pt,minimum size=2*\cellspacing] {} -- ([xshift=-(\tilesize),yshift=-2*(\tilesize)]ris.west) node [left] {unit cell};
    \end{tikzpicture}
    \caption{Tile-based element split of the \glsxtrshort{ris}.}
    \label{fig:element_split}
\end{figure}
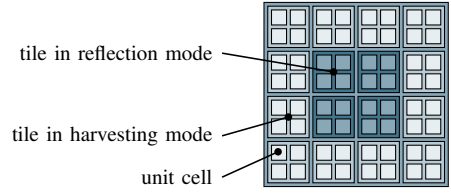

The \gls{ElementSplit} scheme does not use stage \(\stage = \stageEh\), which leads to \(\gls{SetTiles}_{\reflecting,\stageEh}^\policyEs = \gls{SetTiles}_{\harvesting,\stageEh}^\policyEs = \emptyset\) and \(\gls{TimeDuration}_\stageEh^\policyEs = 0\).
In contrast, the split of the incident signal at the \gls{ris} is realized by assigning a particular subset of tiles to signal reflection and energy harvesting, respectively, in stages \(\stage \in \{\stageBt, \stageDt\}\).
The subset of tiles that contribute to signal reflection is inherently defined by the splitting ratio \(\rho_\policyEs = \frac{\abs*{\gls{SetTiles}_{\reflecting,\stage}^\policyEs}}{\abs*{\gls{SetTiles}}}\), \(\stage \in \{\stageBt, \stageDt\}\).
Here, we assume the same subset during beam training and data transmission, i.e., \(\gls{SetTiles}_{\reflecting,\stageBt}^\policyEs = \gls{SetTiles}_{\reflecting,\stageDt}^\policyEs\).
Since each tile is used for either energy harvesting or signal reflection, we have \(\gls{SetTiles}_{\harvesting,\stage}^\policyEs = \gls{SetTiles} \setminus \gls{SetTiles}_{\reflecting,\stage}^\policyEs\), \(\stage \in \{\stageBt, \stageDt\}\).
In addition, we assume that the tiles used for signal reflection form a continuous square-shaped surface at the \gls{ris}, see Fig.~\ref{fig:element_split}, such that \eqref{eq:codebook-size} still applies with \(\gls{Number}\Ucr^\policyEs = \abs*{\gls{SetCells}_m} \abs*{\gls{SetTiles}_{\reflecting,\stage}^\policyEs} = \rho_\policyEs \gls{Number}_\cell \leq \gls{Number}_\cell\), \(\stage \in \{\stageBt, \stageDt\}\).
The assumption of the square-shaped surface for signal reflection restricts the feasible set for the splitting ratio as follows
\begin{equation}\label{eq:feasible_set_reflecting}
    \rho_\policyEs \in \left\{\frac{n^2}{\gls{Number}_\tile} \mathrel{\Big|} n = 1, 2, \dotsc, \sqrt{\gls{Number}_\tile}\right\} \subset (0, 1].
\end{equation}
We note that \(\rho_\policyEs\) determines the effective \gls{ris} size used for signal reflection, which has an impact on the beam width and thus on the codebook size and the beam training overhead.
Moreover, since the \gls{ElementSplit} scheme implements simultaneous energy harvesting and signal reflection, the same incident signal is used for both purposes.
Hence, the adopted transmit strategy and transmit power at the \gls{bs} always have an impact on both the harvested power and the achievable data rate.

\subsubsection{POWER SPLITTING}\label{sec:power-split}
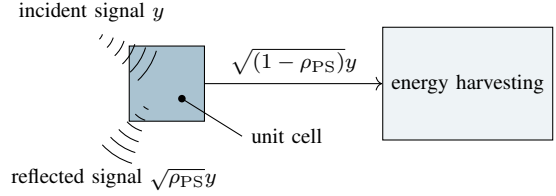
\begin{figure}[!t]
    \centering
    \begin{tikzpicture}[font=\footnotesize]    
        \node [draw,rectangle,minimum size=1cm,fill=BaseColorA] (cell) at (0cm, 0cm) {};
        \node [draw,rectangle,minimum size=1.5cm,fill=BaseColorD] (eh) at (4cm,0cm) {energy harvesting};

        \draw [->] (cell.east) -- (eh.west) node [midway,above] {\(\sqrt{(1-\rho_\policyPs)}y\)};
        \draw [decorate,decoration={expanding waves,angle=15,segment length=0.15cm}] ([xshift=-1cm,yshift=0.7cm]cell.center) node [above] {incident signal \(y\)} -- ([xshift=-0.2cm,yshift=0.2cm]cell.center);
        \draw [decorate,decoration={expanding waves,angle=15,segment length=0.15cm}] ([xshift=-0.2cm,yshift=-0.2cm]cell.center) -- ([xshift=-0.7cm,yshift=-1cm]cell.center) node [below] {reflected signal \(\sqrt{\rho_\policyPs}y\)};

        \draw ([xshift=0.2cm,yshift=-0.2cm]cell.center) node[circle,fill=black,inner sep=0pt,minimum size=1mm] {} -- +(0.8cm,-0.5cm) node [right] {unit cell};
    \end{tikzpicture}
    \caption{Illustration of power splitting at a \glsxtrshort{ris} unit cell.}
    \label{fig:power_split}
\end{figure}

The \gls{PowerSplit} scheme is similar to the \gls{ElementSplit} scheme, where energy is harvested during beam training and data transmission.
Thus, we have \(\gls{SetTiles}_{\reflecting,\stageEh}^\policyPs = \gls{SetTiles}_{\harvesting,\stageEh}^\policyPs = \emptyset\) and \(\gls{TimeDuration}_\stageEh^\policyPs = 0\).
However, as illustrated in Fig.~\ref{fig:power_split}, the split of the incident signal is realized by a power split at each unit cell of each tile, specified by splitting ratio \(\rho_\policyPs \in [0, 1]\).
Here, \(\rho_\policyPs\) determines the effective gains for energy harvesting and signal reflection, while all tiles are used simultaneously for both purposes.
Consequently, we have \(\gls{SetTiles} = \gls{SetTiles}_{\reflecting,\stageBt}^\policyPs = \gls{SetTiles}_{\harvesting,\stageBt}^\policyPs = \gls{SetTiles}_{\reflecting,\stageDt}^\policyPs = \gls{SetTiles}_{\harvesting,\stageDt}^\policyPs\) and \(\gls{Number}\Ucr^\policyPs = \gls{Number}_\cell\).
Moreover, similar to the \gls{ElementSplit} scheme, the same incident signal has an impact on both the harvested power and the achievable data rate.

\subsection{PROBLEM STATEMENT}
\noindent
For each splitting scheme \(\policy\), we aim to minimize the average transmit power at the \gls{bs} while ensuring a minimum effective rate \(\gls{RateEffective}_\minimum\) at the user, a minimum \gls{snr} \(\gls{Snr}_{\stageBt,\minimum}\) for beam training, and self-sustainable operation of the \gls{ris}.
Thereby, our goal is to find the optimal power allocation for each codeword and each stage of the transmission frame, as well as the optimal split of the incident signal at the \gls{ris} for each splitting scheme.

For a particular frame where codeword \(b^\prime\) is selected after beam training, the average transmit power at the \gls{bs} is given by
\begin{equation}
    \gls{PowerAverage}_{\bs,\fr}(b^\prime) = \frac{\gls{TimeDuration}_\stageEh^\policy}{\gls{TimeDuration}_\fr^\policy}
    \gls{PowerAverage}_\stageEh^\policy
    + \frac{\gls{TimeDuration}_\sym}{\gls{TimeDuration}_\fr^\policy}
    \sum_{b \in \gls{Codebook}_\policy}
    \gls{PowerAverage}_\stageBt^\policy(b)
    + \frac{\gls{TimeDuration}_\stageDt}{\gls{TimeDuration}_\fr^\policy}
    \gls{PowerAverage}_\stageDt^\policy(b^\prime).
\end{equation}
Assuming that the mobile user passes each subarea with equal probability, the average transmit power for all frames is given by
\begin{equation}\label{eq:power-bs-average}
    \begin{split}
        \gls{PowerAverage}_\bs^\policy &= \frac{1}{\abs*{\gls{Codebook}_\policy}}
        \sum_{b^\prime \in \gls{Codebook}_\policy}
        \gls{PowerAverage}_{\bs,\fr}(b^\prime)\\
        &= \frac{1}{\abs*{\gls{Codebook}_\policy}}
        \sum_{b \in \gls{Codebook}_\policy}\left(
        \frac{\gls{TimeDuration}_\stageEh^\policy}{\gls{TimeDuration}_\fr^\policy}
        \gls{PowerAverage}_\stageEh^\policy
        + \abs*{\gls{Codebook}_\policy} \frac{\gls{TimeDuration}_\sym}{\gls{TimeDuration}_\fr^\policy}
        \gls{PowerAverage}_\stageBt^\policy(b)
        + \frac{\gls{TimeDuration}_\stageDt}{\gls{TimeDuration}_\fr^\policy}
        \gls{PowerAverage}_\stageDt^\policy(b)
        \right).
    \end{split}
\end{equation}
Thus, the problem statement for splitting scheme \(\policy\) can be formally written as follows
\begin{subequations}\label{eq:problem}
    \begin{align}
        &\min_{
            \rho_\policy, \gls{PowerAverage}_\stageEh^\policy, \gls{PowerAverage}_\stageBt^\policy(b), \gls{PowerAverage}_\stageDt^\policy(b), \forall\, b \in \gls{Codebook}_\policy
        } \quad\gls{PowerAverage}_\bs^\policy\label{eq:problem-objective}\\
        \mathrm{s.t.}
        &\quad
        \gls{PowerAverage}_\stageEh^\policy \leq \gls{PowerAverage}_{\stageEh,\maximum}^\policy\label{eq:constraint-power-max-harvesting}\\
        &\quad
        \gls{PowerAverage}_\stage^\policy(b) \leq \gls{PowerAverage}_{\stage,\maximum}^\policy, \forall\, \stage \in \{\stageBt, \stageDt\}, \forall\, b \in \gls{Codebook}_\policy\label{eq:constraint-power-max-others}\\
        &\quad
        \gls{Snr}_{\stageBt,\minimum} \leq \gls{Snr}_\stageBt^\policy(l), \forall\, l \in \gls{SetLocations}\label{eq:constraint-training}\\
        &\quad
        \gls{RateEffective}_\minimum \leq \gls{RateEffective}_\policy(l), \forall\, l \in \gls{SetLocations}\label{eq:constraint-rate}\\
        &\quad
        \gls{Power}_{\consuming}^\policy(t) \leq \gls{Power}_{\harvesting}^\policy(t).\label{eq:constraint-power}
    \end{align}
\end{subequations}
In \eqref{eq:problem}, constraints \eqref{eq:constraint-power-max-harvesting} and \eqref{eq:constraint-power-max-others} account for the maximum transmit power at the \gls{bs}, which is further specified in Section~\ref{sec:power_constraints}.
A minimum \gls{snr} for beam training and a minimum effective data rate are enforced by constraints \eqref{eq:constraint-training} and \eqref{eq:constraint-rate}, respectively.
Moreover, constraint \eqref{eq:constraint-power} guarantees self-sustainable operation of the \gls{ris} at any time \(t\), where \(\gls{Power}_{\consuming}^\policy\) and \(\gls{Power}_{\harvesting}^\policy\) in \eqref{eq:constraint-power} denote the power consumed by the \gls{ris} and the power supplied to the \gls{ris} by energy harvesting, respectively.
\(\gls{Power}_{\consuming}^\policy\) and \(\gls{Power}_{\harvesting}^\policy\) are further specified in Section~\ref{sec:optimization}.

In the following, we specialize problem \eqref{eq:problem} for each considered splitting scheme and propose an algorithm for obtaining the respective optimal solutions.
As a start, we provide a comprehensive model for the considered energy harvesting at the \gls{ris}.

%% file: sections/energy_harvesting.tex
\noindent
This section describes the considered energy harvesting for self-sustainable operation of the \gls{ris}. In particular, we present the proposed tile-based architecture of the \gls{ris} for hybrid combining and the rectifier characteristics for each stage of the transmission frame.
Finally, we discuss the model for the energy consumption of the \gls{ris}.

\subsection{TILE-BASED ARCHITECTURE}\label{sec:harvesting-architecture}
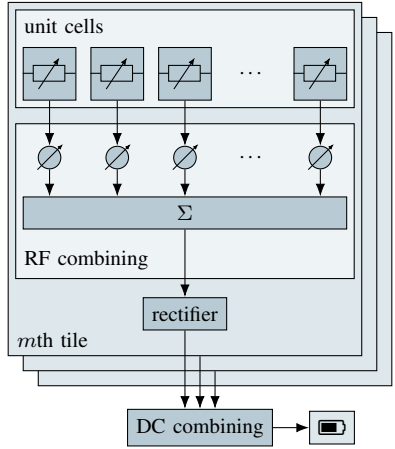
\begin{figure}[!t]
    \centering
    \begin{tikzpicture}[font=\footnotesize]
        \def\cellsize{7mm}
        \def\cellspacing{2mm}
        \def\tileshift{2mm}
        \def\componentspacing{6mm}
        \def\phaseshiftersize{3mm}
        \def\numcellsA{3}
        \def\numcellsB{4}
        \def\numcellsC{5}
        \def\tilewidth{5*\cellsize + 6*\cellspacing}
        \def\tileheight{4.3cm}
        \def\rfwidth{5*\cellsize + 5*\cellspacing}
        \def\rfheight{0.5*\componentspacing + \phaseshiftersize + 0.6*\componentspacing + 11mm}
        \def\adderwidth{5*\cellsize + 4*\cellspacing}
        \def\spaceforlabel{4mm}
        \coordinate (tilepos) at (0cm,0cm);

        \foreach \n in {2,1} {
            \node [draw,rectangle,minimum width=\tilewidth,minimum height=\tileheight+\spaceforlabel,fill=BaseColorC] (tileother\n) at ([xshift=\n*\tileshift,yshift=-\n*\tileshift+\spaceforlabel]tilepos) {};
        }
        
        \node [draw,rectangle,minimum width=\tilewidth,minimum height=\tileheight+\spaceforlabel,fill=BaseColorC] (tilemain) at ([yshift=\spaceforlabel]tilepos) {};
        \node [anchor=south west] at (tilemain.south west) {\(m\)th tile};

        \node [draw,rectangle,anchor=north west,minimum height=\cellsize+\cellspacing+\spaceforlabel,minimum width=5*\cellsize+5*\cellspacing,fill=BaseColorD] (unitcells) at ([xshift=0.5*\cellspacing,yshift=-0.5*\cellspacing]tilemain.north west) {};

        \node [anchor=north west] at (unitcells.north west) {unit cells};

        \foreach \n in {1,2,...,\numcellsA,\numcellsC} {
            \node [draw,rectangle,minimum size=\cellsize,anchor=north west,fill=BaseDarkColorB] (cell\n) at ([xshift=\n*\cellspacing+(\n-1)*\cellsize,yshift=-\cellspacing-\spaceforlabel]tilemain.north west) {};
            \draw ([xshift=\n*\cellspacing+(\n-1)*\cellsize,yshift=-\cellspacing-0.5*\cellsize-\spaceforlabel]tilemain.north west) -- +(0.2*\cellsize,0mm) +(0.8*\cellsize,0mm) -- +(\cellsize,0mm);
            \draw ([xshift=\n*\cellspacing+(\n-1)*\cellsize+0.2*\cellsize,yshift=-\cellspacing-0.65*\cellsize-\spaceforlabel]tilemain.north west) rectangle +(0.6*\cellsize,0.3*\cellsize);
            \draw [-{Latex[length=1mm]}] ([xshift=\n*\cellspacing+(\n-1)*\cellsize+0.3*\cellsize,yshift=-\cellspacing-0.8*\cellsize-\spaceforlabel]tilemain.north west) -- +(0.4*\cellsize,0.6*\cellsize);
        }
        \node [rectangle,minimum size=\cellsize] (celldots) at ([xshift=\numcellsB*\cellspacing+(\numcellsB-1)*\cellsize+0.5*\cellsize,yshift=-\cellspacing-0.5*\cellsize-\spaceforlabel]tilemain.north west) {\dots};

        \node [draw,rectangle,anchor=north west,minimum height=\rfheight,minimum width=\rfwidth,fill=BaseColorD] (rfcombining) at ([xshift=-0.5*\cellspacing,yshift=-0.5*\componentspacing]cell1.south west) {};
        
        \foreach \n in {1,2,...,\numcellsA,\numcellsC} {
            \draw [-Latex] (cell\n.south) -- +(0mm,-\componentspacing) node [draw,circle,inner sep=0mm,minimum size=\phaseshiftersize,anchor=north,fill=BaseDarkColorB] (shifter\n) {};
            \draw [-{Latex[length=0.8mm]}] ([xshift=-0.5mm,yshift=-0.5mm]shifter\n.south west) -- ([xshift=0.7mm,yshift=0.7mm]shifter\n.north east);
        }
        \node [inner sep=0mm,minimum size=\phaseshiftersize,anchor=north] at ([yshift=-\componentspacing]celldots.south) {\dots};

        \node [draw,rectangle,minimum width=\adderwidth,anchor=north,fill=BaseDarkColorB] (adder) at ([yshift=-0.6*\componentspacing]shifter3.south) {\(\Sigma\)};

        \foreach \n in {1,2,...,\numcellsA,\numcellsC} {
            \draw [-Latex] (shifter\n.south) -- ([xshift=(\n-3)*(\cellspacing+\cellsize)]adder.north);
        }
        \node [anchor=south west] at (rfcombining.south west) {\gls{rf} combining};

        \draw [-Latex] (adder.south) -- +(0mm,-1.5*\componentspacing) node [draw,rectangle,anchor=north,fill=BaseDarkColorB] (rectifier) {rectifier};

        \node [draw,rectangle,anchor=north,fill=BaseDarkColorB] (dccombining) at ([xshift=\tileshift,yshift=-3mm-2*\tileshift]tilemain.south) {\gls{dc} combining};
        \draw [-Latex] (rectifier.south) -- ([xshift=-\tileshift]dccombining.north);
        \draw [-Latex] ([xshift=\tileshift]tilemain.south) -- (dccombining.north);
        \draw [-Latex] ([xshift=\tileshift]tileother1.south) -- ([xshift=\tileshift]dccombining.north);

        \draw [-Latex] (dccombining.east) -- +(0.8*\componentspacing,0mm) node [draw,anchor=west,fill=BaseColorC] {\faBatteryThreeQuarters};
    \end{tikzpicture}
    \caption{Illustration of tile-based hybrid combining.}
    \label{fig:structure-tile}
\end{figure}

\noindent
As described in Section~\ref{sec:system}, each tile of the \gls{ris} can be used for signal reflection, energy harvesting, or both.
Signal reflection to a desired direction is realized by modifying the phase of the reflected wave at each unit cell of a tile.
For energy harvesting, all unit cells of a tile are connected to an \gls{rf} combining network, and the combined signal is fed into a rectifier circuit for \gls{rf}-\gls{dc} conversion, see Fig.~\ref{fig:structure-tile}.
The output of the rectifiers, i.e., of all tiles, is combined in the \gls{dc} domain and delivered to a central energy storage that provides power supply for the \gls{ris}.
Hence, the \gls{ris} implements hybrid combining~\cite{khan2024rfenergyharvesting}, which strikes a balance between hardware complexity for signal combining and \gls{rf} input power at the rectifiers.
For example, tiles comprising few unit cells entail a low complexity for the \gls{rf} combining network, but also yield a low \gls{rf} input power which may lead to poor \gls{rf}-\gls{dc} conversion efficiency~\cite{clerckx2022foundationswirelessinformation,ntontin2023wirelessenergyharvesting,ma2023wirelesspoweredintelligent}.
On the other hand, tiles with many unit cells yield a higher \gls{rf} input power at the rectifiers, but also lead to a complex \gls{rf} combining network that may increase losses and power consumption\footnote{
    Such losses are commonly ignored, but may be significant when the power of the incident signals is low, e.g., due to the large pathloss in the \gls{mmwave} bands~\cite{lopez2022dynamicrfcombining}.
}~\cite{yan2019performancepowerarea}.

\subsection{COMBINING LOSSES}\label{sec:harvesting-losses}
\noindent
If a tile contains more than one unit cell, the signals are coherently\footnote{
    Since the \gls{bs} and the \gls{ris} are located at fixed positions with a dominant \gls{los} link, we assume that the \gls{csi} of the \gls{bs}-\gls{ris} link required for coherent combining is available.
} combined using power combiners and discrete phase shifters with a resolution of \(\gls{Number}_\bits\) bit.
For the former, we assume ideal power combiners.
For the latter, each phase shifter typically introduces an insertion loss as the signal passes the phase shifter.
Such loss, for the \(n\)th unit cell of the \(m\)th tile, can be modeled as~\cite{lopez2022dynamicrfcombining}
\begin{equation}\label{eq:insertion_loss}
    \gls{LossInsertion}_n =
    \begin{cases}
        1, &n=1,\\
        \gls{LossInsertion}_0^{\gls{Number}_\bits}, &n \in \{2, \dotsc, \abs*{\gls{SetCells}_m}\},
    \end{cases}
\end{equation}
i.e., the first unit cell of a tile is considered as the reference for coherent combining, and the other unit cells induce an insertion loss of \(\gls{LossInsertion}_0\) per bit.

Furthermore, it is known that \gls{dc} combining is subject to losses caused by the power management network, which includes the routing of the DC signals, voltage conversion, etc.~\cite{hemour2014towardslowpower,gu2022farfieldwireless,lee2017hybridpowercombining}.
Such losses are typically modeled by an efficiency factor, and it is reasonable to assume that the loss increases with the number of signals to combine. Thus, we propose to model the loss by gain factor
\begin{equation}\label{eq:dc-loss}
    \gls{Gain}_{\dc,\stage}^\policy = \frac{1}{\abs*{\gls{SetTiles}_{\harvesting,\stage}^\policy}^{\gls{LossExponent}}},
\end{equation}
where \(\gls{LossExponent} > 0\) denotes a model parameter indicating the rate at which the loss increases as \(\abs*{\gls{SetTiles}_{\harvesting,\stage}^\policy}\) grows.

\subsection{RECTIFIER MODEL}\label{sec:rectifier}
\noindent
For a particular symbol with transmit power \(\gls{Power}_\stage^\policy(b)\), let 
\begin{equation}\label{eq:power-rf}
    \gls{Power}\RfVarVar(b) = \glsSubIncident{Gain} \glsSubTx{Gain} \gls{Gain}_{\tile,m}^\policy \gls{Power}_\stage^\policy(b)
\end{equation}
and
\begin{equation}\label{eq:power-average-rf}
    \gls{PowerAverage}\RfVarVar(b) = \E{\gls{Power}\RfVarVar(b)} = \glsSubIncident{Gain} \glsSubTx{Gain} \gls{Gain}_{\tile,m}^\policy \gls{PowerAverage}_\stage^\policy(b)
\end{equation}
denote the instantaneous and average \gls{rf} input power, respectively, at the \(m\)th rectifier.
In \eqref{eq:power-rf} and \eqref{eq:power-average-rf}, \(\gls{Gain}_{\tile,m}^\policy\) denotes the gain of the \(m\)th tile, which accounts for the insertion loss of the phase shifters and the gain of \gls{rf} combining.
A detailed characterization of \(\gls{Gain}_{\tile,m}^\policy\) for each splitting scheme is provided in Section~\ref{sec:optimization}.

The efficiency of the \gls{rf}-\gls{dc} conversion can be described by a non-linear function \(\gls{FunctionModel}: \gls{Power}\RfVarVar \mapsto \gls{Power}\DcVarVar\), which maps \(\gls{Power}\RfVarVar\) to \gls{dc} output power \(\gls{Power}\DcVarVar\).
The non-linearity of \(\gls{FunctionModel}\) is due to the non-linear I-V characteristic and the breakdown voltage of the diode, which is typically used in rectifier circuits.
The rectifier model \(\gls{FunctionModel}\) can be obtained from circuit analysis~\cite{morsi2020conditionalcapacitytransmit} or from measurements of practical rectifier implementations such as~\cite{eid2020mmwavetunnel,singh2020compactbroadbandgfet,riaz2021tribandrectifiermillimeter}, which are fitted to a suitable parametric function. To this end, the logistic function has been adopted in many works, e.g., see~\cite{hu2021robustsecuresum,wang2024multifunctionalreconfigurable,ntontin2023wirelessenergyharvesting}.

In this work, we follow this approach and fit the rectifier model using the measurement results from \cite{eid2020mmwavetunnel}, where a rectifier with a very low turn-on power is proposed, which is particularly relevant for the strong pathloss in \gls{mmwave} applications.
The fitting model is given by~\cite{boshkovska2015practicalnonlinear}
\begin{equation}\label{eq:model-eh}
    \gls{Power}\DcVarVar = \gls{FunctionModel}(\gls{Power}\RfVarVar) =
    \frac{\Psi - \glsSubSaturation{Power} \Omega}{1 - \Omega},
\end{equation}
where
\begin{align}
    \Psi &= \frac{\glsSubSaturation{Power}}{1 + e^{-c(\gls{Power}\RfVarVar - d)}},\\
    \Omega &= \frac{1}{1 + e^{cd}},
\end{align}
and \(c\), \(d\), and \(P_\mathrm{sat}\) are the parameters to be fitted.

Moreover, we are interested in the average harvested power for the employed codeword \(b\) as a function of the average \gls{rf} input power, given by
\begin{equation}\label{eq:power_dc}
    \gls{PowerAverage}\DcVarVar(\gls{PowerAverage}\RfVarVar(b)) = \E{\gls{FunctionModel}(\gls{Power}\RfVarVar(b))},
\end{equation}
where the expectation is over \(\gls{Power}\RfVarVar\).
Here, we note that \(\gls{PowerAverage}\DcVarVar\) is different for \(\stage \in \{\stageEh, \stageBt\}\) and \(\stage = \stageDt\), which can be seen as follows.
Since \(\gls{Power}_{\rf,\stageEh,m}^\policy(b)\) and \(\gls{Power}_{\rf,\stageBt,m}^\policy(b)\) are deterministic, we have \(\gls{PowerAverage}\DcVarVar(\gls{PowerAverage}\RfVarVar(b)) = \gls{FunctionModel}(\gls{Power}_{\rf,\stage,m}^\policy(b))\) for \(\stage \in \{\stageEh, \stageBt\}\).
In contrast, \(\gls{Power}_{\rf,\stageDt,m}^\policy(b)\) follows the exponential distribution because Gaussian symbols are assumed for data transmission.
As a result, we find
\begin{equation}\label{eq:power-dc-average-data}
    \begin{split}
        &\gls{PowerAverage}_{\dc,\stageDt,m}^\policy(\gls{PowerAverage}_{\rf,\stageDt,m}(b)) = \E{\gls{FunctionModel}(\gls{Power}_{\rf,\stageDt,m}^\policy(b))}\\
        &= \int_0^\infty \PdfRfDt\\
        &\quad\quad\quad \times \gls{FunctionModel}(\gls{Power}_{\rf,\stageDt,m}^\policy(b))\,\mathrm{d}\gls{Power}_{\rf,\stageDt,m}^\policy(b)\\
        &= \int_0^\infty \PdfDt\\
        &\quad\quad\quad \times \gls{FunctionModel}(\glsSubIncident{Gain} \glsSubTx{Gain} \gls{Gain}_{\tile,m}^\policy \gls{Power}_\stageDt^\policy(b))\,\mathrm{d}\gls{Power}_\stageDt^\policy(b),
    \end{split}
\end{equation}
where \(\PdfX = \frac{1}{\bar{x}} e^{-\frac{x}{\bar{x}}}\) denotes the PDF of the exponentially distributed random variable \(x\) with mean value \(\bar{x}\).
Since the integral in \eqref{eq:power-dc-average-data} is difficult to solve analytically, it is evaluated numerically whenever required for the remainder of this paper.
Furthermore, we simplify notation and denote \(\bar{\gls{FunctionModel}}_\stage(\gls{PowerAverage}\RfVarVar(b)) = \E{\gls{FunctionModel}(\gls{Power}\RfVarVar(b))}\) as the rectifier characteristic for average powers.

\begin{figure}[!t]
    \centering
    \begin{tikzpicture}
        \begin{scope}
            \begin{axis}[
                line plot style,
                xlabel=Average \gls{rf} input power \(\gls{PowerAverage}\RfVarVar\) (\unit{\milli\watt}),
                ylabel style={align=center},
                ylabel={Average harvested power\\\(\gls{PowerAverage}\DcVarVar\) (\unit{\micro\watt})},
                legend pos=south east,
                xmax=3,
                width=6.5cm,
                height=3.66cm,
                scale only axis,
            ]
                \addplot [black,mark=x,only marks]
                table [col sep=comma] {data/rectifier_measurements-00-01.csv};
                \addlegendentry{measured \cite{eid2020mmwavetunnel}}
    
                \addplot [color1]
                table [col sep=comma] {data/rectifier_measurements-00-00.csv};
                \addlegendentry{\(\stage \in \{\stageEh, \stageBt\}\)}
    
                \addplot [color2]
                table [col sep=comma] {data/rectifier_measurements-00-04.csv};
                \addlegendentry{\(\stage = \stageDt\)}
                
                \draw [<-] (0.2,2) -- (0.6,5);
            \end{axis}
        \end{scope}

        \begin{scope}[shift={(1.5,0.7)}]
            \begin{axis}[
                line plot style,
                width=3.3cm,
                height=2.6cm,
                xmode=log,
                xmin=0.005,
                xmax=0.15,
                axis background/.style={fill=white},
            ]
                \addplot [black,mark=x,only marks]
                table [col sep=comma] {data/rectifier_measurements-00-01.csv};
    
                \addplot [color1]
                table [col sep=comma] {data/rectifier_measurements-00-00.csv};

                \addplot [color2]
                table [col sep=comma] {data/rectifier_measurements-00-04.csv};
            \end{axis}
        \end{scope}
    \end{tikzpicture}
    \caption{Rectifier characteristics based on measurements in \cite{eid2020mmwavetunnel}.}
    \label{fig:rectifier}
\end{figure}
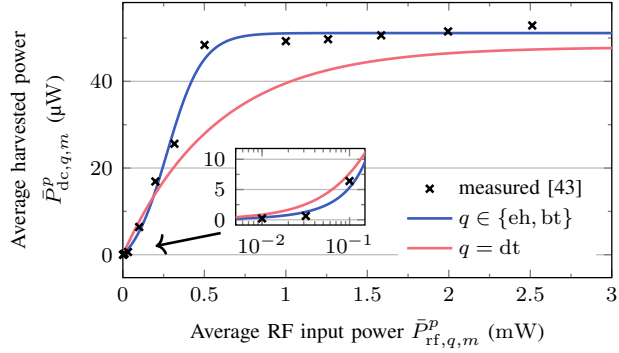

In Fig.~\ref{fig:rectifier}, we show \(\bar{\gls{FunctionModel}}_\stage\) after curve fitting based on measurements from \cite{eid2020mmwavetunnel}.
For \(\gls{PowerAverage}\RfVarVar > \qty{0.25}{\milli\watt}\), one can observe significant differences between the stages \(\stage \in \{\stageEh, \stageBt\}\) and \(\stage = \stageDt\), which originate from the different transmit strategies based on deterministic and random transmit signals.
These results highlight the importance of adopting the correct \gls{rf}-\gls{dc} characteristic for evaluating the harvested power.

\subsection{TRANSMIT POWER CONSTRAINTS}\label{sec:power_constraints}
\noindent
The maximum transmit power at the \gls{bs} is subject to two constraints.
On the one hand, \(\gls{Power}_\stage^\policy(b)\) is upper bounded by the maximum output of the power amplifier at the \gls{bs}, or by regulation authorities.
We denote this upper bound by \(\gls{Power}_\maximum\).
On the other hand, the \gls{rf} input power \(\gls{Power}\RfVarVar(b)\) should not exceed a threshold power \(\gls{Power}_\threshold\) in order to avoid the saturation regime of the rectifiers~\cite{clerckx2019fundamentalswirelessinformation}.
Based on \eqref{eq:power-rf}, the latter holds for all tiles if \(\gls{Power}_\stage^\policy(b) \leq \frac{
        \gls{Power}_\threshold
    }{
        \glsSubIncident{Gain} \glsSubTx{Gain} \TileGainMax^\policy
    }\),
where \(\TileGainMax^\policy = \max_{m\in \gls{SetTiles}_{\harvesting,\stage}^\policy}\gls{Gain}_{\tile,m}^\policy\).
Thus, both constraints are satisfied if
\begin{equation}\label{eq:power-upper-bound}
    \gls{Power}_\stage^\policy(b) \leq \min\left\{\gls{Power}_\maximum, \frac{
    \gls{Power}_\threshold
}{
    \glsSubIncident{Gain} \glsSubTx{Gain} \TileGainMax^\policy
}\right\}.
\end{equation}

Since the optimization problem in \eqref{eq:problem} is formulated in terms of average powers, \eqref{eq:power-upper-bound} is reformulated as follows.
Recall that \(\gls{Power}_\stageEh^\policy(b)\) and \(\gls{Power}_\stageBt^\policy(b)\) are deterministic, which leads to \(\gls{PowerAverage}_\stageEh^\policy(b) \leq \gls{PowerAverage}_{\stageEh,\maximum}^\policy\) and \(\gls{PowerAverage}_\stageBt^\policy(b) \leq \gls{PowerAverage}_{\stageBt,\maximum}^\policy\) for
\begin{equation}\label{eq:power-upper-bound-training}
    \gls{PowerAverage}_{\stageEh,\maximum}^\policy
    = \gls{PowerAverage}_{\stageBt,\maximum}^\policy
    = \min\left\{
        \gls{Power}_\maximum,
        \frac{
            \gls{Power}_\threshold
        }{
            \glsSubIncident{Gain} \glsSubTx{Gain}
            \TileGainMax^\policy
        }
    \right\}.
\end{equation}
For the data transmission stage, where \(\gls{Power}_\stageDt^\policy(b)\) is random, \eqref{eq:power-upper-bound} is satisfied with probability at least \(\epsilon \in (0, 1)\) if
\begin{equation}\label{eq:probability-upper-bound-data}
    \mathrm{Pr} \left\{\gls{Power}_\stageDt^\policy(b)
    \leq \min\left\{
        \gls{Power}_\maximum,
        \frac{
            \gls{Power}_\threshold
        }{
            \glsSubIncident{Gain} \glsSubTx{Gain}
            \TileGainMax^\policy
        }
    \right\} \right\} \geq \epsilon.
\end{equation}
Using the CDF of \(\gls{Power}_\stageDt^\policy(b)\), given by \(\mathrm{Pr}\{\gls{Power}_\stageDt^\policy(b) \leq p\} = 1-e^{-\frac{p}{\gls{PowerAverage}_\stageDt^\policy(b)}}\), \eqref{eq:probability-upper-bound-data} is reformulated as follows
\begin{equation}\label{eq:power-upper-bound-data}
    \gls{PowerAverage}_\stageDt^\policy(b) \leq \gls{PowerAverage}_{\stageDt,\maximum}^\policy = \frac{
        \min\left\{
            \gls{Power}_\maximum,
            \frac{
                \gls{Power}_\threshold
            }{
                \glsSubIncident{Gain} \glsSubTx{Gain}
                \TileGainMax^\policy
            }
        \right\}
    }{
        \ln\left(\frac{1}{1-\epsilon}\right)
    }.
\end{equation}

\subsection{RIS POWER CONSUMPTION}\label{sec:ris-power-consumption}
\noindent
Self-sustainable operation of the \gls{ris} can be achieved if the energy consumption of the \gls{ris} does not exceed the energy harvested by the \gls{ris}.
In the following, we review the state-of-the-art models for \gls{ris} power consumption and propose a suitable model for this work.

It is difficult to derive a general model for \gls{ris} power consumption because the model strongly depends on the hardware implementation of the \gls{ris} \cite{albanese2024aresautonomousris} and current \gls{ris} prototypes typically do not focus on energy efficiency.
However, there are different approaches in the literature for modeling the power consumption, focusing on both the static and dynamic power consumption of the \gls{ris}.
In the simplest static model, the power consumption scales linearly with the number of unit cells.
This model can be extended by a dynamic component that takes into account the energy required for reconfiguration of the unit cells~\cite{ntontin2022millimeterwavevs,ntontin2023timeunitcell,ntontin2023wirelessenergyharvesting}.
Another approach for modeling the dynamic component is to adopt the power consumption of the phase manipulating hardware components, e.g., the PINs diodes, at each unit cell~\cite{albanese2024aresautonomousris}.
Depending on the desired phase shift for the reflected wave, each PIN diode is switched either \emph{on} or \emph{off}.
Assuming \(\gls{Number}_\bits\) PIN diodes per unit cell, realizing \(2^{\gls{Number}_\bits}\) different phase shifts, let \(i_{b,n} \in \{0, 1, \dotsc, 2^{\gls{Number}_\bits} - 1\}\) denote the index of the phase shift at the \(n\)th unit cell applied for codebook \(b\).
Then, the number of active (\emph{on} state) PIN diodes at the \(n\)th unit cell is given by~\cite{albanese2024aresautonomousris}
\begin{equation}\label{eq:number-on-unit-cells}
    \gls{Number}_\diode(b,n) = \left(i_{b,n} - \sum_{k=1}^{\gls{Number}_\bits} \left\lfloor\frac{i_{b,n}}{2^k}\right\rfloor\right).
\end{equation}
Hence, the total power consumption of all unit cells at the \(m\)th tile is given by
\begin{equation}
    \gls{PowerConsumed}_{\reflecting,m}(b) = \sum_{n\in\gls{SetCells}_m} \gls{PowerConsumed}_\cell \gls{Number}_\diode(b,n),
\end{equation}
where \(\gls{PowerConsumed}_\cell\) denotes the power consumption of an active PIN diode.

In this work, we adopt the above model for the \gls{ris} unit cells and, in order to take the \gls{rf} combining network into account, extend it by the power consumption of the \gls{rf} phase shifters.
In particular, we have \(\abs*{\gls{SetCells}_m} - 1\) phase shifters with an \(\gls{Number}_\bits\)-bit resolution at the \(m\)th tile, each consuming a power of \(\gls{PowerConsumed}_\shifter\) per bit \cite{lopez2022dynamicrfcombining}.
Thus, the power consumption of the \gls{rf} combining network at the \(m\)th tile is given by \(\gls{PowerConsumed}_{\harvesting,m} = (\abs*{\gls{SetCells}_m} - 1) \gls{Number}_\bits \gls{PowerConsumed}_\shifter\).

In addition, we assume that the RIS power consumption has a static component, denoted by \(\gls{PowerConsumed}_\static\), which accounts for the power consumption of a central RIS controller.
As a result, the total power consumption of the \gls{ris} for codeword \(b\) is given by
\begin{equation}\label{eq:power-consumed-dynamic-harvesting}
    \gls{Power}_{\consuming,\stage}^\policy(b) = \gls{PowerConsumed}_\static + \sum_{m \in \gls{SetTiles}_{\harvesting,\stage}^\policy} \gls{PowerConsumed}_{\harvesting,m} + \sum_{m \in \gls{SetTiles}_{\reflecting,\stage}^\policy} \gls{PowerConsumed}_{\reflecting,m}(b).
\end{equation}

%% file: sections/optimization_problem.tex
\noindent
We aim to minimize the average transmit power at the \gls{bs} while ensuring self-sustainable operation of the \gls{ris}, a minimum \gls{snr} for beam training, and a minimum effective data rate, cf.~\eqref{eq:problem}.
To this end, the goal is to determine the optimal splitting ratio \(\rho_\policy\) and the optimal transmit powers \(\gls{PowerAverage}_\stage^\policy(b)\) for each codeword and each stage of the transmission frame.
In the following, we specialize the optimization problem in \eqref{eq:problem} to each splitting scheme and propose an algorithm for obtaining the respective solution.
We begin with general definitions and reformulations that hold for all splitting schemes.

\subsection{RIS GAIN AND TILE GAIN}\label{sec:gains}
\noindent
In order to solve the optimization problem in \eqref{eq:problem}, we need to define \gls{ris} gain \(\gls{Gain}_\ris^\policy\) in \eqref{eq:snr} and tile gain \(\gls{Gain}_{\tile,m}^\policy\) in \eqref{eq:power-rf}.
The latter influences \(\gls{PowerAverage}\DcVarVar\) in \eqref{eq:power_dc} and thus the harvested power.
The former is used in \eqref{eq:constraint-training} and \eqref{eq:constraint-rate}, and is specified as follows, assuming a unit amplitude factor for the \gls{ris} unit cells~\cite{xie2023sumratemaximization,tyrovolas2023zeroenergyreconfigurablea,ntontin2023timeunitcell}.
For \(\policy \in \{\policyTs, \policyEs\}\), we have
\begin{equation}\label{eq:gain-ris-time-element}
    \gls{Gain}_\ris^\policy(l) = \abs*{
        \sum_{m \in \gls{SetTiles}_{\reflecting,\stage}^\policy} \sum_{n \in \gls{SetCells}_m} e^{j(\gls{PhaseIncident} + \gls{PhaseReflected}(l) + \gls{ShiftUnitCell}(b_l))}
    }^2,
\end{equation}
where \(\gls{PhaseIncident}\) and \(\gls{PhaseReflected}(l)\) denote the relative phase shifts at the \(n\)th unit cell caused by the incident wave and the reflected wave for user location \(l \in \gls{SetLocations}\), respectively, and \(\gls{ShiftUnitCell}(b_l)\) denotes the phase shift applied at the \(n\)th unit cell for codeword \(b_l\) selected for location \(l\).
For \(\policy = \policyPs\), the \gls{ris} gain includes the power splitting factor, leading to
\begin{equation}\label{eq:gain-ris-power}
    \gls{Gain}_\ris^\policyPs(l) = \rho_\policyPs \abs*{
        \sum_{m \in \gls{SetTiles}_{\reflecting,\stage}^\policyPs} \sum_{n \in \gls{SetCells}_m} e^{j(\gls{PhaseIncident} + \gls{PhaseReflected}(l) + \gls{ShiftUnitCell}(b_l))}
    }^2.
\end{equation}

For tile gain \(\gls{Gain}_{\tile,m}^\policy\), we take into account the insertion loss of the phase shifters in the \gls{rf} combining network, cf. \eqref{eq:insertion_loss}, that apply phase shift \(\gls{ShiftCombining}\) at the \(n\)th unit cell in the \(m\)th tile.
Hence, the tile gain for \(\policy \in \{\policyTs, \policyEs\}\) is given by
\begin{equation}\label{eq:gain-tile-time-element}
    \gls{Gain}_{\tile,m}^\policy =
    \frac{1}{\abs*{\gls{SetCells}_m}}
    \abs*{
        \sum_{n \in \gls{SetCells}_m}
        \sqrt{\gls{LossInsertion}_n} e^{j\gls{ShiftCombining}}
        e^{j\gls{PhaseIncident}}
    }^2, m \in \gls{SetTiles}_{\harvesting,\stage}^\policy.
\end{equation}
The division by \(\abs*{\gls{SetCells}_m}\) is required to satisfy the law of conservation of energy, i.e., the combined power at the \(m\)th tile cannot be larger than the sum power of all incident signals at the \(m\)th tile.
Similar to \eqref{eq:gain-ris-power}, the tile gain for \(\policy = \policyPs\) includes the power splitting factor and is given by
\begin{equation}\label{eq:gain-tile-power}
    \gls{Gain}_{\tile,m}^\policyPs =
    \frac{1 - \rho_\policyPs}{\abs*{\gls{SetCells}_m}}
    \abs*{
        \sum_{n \in \gls{SetCells}_m}
        \sqrt{\gls{LossInsertion}_n} e^{j\gls{ShiftCombining}}
        e^{j\gls{PhaseIncident}}
    }^2, m \in \gls{SetTiles}_{\harvesting,\stage}^\policyPs.
\end{equation}

Furthermore, we assume \(\gls{PhaseIncident}\) to be known because of the fixed locations of the \gls{bs} and the \gls{ris} and due to the \gls{los}-dominated \gls{bs}-\gls{ris} channel.
Thus, the \gls{rf} combining network for energy harvesting realizes coherent combining up to a quantization error caused by the quantization of \(\gls{ShiftCombining}\) based on \(\gls{Number}_\bits\) bits.
However, as shown in Appendix~\ref{sec:impact-phase-quantization}, the quantization error is usually small if \(\gls{Number}_\bits \geq 3\).
In this case, one can assume that \(\TileGainMin^\policy = \min_{m \in \gls{SetTiles}_{\harvesting,\stage}^\policy} \gls{Gain}_{\tile,m}^\policy\) is a tight lower bound on \(\gls{Gain}_{\tile,m}^\policy\).
Hence,
\begin{equation}\label{eq:tile-gain-min}
    \abs*{\gls{SetTiles}_{\harvesting,\stage}^\policy}
    \bar{\gls{FunctionModel}}_\stage(\glsSubIncident{Gain} \glsSubTx{Gain} \TileGainMin^\policy \gls{PowerAverage}_\stage^\policy)
    \lessapprox \negthinspace \sum_{m \in \gls{SetTiles}_{\harvesting,\stage}^\policy} \negthinspace \negthinspace
    \bar{\gls{FunctionModel}}_\stage(\glsSubIncident{Gain} \glsSubTx{Gain} \gls{Gain}_{\tile,m}^\policy \gls{PowerAverage}_\stage^\policy),
\end{equation}
which will be used for solving problem \eqref{eq:problem} in Sections~\ref{sec:optimal_es_ps} and~\ref{sec:optimal_ts}.

\subsection{MINIMUM SNRS FOR BEAM TRAINING AND DATA TRANSMISSION}
\noindent
Constraints \eqref{eq:constraint-training} and \eqref{eq:constraint-rate} must hold for all locations within the coverage area.
In other words, for a given codeword \(b \in \gls{Codebook}_\policy\), the transmit power \(\gls{PowerAverage}_\stage^\policy(b)\) must be sufficiently large to satisfy \eqref{eq:constraint-training} and \eqref{eq:constraint-rate} for all locations in the subarea illuminated by \(b\).
Let us denote these locations by \(\gls{SetLocations}_b\).
Then, using \(\gls{Snr}_{\stageDt,\minimum}^\policy = 2^{\gls{RateEffective}_\minimum/(\bar{\gls{TimeDuration}}_\stageDt^\policy / \gls{TimeDuration}_\fr^\policy)} - 1\), \eqref{eq:constraint-training} and \eqref{eq:constraint-rate} can be written as follows
\begin{align}
    \gls{Snr}_{\stageBt,\minimum}
    &\leq \frac{
        \glsSubRx{Gain}
        \glsSubTx{Gain}
        \glsSubIncident{Gain}
        \gls{Gain}_{\gls{SetLocations}_b}
    }{
        \gls{PowerNoise}
    }
    \gls{PowerAverage}_\stageBt^\policy(b),
    \ \forall b \in \gls{Codebook}_\policy\label{eq:constraint-training-subareas}\\
    \gls{Snr}_{\stageDt,\minimum}
    &\leq \frac{
        \glsSubRx{Gain}
        \glsSubTx{Gain}
        \glsSubIncident{Gain}
        \gls{Gain}_{\gls{SetLocations}_b}
    }{
        \gls{PowerNoise}
    }
    \gls{PowerAverage}_\stageDt^\policy(b),
    \ \forall b \in \gls{Codebook}_\policy\label{eq:constraint-rate-subareas},
\end{align}
where \(
\gls{Gain}_{\gls{SetLocations}_b}
= \min_{l \in \gls{SetLocations}_b} \left(
    \gls{Gain}_\reflected(l)
    \gls{Gain}_\ris^\policy(l)
\right)
\).

\subsection{OPTIMAL ES AND PS SCHEMES}\label{sec:optimal_es_ps}
\noindent
In this section, problem \eqref{eq:problem} is reformulated for the \gls{ElementSplit} and \gls{PowerSplit} schemes.
Thus, in the following, we assume \(\policy \in \{\policyEs, \policyPs\}\) and \(\stage \in \{\stageBt, \stageDt\}\).

In order to solve the optimization problem in \eqref{eq:problem} for the \gls{ElementSplit} and \gls{PowerSplit} schemes, we need to further specify constraint \eqref{eq:constraint-power}, requiring that the harvested power available at the \gls{ris} is not smaller than the \gls{ris} power consumption.
To this end, we note that most intervals of the transmission frame can be considered as \emph{active} intervals, i.e., the \gls{bs} transmits a signal that can be used for energy harvesting.
During these \emph{active} intervals, where a particular codeword \(b \in \gls{Codebook}_\policy\) is employed at the \gls{ris}, the power consumption of the \gls{ris} is given by \(\gls{Power}_{\consuming,\stage}^\policy(b)\) in \eqref{eq:power-consumed-dynamic-harvesting}.
Moreover, based on \eqref{eq:tile-gain-min}, the harvested power is given by
\begin{equation}    
    \gls{PowerAverage}_{\harvesting,\stage}^\policy(b)
    = \gls{Gain}_{\dc,\stage}^\policy
    \abs*{\gls{SetTiles}_{\harvesting,\stage}^\policy}
    \bar{\gls{FunctionModel}}_\stage(
        \glsSubIncident{Gain} \glsSubTx{Gain} \TileGainMin^\policy
        \gls{PowerAverage}_\stage^\policy(b)
    ).
\end{equation}
As a result, during the \emph{active} intervals of the frame, self-sustainable operation of the \gls{ris} is guaranteed if 
\begin{equation}\label{eq:power-harvested-bound-element-split}
    \gls{Power}_{\consuming,\stage}^\policy(b) \leq \gls{PowerAverage}_{\harvesting,\stage}^\policy(b),\ \forall\, b \in \gls{Codebook}_\policy.
\end{equation}

It remains to account for the \emph{inactive} intervals of the frame, i.e., the intervals where the \gls{ris} changes the phase-shift configuration and where the user transmits training feedback.
These intervals have a total duration of \(\glsSubResponse{TimeDuration}(\abs*{\gls{Codebook}_\policy} + 1) + \glsSubDelay{TimeDuration}\) and the power consumption is only determined by the static component \(\gls{PowerConsumed}_\static\).
Since, compared to the full frame, the duration of the \emph{inactive} intervals is relatively short and the total energy required for these intervals is relatively small, we propose to add\footnote{Technically, adding \(\tilde{\gls{PowerConsumed}}_\static^\policy\) to \(\gls{Power}_{\consuming,\stage}^\policy(b)\) is a heuristic approach that may lead to a suboptimal solution of \eqref{eq:problem}. However, \(\tilde{\gls{Energy}}_\consuming^\policy\) is typically small compared to the total energy consumed in a frame, such that the expected deviation from the optimal solution is small. Moreover, this approach allows to solve the optimization problem via an efficient algorithm, see Section~\ref{sec:optimal-solution}.} a small static component, denoted by \(\tilde{\gls{PowerConsumed}}_\static^\policy\), to \(\gls{Power}_{\consuming,\stage}^\policy(b)\) in \eqref{eq:power-harvested-bound-element-split}.
As a result, the additional power \(\tilde{\gls{PowerConsumed}}_\static^\policy\) harvested during the \emph{active} intervals can be fed into a short-term energy storage, providing power supply during the \emph{inactive} intervals.
The additional energy required is given by \(\tilde{\gls{Energy}}_\consuming^\policy = \gls{PowerConsumed}_\static \left(\glsSubResponse{TimeDuration}(\abs*{\gls{Codebook}_\policy} + 1) + \glsSubDelay{TimeDuration}\right)\).
In order to keep \(\tilde{\gls{PowerConsumed}}_\static^\policy\) as small as possible, \(\tilde{\gls{Energy}}_\consuming^\policy\) is divided by the total duration of the \emph{active} intervals, resulting in
\begin{equation}
    \tilde{\gls{PowerConsumed}}_\static^\policy
    = \frac{\tilde{\gls{Energy}}_\consuming^\policy}{
        \gls{TimeDuration}_\sym \abs*{\gls{Codebook}_\policy}
        + \gls{TimeDuration}_\stageDt
    }.
\end{equation}
Hence, constraint \eqref{eq:constraint-power} for self-sustainable operation of the \gls{ris} is satisfied if
\begin{equation}\label{eq:power-harvested-bound-element-split-added}
    \tilde{\gls{Power}}_{\consuming,\stage}^\policy(b) = \gls{Power}_{\consuming,\stage}^\policy(b) + \tilde{\gls{PowerConsumed}}_\static^\policy \leq \gls{PowerAverage}_{\harvesting,\stage}^\policy(b),\ \forall\, b \in \gls{Codebook}_\policy.
\end{equation}

Solving \eqref{eq:constraint-training-subareas}, \eqref{eq:constraint-rate-subareas}, and \eqref{eq:power-harvested-bound-element-split-added} for \(\gls{PowerAverage}_\stageBt^\policy(b)\) and \(\gls{PowerAverage}_\stageDt^\policy(b)\) results in
\begin{align}
    \gls{PowerAverage}_\stageBt^\policy(b) &\geq \gls{PowerAverage}_{\stageBt,\minimum}^\policy(b)\nonumber\\
    &= \max\left\{
        \frac{
            \bar{\gls{FunctionModel}}_\stageBt^{-1}\left(
                \frac{
                    \tilde{\gls{Power}}_{\consuming,\stageBt}^\policy(b)
                }{
                    \gls{Gain}_{\dc,\stageBt}^\policy
                    \abs*{\gls{SetTiles}_{\harvesting,\stageBt}^\policy}
                }
            \right)
        }{
            \glsSubIncident{Gain} \glsSubTx{Gain} \TileGainMin^\policy
        },
        \frac{
            \gls{Snr}_{\stageBt,\mathrm{min}} \sigma^2
        }{
            \glsSubRx{Gain}
            \glsSubTx{Gain}
            \glsSubIncident{Gain}
            \gls{Gain}_{\gls{SetLocations}_b}
        }
    \right\}\label{eq:element-lower-bound-training}\\
    \gls{PowerAverage}_\stageDt^\policy(b) &\geq \gls{PowerAverage}_{\stageDt,\minimum}^\policy(b)\nonumber\\
    &= \max\left\{
        \frac{
            \bar{\gls{FunctionModel}}_\stageDt^{-1}\left(
                \frac{
                    \tilde{\gls{Power}}_{\consuming,\stageDt}^\policy(b)
                }{
                    \gls{Gain}_{\dc,\stageDt}^\policy
                    \abs*{\gls{SetTiles}_{\harvesting,\stageDt}^\policy}
                }
            \right)
        }{
            \glsSubIncident{Gain} \glsSubTx{Gain} \TileGainMin^\policy
        },
        \frac{
            \gls{Snr}_{\stageDt,\minimum}^\policy \sigma^2
        }{
            \glsSubRx{Gain}
            \glsSubTx{Gain}
            \glsSubIncident{Gain}
            \gls{Gain}_{\gls{SetLocations}_b}
        }
    \right\}.\label{eq:element-lower-bound-data}
\end{align}
In addition, adopting \(\gls{TimeDuration}_\stageEh^\policy = 0\) and \(\gls{PowerAverage}_\stageEh^\policy = \gls{PowerAverage}_{\stageEh,\minimum}^\policy = 0\) for \(\policy \in \{\policyEs, \policyPs\}\), one can substitute \(\gls{PowerAverage}_{\stage,\minimum}^\policy\) for \(\gls{PowerAverage}_\stage^\policy\) in \eqref{eq:power-bs-average}.
Then, the optimization problem for the \gls{ElementSplit} and \gls{PowerSplit} schemes simplifies to
\begin{subequations}\label{eq:problem-element-power}
    \begin{align}
        \min_{
            \rho_\policy
        }
        &\quad
        \frac{1}{\abs*{\gls{Codebook}_\policy}}
        \sum_{b \in \gls{Codebook}_\policy} \left(
            \abs*{\gls{Codebook}_\policy}
            \frac{\gls{TimeDuration}_\sym}{\gls{TimeDuration}_\fr^\policy}
            \gls{PowerAverage}_{\stageBt,\minimum}^\policy(b)
            + \frac{
                \gls{TimeDuration}_\stageDt^\policy
            }{
                \gls{TimeDuration}_\fr^\policy
            }
            \gls{PowerAverage}_{\stageDt,\minimum}^\policy(b)
        \right)\\
        \mathrm{s.t.}
        &\quad
        \gls{PowerAverage}_{\stage,\minimum}^\policy(b)
        \leq \gls{PowerAverage}_{\stage,\mathrm{max}}^\policy,\ \forall\, \stage \in \{\stageBt, \stageDt\},\ \forall\, b \in \gls{Codebook}_\policy\label{eq:problem_element-constraint}.
    \end{align}
\end{subequations}
Note that constraints \cref{eq:constraint-training,eq:constraint-rate,eq:constraint-power} are captured by \(\gls{PowerAverage}_{\stageBt,\minimum}^\policy(b)\) and \(\gls{PowerAverage}_{\stageDt,\minimum}^\policy(b)\).
An algorithm for solving \eqref{eq:problem-element-power} is provided in Section~\ref{sec:optimal-solution}.

\subsection{OPTIMAL TS SCHEME}\label{sec:optimal_ts}
\noindent
In order to solve the optimization problem in \eqref{eq:problem} for the \gls{TimeSplit} scheme, constraint \eqref{eq:constraint-power} for self-sustainable operation of the \gls{ris} is satisfied as follows.
Since power is only harvested during stage \(\stage = \stageEh\), we assume a short-term energy storage and require that the energy harvested per transmission frame is not smaller than the energy consumed per transmission frame.
For a particular frame where codeword \(b^\prime\) is selected for data transmission, based on \eqref{eq:power-consumed-dynamic-harvesting}, the consumed energy at the \gls{ris} for the \gls{TimeSplit} scheme is given by
\begin{multline}\label{eq:power-consumed-average}
    \gls{Energy}_{\consuming}^\policyTs(b^\prime)
    = \gls{TimeDuration}_\fr^\policyTs \gls{PowerConsumed}_\static
    + \gls{TimeDuration}_\stageEh^\policyTs
    \sum_{m \in \gls{SetTiles}_{\harvesting,\stageEh}^\policyTs}
    \gls{PowerConsumed}_{\harvesting,m}\\
    + \gls{TimeDuration}_\sym
    \sum_{b \in \gls{Codebook}_\policyTs}
    \sum_{m \in \gls{SetTiles}_{\reflecting,\stageBt}^\policyTs}
    \gls{PowerConsumed}_{\reflecting,m}(b)
    + \gls{TimeDuration}_\stageDt^\policyTs
    \sum_{m \in \gls{SetTiles}_{\reflecting,\stageDt}^\policyTs}
    \gls{PowerConsumed}_{\reflecting,m}(b^\prime).
\end{multline}
Moreover, the energy harvested in a frame for the \gls{TimeSplit} scheme is given by
\begin{equation}\label{eq:power-harvested-ts}
    \gls{Energy}_\harvesting^\policyTs
    = \gls{TimeDuration}_\stageEh^\policyTs
    \gls{Gain}_{\dc,\stageEh}^\policyTs
    \abs*{\gls{SetTiles}_{\harvesting,\stageEh}^\policyTs}
    \bar{\gls{FunctionModel}}_\stageEh(
        \glsSubIncident{Gain} \glsSubTx{Gain} \TileGainMin^\policyTs
        \gls{PowerAverage}_\stageEh^\policyTs
    ),
\end{equation}
which is independent of the codeword employed at the \gls{ris}.
Thus, based on \eqref{eq:power-consumed-average} and \eqref{eq:power-harvested-ts}, constraint \eqref{eq:constraint-power} is satisfied if \(
\gls{Energy}_\consuming^\policyTs(b^\prime)/\gls{TimeDuration}_\stageEh^\policyTs
\leq \gls{Gain}_{\dc,\stageEh}^\policyTs \abs*{\gls{SetTiles}_{\harvesting,\stageEh}^\policyTs}
\bar{\gls{FunctionModel}}_\stageEh(
    \glsSubIncident{Gain} \glsSubTx{Gain} \TileGainMin^\policyTs
    \gls{PowerAverage}_\stageEh^\policyTs
)
\), which can be solved for \(\gls{PowerAverage}_\stageEh^\policyTs\) as follows
\begin{equation}\label{eq:time-lower-bound-harvesting}
    \gls{PowerAverage}_\stageEh^\policyTs
    \geq \gls{PowerAverage}_{\stageEh,\minimum}^\policyTs(b^\prime)
    = \frac{
        \bar{\gls{FunctionModel}}_\stageEh^{-1}\left(
            \frac{
                \gls{Energy}_\consuming^\policyTs(b^\prime)
            }{
                \gls{TimeDuration}_\stageEh^\policyTs
                \gls{Gain}_{\dc,\stageEh}^\policyTs
                \abs*{\gls{SetTiles}_{\harvesting,\stageEh}^\policyTs}
            }
        \right)
    }{
        \glsSubIncident{Gain} \glsSubTx{Gain} \TileGainMin^\policyTs
    }.
\end{equation}
Moreover, solving \eqref{eq:constraint-training-subareas} and \eqref{eq:constraint-rate-subareas} for \(\gls{PowerAverage}_\stageBt^\policyTs(b)\) and \(\gls{PowerAverage}_\stageDt^\policyTs(b)\), respectively, leads to
\begin{align}
    \gls{PowerAverage}_\stageBt^\policyTs(b)
    &\geq \gls{PowerAverage}_{\stageBt,\minimum}^\policyTs(b)
    = \frac{
        \gls{Snr}_{\stageBt,\mathrm{min}} \sigma^2
    }{
        \glsSubRx{Gain}
        \glsSubTx{Gain}
        \glsSubIncident{Gain}
        \gls{Gain}_{\gls{SetLocations}_b}
    }\label{eq:time-lower-bound-training}\\
    \gls{PowerAverage}_\stageDt^\policyTs(b)
    &\geq \gls{PowerAverage}_{\stageDt,\minimum}^\policyTs(b)
    = \frac{
        \gls{Snr}_{\stageDt,\minimum}^\policyTs \sigma^2
    }{
        \glsSubRx{Gain}
        \glsSubTx{Gain}
        \glsSubIncident{Gain}
        \gls{Gain}_{\gls{SetLocations}_b}
    }.\label{eq:time-lower-bound-data}
\end{align}
As a result, substituting \(\gls{PowerAverage}_{\stage,\minimum}^\policyTs\) in \eqref{eq:time-lower-bound-harvesting}, \eqref{eq:time-lower-bound-training}, and \eqref{eq:time-lower-bound-data} for \(\gls{PowerAverage}_\stage^\policyTs\) in \eqref{eq:power-bs-average} simplifies the optimization problem for the \gls{TimeSplit} scheme to
\begin{subequations}\label{eq:problem-time}
    \begin{align}
        \min_{
            \rho_\policyTs
        }
        &\quad
        \gls{PowerAverage}_{\bs,\minimum}^\policyTs
        \label{eq:problem-time-objective}\\
        \mathrm{s.t.}
        &\quad
        \gls{PowerAverage}_{\stage,\minimum}^\policyTs(b)
        \leq \gls{PowerAverage}_{\stage,\mathrm{max}}^\policyTs \forall \stage \in \{\stageEh, \stageBt, \stageDt\}, b \in \gls{Codebook}_\policyTs,
        \label{eq:problem-time-constraint}
    \end{align}
\end{subequations}
where
\begin{multline}\label{eq:problem-simplified-objective}
    \gls{PowerAverage}_{\bs,\minimum}^\policy
    = \frac{1}{\abs*{\gls{Codebook}_\policy}}
    \sum_{b \in \gls{Codebook}_\policy} \left(
        \frac{
            \gls{TimeDuration}_\stageEh^\policy
        }{
            \gls{TimeDuration}_\fr^\policy
        }
        \gls{PowerAverage}_{\stageEh,\minimum}^\policy(b)\right.\\
        \left.+ \abs*{\gls{Codebook}_\policy}
        \frac{\gls{TimeDuration}_\sym}{\gls{TimeDuration}_\fr^\policy}
        \gls{PowerAverage}_{\stageBt,\minimum}^\policy(b)
        + \frac{
            \gls{TimeDuration}_\stageDt^\policy
        }{
            \gls{TimeDuration}_\fr^\policy
        }
        \gls{PowerAverage}_{\stageDt,\minimum}^\policy(b)
    \right).
\end{multline}
An algorithm for solving \eqref{eq:problem-time} is provided in Section~\ref{sec:optimal-solution}.

\subsection{OPTIMAL SOLUTION}\label{sec:optimal-solution}
\noindent
The simplified optimization problems in \eqref{eq:problem-element-power} and \eqref{eq:problem-time} are still not straightforward to solve analytically.
In addition, due to the non-convex rectifier model, these problems cannot be solved numerically using frameworks for convex optimization.
However, the problems in \eqref{eq:problem-element-power} and \eqref{eq:problem-time} follow the same structure, in particular, both problems have only one feasibility constraint and one optimization variable defined in the interval \([0, 1]\).
Moreover, using \(\gls{PowerAverage}_{\stageEh,\minimum}^\policy(b) = 0\) for \(\policy \in \{\policyEs, \policyPs\}\), \eqref{eq:problem-element-power} and \eqref{eq:problem-time} are equivalent and can be written as
\begin{subequations}\label{eq:problem-reformulated}
    \begin{align}
        \min_{
            \rho_\policy
        }
        &\quad
        \gls{PowerAverage}_{\bs,\minimum}^\policy
        \label{eq:problem-reformulated-objective}\\
        \mathrm{s.t.}
        &\quad
        \gls{PowerAverage}_{\stage,\minimum}^\policy(b)
        \leq \gls{PowerAverage}_{\stage,\mathrm{max}}^\policy \forall \stage \in \{\stageEh, \stageBt, \stageDt\}, b \in \gls{Codebook}_\policy,
        \label{eq:problem-reformulated-constraint}
    \end{align}
\end{subequations}
where \(\gls{PowerAverage}_{\bs,\minimum}^\policy\) is given in \eqref{eq:problem-simplified-objective}.
Thus, \eqref{eq:problem-reformulated} solves the original problem in \eqref{eq:problem} for each considered splitting scheme.
Since there is only one optimization variable and one feasibility constraint, the optimal solution of \eqref{eq:problem-reformulated} can be found efficiently via grid search.
More specifically, as summarized in Algorithm~\ref{alg:grid-search}, we propose the following procedure to solve \eqref{eq:problem-reformulated}.
Let \(\mathcal{I}\) denote the set of equally spaced samples of \(\rho_\policy \in [0, 1]\), where the sampling distance is denoted by \(\gls{Accuracy} \in \mathbb{R}\).
Then, the set of feasible samples, denoted by \(\mathcal{F}\), is given by each value of \(\rho_\policy \in \mathcal{I}\) that satisfies \eqref{eq:problem-reformulated-constraint}, or, if \(\policy = \policyEs\), \eqref{eq:problem-reformulated-constraint} and \eqref{eq:feasible_set_reflecting}.
Finally, the objective function \eqref{eq:problem-reformulated-objective} is computed for each value of \(\rho_\policy \in \mathcal{F}\), and the minimum value is returned.

\begin{algorithm}[t]
    \caption{Grid search over \(\rho_\policy\)}\label{alg:grid-search}
    \DontPrintSemicolon
    \KwIn{\(\text{Sampling distance } \gls{Accuracy}\), \(\text{splitting scheme } \policy\)}
    \KwOut{\(\text{Optimal } \rho_\policy \text{ that solves problem \eqref{eq:problem-reformulated}}\)}
    \(\mathcal{I} \gets {\text{Samples of } [0, 1] \text{ based on } \gls{Accuracy}}\)\;
    \eIf{\(\policy = \policyEs\)}{
        \(\mathcal{F} \gets \{\rho_\policy \in \mathcal{I} \mid \eqref{eq:problem-reformulated-constraint}, \eqref{eq:feasible_set_reflecting} \text{ true}\}\)\;
    }{
        \(\mathcal{F} \gets \{\rho_\policy \in \mathcal{I} \mid \eqref{eq:problem-reformulated-constraint} \text{ true}\}\)\;
    }
    \KwRet{\(\argmin_{\rho_\policy \in \mathcal{F}} \gls{PowerAverage}_{\bs,\minimum}^\policy\)}
\end{algorithm}

The proposed algorithm involves the computation of \(\gls{PowerAverage}_{\bs,\minimum}^\policy\), \(\gls{PowerAverage}_{\stage,\minimum}^\policy(b)\), and \(\gls{PowerAverage}_{\stage,\mathrm{max}}^\policy\) for each \(\rho_\policy \in \mathcal{F}\).
Based on the analytical expressions provided, these computations are straightforward, except for the integral in \eqref{eq:power-dc-average-data}, which is difficult to solve analytically.
However, \eqref{eq:power-dc-average-data} can be solved efficiently using standard tools for numerical integration.
Thus, the computational complexity of Algorithm~\ref{alg:grid-search} mainly depends on the sampling distance \(\gls{Accuracy}\) that determines the number of computations of \(\gls{PowerAverage}_{\bs,\minimum}^\policy\), \(\gls{PowerAverage}_{\stage,\minimum}^\policy(b)\), and \(\gls{PowerAverage}_{\stage,\mathrm{max}}^\policy\).
Therefore, the complexity is independent of the number of unit cells and the proposed scheme is suitable for large-scale \glspl{ris}.

%% file: sections/performance_evaluation.tex
\noindent
In this section, we evaluate the optimal solutions of problems \eqref{eq:problem-element-power} and \eqref{eq:problem-time}, i.e., the optimal splits and power allocations for the \gls{PowerSplit}, \gls{ElementSplit}, and \gls{TimeSplit} schemes.
First, we provide insight into the minimum powers required for energy harvesting, beam training, and data transmission, and their respective scaling with the splitting ratio.
Afterwards, we present the optimized solutions based on Algorithm~\ref{alg:grid-search} for three exemplary characteristics of the \gls{ris} power consumption, which highlight the advantages and disadvantages of each splitting scheme.
The adopted values of the system parameters are provided in Table~\ref{tab:parameters}, and the parameters for the power consumption model are given in the captions of the individual figures.

Various assumptions about \gls{ris} power consumption can be found in the literature, typically specified based on the power consumption of a unit cell, and ranging from nanowatts to milliwatts~\cite{ntontin2022millimeterwavevs,ma2022reconfigurableintelligentsurface,hu2021robustsecuresum,tyrovolas2023zeroenergyreconfigurablea,xu2022selfsustainablewireless,zou2022robustbeamformingoptimization,albanese2024aresautonomousris,xie2023sumratemaximization,zheng2023zeroenergydevice,ma2023wirelesspoweredintelligent}.
In this work, our goal is to demonstrate the impact of the assumptions made regarding the \gls{ris} power consumption. Hence, we consider \(\gls{PowerConsumed}_\static \in \{\qty{1}{\micro\watt}, \qty{80}{\micro\watt}\}\), \(\gls{PowerConsumed}_\cell = \{\qty{0}{\nano\watt}, \qty{10}{\nano\watt}, \qty{80}{\nano\watt}\}\), \(\gls{PowerConsumed}_\shifter = \{\qty{0}{\nano\watt}, \qty{10}{\nano\watt}, \qty{30}{\nano\watt}\}\), cf. Section~\ref{sec:evaluation-optimized}.
The main reason for considering relatively low values is that the adopted rectifier model has a low \gls{rf}-to-\gls{dc} efficiency of about \qty{10}{\percent}.
We note that this model ensures practical relevance because it is adapted to the considered operating frequency in the \gls{mmwave} band.
In contrast, other works assume a higher efficiency of, e.g., \qty{65}{\percent}~\cite{tyrovolas2023zeroenergyreconfigurablea} or \qty{80}{\percent}~\cite{liu2022jointtrajectorydesign}, but do not provide the details of the underlying rectifier.

Furthermore, we note that the proposed framework models practical imperfections, including quantized phase shifts, insertion losses, and a non-ideal \gls{dc} power management network.
Although a comprehensive evaluation of such imperfections is beyond the scope of this paper, an illustrative example is provided in Appendix~\ref{sec:imperfections}.
Moreover, the proposed framework can also be applied if further imperfections are considered, as long as \(\gls{PowerAverage}_{\bs,\minimum}^\policy\) and \(\gls{PowerAverage}_{\stage,\minimum}^\policy\) are available in analytical form or can be obtained by numerical simulations.

\subsection{POWER LOWER BOUNDS FOR HARVESTING AND REFLECTION}
\noindent
As shown in Section~\ref{sec:optimization}, the constraints of the optimization problem in \eqref{eq:problem} can be rewritten as lower bounds for the transmit power at the \gls{bs}, cf. \eqref{eq:element-lower-bound-training}, \eqref{eq:element-lower-bound-data}, \eqref{eq:time-lower-bound-harvesting}, \eqref{eq:time-lower-bound-training}, and \eqref{eq:time-lower-bound-data}.
One can gain interesting insight into the solution space of the optimization problem by plotting these lower bounds as a function of the splitting ratio.

\begin{table}
    \begin{center}
    \caption{System parameters.}
    \label{tab:parameters}
    \begin{tabular}{| l | l | l | l |}
        \hline
        Symbol & Value & Symbol & Value\\
        \hline
        \gls{Accuracy} & 0.001 & \((x_c, y_c, z_c)\) & \((\qty{-10}{\metre}, \qty{-50}{\metre}, \qty{75}{\metre})\)\\
        \(\gls{Number}_\cell\) & 900 & \((A_y, A_z)\) & \((\qty{100}{\metre}, \qty{50}{\metre})\)\\
        \(\gls{Power}_\threshold\) & \(\gls{FunctionModel}^{-1}(0.99 \glsSubSaturation{Power})\) & \(\gls{Velocity}\) & \qty{3}{\kilo\meter/\hour}\\
        \(\gls{Snr}_{\stageBt,\minimum}\) & \qty{10}{\deci\bel} & \(\gls{ParameterFrame}\) & 0.2\\
        \(\glsSubTx{Gain}\) & \qty{20}{\deci\bel} \cite{hu2021robustsecuresum} & \(\glsSubResponse{TimeDuration}\) & \qty{0.1}{\milli\second}\\
        \(\epsilon\) & 0.5 & \(\gls{Power}_\maximum\) & \qty{10}{\watt} \cite{pan2022selfsustainablereconfigurable}\\
        \(\gls{Wavelength}\) & \qty{0.01}{\metre} & \(\gls{LossInsertion}_0\) & \qty{-0.5}{\deci\bel}\cite{lopez2022dynamicrfcombining}\\
        \(\gls{Number}_\bits\) & 3 \cite{hu2021robustsecuresum} & \(\gls{RateEffective}_\minimum\) & \qty{3}{\bit/\second/\hertz}\\
        \(\gls{LossExponent}\) & 0.1 & \(\glsSubDelay{TimeDuration}\) & \qty{1}{\milli\second} \cite{ji2021severalkeytechnologies}\\
        \hline
    \end{tabular}
    \end{center}
\end{table}

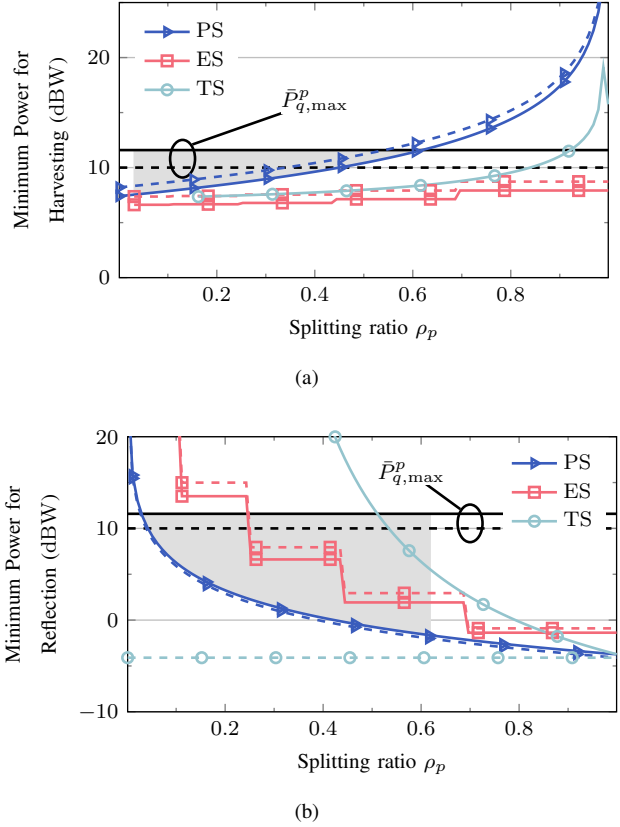
\begin{figure}[!t]
    \centering
    \subfloat[]{
        \begin{tikzpicture}
            \begin{axis}[
                line plot style,
                xlabel=Splitting ratio \(\rho_\policy\),
                ylabel style={align=center},
                ylabel={Minimum Power for\\Harvesting (\unit[qualifier-mode=combine]{\deci\bel\of{W}})},
                ymin=0,
                ymax=25,
                mark repeat=15,
                legend pos=north west,
                width=6.5cm,
                height=3.66cm,
                scale only axis,
            ]
                \addplot [black,dashed,forget plot,line width=1pt]
                table [col sep=comma] {data/upper_bounds-00-00.csv};
                \addplot [name path=setupper,black,line width=1pt]
                table [col sep=comma,forget plot] {data/upper_bounds-00-02.csv};

                \draw [line width=1pt] (0.15, 12.1) node [
                    ellipse,draw,anchor=north east,minimum width=0.3cm,minimum height=0.5cm
                ] {} -- (0.4, 16) node [fill=white,inner sep=0] {
                    \(\gls{PowerAverage}_{\stage,\maximum}^\policy\)
                };

                \addplot [
                    color1,mark=triangle,mark options={rotate=270,solid},
                    dashed,forget plot,line width=1pt
                ] table [col sep=comma] {data/tradeoffs-00-01.csv};
                \addplot [
                    name path=setlower,color1,mark=triangle,
                    mark options={rotate=270},line width=1pt
                ]
                table [col sep=comma] {data/tradeoffs-00-02.csv};
                \addlegendentry{\(\policyPs\)}
                
                \addplot [
                    color2,mark=square,dashed,forget plot,
                    line width=1pt,mark options={solid}
                ] table [col sep=comma] {data/tradeoffs-00-06.csv};
                \addplot [color2,mark=square,line width=1pt]
                table [col sep=comma] {data/tradeoffs-00-07.csv};
                \addlegendentry{\(\policyEs\)}
                
                \addplot [color3,mark=o,line width=1pt]
                table [col sep=comma] {data/tradeoffs-00-10.csv};
                \addlegendentry{\(\policyTs\)}

                \addplot [fill=lightgray!50] fill between [
                    of=setupper and setlower,
                    soft clip={domain=0.03:0.62}
                ];
            \end{axis}
        \end{tikzpicture}%
        \label{fig:tradeoff-harvesting-bounds}
    }\\
    \subfloat[]{
        \begin{tikzpicture}
            \begin{axis}[
                line plot style,
                xlabel=Splitting ratio \(\rho_\policy\),
                ylabel style={align=center},
                ylabel={Minimum Power for\\Reflection (\unit[qualifier-mode=combine]{\deci\bel\of{W}})},
                ymin=-10,
                ymax=20,
                mark repeat=15,
                width=6.5cm,
                height=3.66cm,
                scale only axis,
            ]
                \addplot [black,dashed,forget plot,line width=1pt]
                table [col sep=comma] {data/upper_bounds-00-00.csv};
                \addplot [name path=setupper,black,line width=1pt]
                table [col sep=comma,forget plot] {data/upper_bounds-00-02.csv};

                \draw [line width=1pt] (0.68, 12.1) node [
                    ellipse,draw,anchor=north west,minimum width=0.3cm,minimum height=0.5cm
                ] {} -- (0.58, 16) node [fill=white,inner sep=0] {
                    \(\gls{PowerAverage}_{\stage,\maximum}^\policy\)
                };

                \addplot [
                    color1,mark=triangle,mark options={rotate=270,solid},
                    dashed,forget plot,line width=1pt
                ] table [col sep=comma] {data/tradeoffs-00-03.csv};
                \addplot [name path=setlower,color1,mark=triangle,mark options={rotate=270},line width=1pt]
                table [col sep=comma] {data/tradeoffs-00-04.csv};
                \addlegendentry{\(\policyPs\)}

                \addplot [
                    color2,mark=square,dashed,forget plot,
                    mark options={solid},line width=1pt
                ] table [col sep=comma] {data/tradeoffs-00-08.csv};
                \addplot [color2,mark=square,line width=1pt]
                table [col sep=comma] {data/tradeoffs-00-09.csv};
                \addlegendentry{\(\policyEs\)}

                \addplot [color3,mark=o,dashed,forget plot,line width=1pt]
                table [col sep=comma] {data/tradeoffs-00-13.csv};
                \addplot [color3,mark=o,line width=1pt]
                table [col sep=comma] {data/tradeoffs-00-14.csv};
                \addlegendentry{\(\policyTs\)}

                \addplot [fill=lightgray!50] fill between [
                    of=setupper and setlower,
                    soft clip={domain=0.03:0.62}
                ];
            \end{axis}
        \end{tikzpicture}%
        \label{fig:tradeoff-data-bounds}
    }
    \caption{Power lower bounds for (a) energy harvesting and (b) signal reflection, averaged over all codewords. The bounds in (a) are given by \eqref{eq:time-lower-bound-harvesting} (\gls{TimeSplit}, solid line) and the first arguments of the \(\max\) operators in \eqref{eq:element-lower-bound-training} (\gls{PowerSplit}, \gls{ElementSplit}, dashed lines) and \eqref{eq:element-lower-bound-data} (\gls{PowerSplit}, \gls{ElementSplit}, solid lines). In (b), the bounds are given by \eqref{eq:time-lower-bound-training} (\gls{TimeSplit}, dashed line), \eqref{eq:time-lower-bound-data} (\gls{TimeSplit}, solid line), and the second arguments of the \(\max\) operators in \eqref{eq:element-lower-bound-training} (\gls{PowerSplit}, \gls{ElementSplit}, dashed lines) and \eqref{eq:element-lower-bound-data} (\gls{PowerSplit}, \gls{ElementSplit}, solid lines). The upper bounds \(\gls{PowerAverage}_{\stage,\maximum}^\policy\) are given by \eqref{eq:power-upper-bound-training} (dashed lines) and \eqref{eq:power-upper-bound-data} (solid lines). The shaded areas, as an example, indicate the joint feasible set for \(\rho_\policyPs\) and \(\gls{PowerAverage}_\stageDt^\policyPs(b)\) based on \eqref{eq:element-lower-bound-data}. \(\gls{PowerConsumed}_\static = \qty{1}{\micro\watt}\), \(\gls{PowerConsumed}_\cell = \qty{10}{\nano\watt}\), \(\gls{PowerConsumed}_\shifter = \qty{30}{\nano\watt}\), \(\gls{Number}_\tile=36\).}
    \label{fig:tradeoff}
\end{figure}

In Fig.~\ref{fig:tradeoff}, we show these lower bounds for both energy harvesting and signal reflection for each stage of the transmission frame.
In particular, Fig.~\ref{fig:tradeoff-harvesting-bounds} depicts the minimum powers required for energy harvesting, given by \eqref{eq:time-lower-bound-harvesting} (\gls{TimeSplit}, solid line) and the first arguments of the \(\max\) operators in \eqref{eq:element-lower-bound-training} (\gls{PowerSplit}, \gls{ElementSplit}, dashed lines) and \eqref{eq:element-lower-bound-data} (\gls{PowerSplit}, \gls{ElementSplit}, solid lines).
Similarly, Fig.~\ref{fig:tradeoff-data-bounds} shows the minimum powers required for signal reflection, i.e., for achieving the minimum \gls{snr} for beam training and the desired minimum data rate at the user, given by the second arguments of the \(\max\) operators in \eqref{eq:element-lower-bound-training} (\gls{PowerSplit}, \gls{ElementSplit}, dashed lines) and \eqref{eq:element-lower-bound-data} (\gls{PowerSplit}, \gls{ElementSplit}, solid lines) as well as \eqref{eq:time-lower-bound-training} (\gls{TimeSplit}, dashed lines) and \eqref{eq:time-lower-bound-data} (\gls{TimeSplit}, solid lines).
Moreover, the black lines in Fig.~\ref{fig:tradeoff} show the upper bounds for the transmit power, i.e., \(\gls{PowerAverage}_{\stage,\maximum}^\policy\), as defined in \eqref{eq:power-upper-bound-training} (dashed lines) and \eqref{eq:power-upper-bound-data} (solid lines).
The areas between the lower bounds and the upper bounds define the joint feasible set for \(\rho_\policy\) and \(\gls{PowerAverage}_\stage^\policy(b)\).
As an example, this is visualized by the shaded areas in Fig.~\ref{fig:tradeoff}, indicating the feasible set for \(\rho_\policyPs\) and \(\gls{PowerAverage}_\stageDt^\policyPs(b)\) based on \eqref{eq:element-lower-bound-data}. Note that the left and right boundaries of the shaded areas are specified by the intersections of the blue solid lines and the black solid lines in Fig.~\ref{fig:tradeoff-data-bounds} and Fig.~\ref{fig:tradeoff-harvesting-bounds}, respectively.

In general, one can see that the power lower bounds for energy harvesting increase with \(\rho_\policy\), whereas the bounds for signal reflection decrease.
This is the expected behavior because a larger splitting ratio allocates more resources to signal reflection, which reduces the transmit power required for beam training and data transmission, but increases the transmit power required for energy harvesting.

We note that the results in Fig.~\ref{fig:tradeoff} depend on the considered system parameters, i.e., each parameter can have an impact on the exact shape of the curves in Fig.~\ref{fig:tradeoff}.
For example, a larger minimum data rate increases the power bounds for data transmission as they are proportional to \(2^{\gls{RateEffective}_\minimum}\).
Such dependence was also observed in~\cite{tyrovolas2023zeroenergyreconfigurablea}.
However, it is worth noting that the lowest bound for data transmission is typically observed for the \gls{PowerSplit} scheme, followed by the \gls{ElementSplit} scheme and the \gls{TimeSplit} scheme.
The reason is that, assuming a relatively short beam training stage, as typically desired, the scalings of these bounds are dominated by \(1/\rho_\policyPs\), \(1/\rho_\policyEs^2\), and \(2^{\gls{RateEffective}_\minimum/\rho_\policyTs}\), which can be seen by writing \(
\frac{
    \gls{Snr}_{\stageDt,\minimum}^\policy \sigma^2
}{
    \glsSubRx{Gain}
    \glsSubTx{Gain}
    \glsSubIncident{Gain}
    \gls{Gain}_{\gls{SetLocations}_b}
}\) in \eqref{eq:element-lower-bound-data} and \eqref{eq:time-lower-bound-data} as functions of \(\rho_\policy\).
This observation is consistent with the optimal values of the splitting ratios for \gls{PowerSplit}, \gls{ElementSplit}, and \gls{TimeSplit}, as will be shown in the next section.

\subsection{OPTIMIZED SPLITTING SCHEMES}\label{sec:evaluation-optimized}
\noindent
In this section, we show the system performance for the optimized splitting schemes for three exemplary system configurations.
Although all system parameters can have an impact on the optimal solution of \eqref{eq:problem}, we focus on the parameters of the power consumption model, because the power consumption is usually the bottleneck in energy harvesting-based systems.
In particular, we consider three use cases where \(\gls{PowerConsumed}_\static\), \(\gls{PowerConsumed}_\cell\), and \(\gls{PowerConsumed}_\shifter\) take different values.
In the first case, by choosing \(\gls{PowerConsumed}_\cell < \gls{PowerConsumed}_\shifter\), we assume that a phase shifter in the \gls{rf} combining network consumes more power than the phase-shifting components at a unit cell used for signal reflection.
The second case assumes by the opposite, i.e., we choose \(\gls{PowerConsumed}_\cell > \gls{PowerConsumed}_\shifter\).
In the third case, we assume that the total power consumption is dominated by the static power consumption \(\gls{PowerConsumed}_\static\), i.e., the impact of \(\gls{PowerConsumed}_\cell\) and \(\gls{PowerConsumed}_\shifter\) is negligible.
Depending on the hardware design, all of these use cases can be relevant in practice.

In the following, for each of the above use cases, we present the results for the optimized splitting schemes in terms of the average transmit power at the \gls{bs}, the splitting ratios, and the average power consumption of the \gls{ris}.
Recall that the considered splitting schemes are optimized with respect to the power allocation \(\gls{PowerAverage}_\stage^\policy\) and the splitting ratio \(\rho_\policy\).
The optimal power allocation is derived analytically and given by \(\gls{PowerAverage}_{\stage,\minimum}^\policy\), whereas the optimal value for \(\rho_\policy\) is determined using Algorithm~\ref{alg:grid-search}.
As a benchmark, we evaluate the average system performance by randomly selecting \(\rho_\policy\) from its feasible domain.

The results are shown as a function of the number of tiles, which demonstrates the impact of the tile-based hybrid combining adopted for energy harvesting.

\subsubsection{RF PHASE SHIFTERS DOMINATE POWER CONSUMPTION}\label{sec:evaluation-ts_best}
\begin{figure}[!t]
    \centering
    \subfloat[]{
        \begin{tikzpicture}
            \begin{axis}[
                line plot style,
                xlabel=Number of tiles \(\gls{Number}_\tile\),
                ylabel={Minimum average transmit\\power \(\gls{PowerAverage}_{\bs,\minimum}^\policy\) (\unit{\watt})},
                ylabel style={align=center},
                xmode=log,
                legend pos=north west,
                xmin=1,
                xmax=1000,
                ymin=0,
                ymax=10,
                width=6.5cm,
                height=3.66cm,
                scale only axis,
            ]
                \addplot [color1,mark=triangle,mark options={rotate=270}] table [col sep=comma]
                {data/ts_best-00-00.csv};
                \addlegendentry{\(\policyPs\)}

                \addplot [color1,dotted,mark=triangle,mark options={solid,rotate=270},forget plot] table [col sep=comma]
                {data/ts_best-00-03.csv};
    
                \addplot [color2,mark=square] table [col sep=comma]
                {data/ts_best-00-01.csv};
                \addlegendentry{\(\policyEs\)}

                \addplot [color2,dotted,mark=square,forget plot] table [col sep=comma]
                {data/ts_best-00-04.csv};
    
                \addplot [color3,mark=o] table [col sep=comma]
                {data/ts_best-00-02.csv};
                \addlegendentry{\(\policyTs\)}

                \addplot [color3,dotted,mark=o,forget plot] table [col sep=comma]
                {data/ts_best-00-05.csv};
            \end{axis}
        \end{tikzpicture}%
        \label{fig:ts_best-powers}
    }\\
    \subfloat[]{
        \begin{tikzpicture}
            \begin{axis}[
                line plot style,
                xlabel=Number of tiles \(\gls{Number}_\tile\),
                ylabel=Splitting ratio \(\rho_\policy\),
                xmode=log,
                legend pos=north west,
                xmin=1,
                xmax=1000,
                ymin=0,
                ymax=1,
                width=6.5cm,
                height=3.66cm,
                scale only axis,
            ]
                \addplot [color1,mark=triangle,mark options={rotate=270}] table [col sep=comma]
                {data/ts_best-01-00.csv};
                \addlegendentry{\(\policyPs\)}

                \addplot [color1,dotted,mark=triangle,mark options={solid,rotate=270},forget plot] table [col sep=comma]
                {data/ts_best-01-03.csv};
    
                \addplot [color2,mark=square] table [col sep=comma]
                {data/ts_best-01-01.csv};
                \addlegendentry{\(\policyEs\)}

                \addplot [color2,dotted,mark=square,forget plot] table [col sep=comma]
                {data/ts_best-01-04.csv};
    
                \addplot [color3,mark=o] table [col sep=comma]
                {data/ts_best-01-02.csv};
                \addlegendentry{\(\policyTs\)}

                \addplot [color3,dotted,mark=o,forget plot] table [col sep=comma]
                {data/ts_best-01-05.csv};
            \end{axis}
        \end{tikzpicture}%
        \label{fig:ts_best-split}
    }\\
    \subfloat[]{
        \begin{tikzpicture}
            \begin{axis}[
                line plot style,
                xlabel=Number of tiles \(\gls{Number}_\tile\),
                ylabel={Average power consumption\\of the RIS (\unit{\micro\watt})},
                ylabel style={align=center},
                legend pos=north west,
                xmode=log,
                xmin=1,
                xmax=1000,
                ymin=0,
                ymax=130,
                width=6.5cm,
                height=3.66cm,
                scale only axis,
            ]
                \addplot [color1,solid,mark=triangle,mark options={solid,rotate=270}] table [col sep=comma,y expr=\thisrow{y}*1000000]
                {data/ts_best-02-00.csv};
                \addlegendentry{\(\policyPs\)}

                \addplot [color1,dotted,mark=triangle,mark options={rotate=270},forget plot] table [col sep=comma,y expr=\thisrow{y}*1000000]
                {data/ts_best-02-03.csv};
    
                \addplot [color2,solid,mark=square] table [col sep=comma,y expr=\thisrow{y}*1000000]
                {data/ts_best-02-01.csv};
                \addlegendentry{\(\policyEs\)}

                \addplot [color2,dotted,mark=square,forget plot] table [col sep=comma,y expr=\thisrow{y}*1000000]
                {data/ts_best-02-04.csv};
    
                \addplot [color3,solid,mark=o] table [col sep=comma,y expr=\thisrow{y}*1000000]
                {data/ts_best-02-02.csv};
                \addlegendentry{\(\policyTs\)}

                \addplot [color3,dotted,mark=o,forget plot] table [col sep=comma,y expr=\thisrow{y}*1000000]
                {data/ts_best-02-05.csv};
            \end{axis}
        \end{tikzpicture}%
        \label{fig:ts_best-energy}
    }
    \caption{Minimum average \gls{bs} transmit powers (a), splitting ratios (b), and average \gls{ris} power consumption (c) for \(\gls{PowerConsumed}_\static = \qty{1}{\micro\watt}\), \(\gls{PowerConsumed}_\cell = \qty{10}{\nano\watt}\), \(\gls{PowerConsumed}_\shifter = \qty{30}{\nano\watt}\). The solid lines represent the optimized solutions based on Algorithm~\ref{alg:grid-search}. The dotted lines indicate the average performance using random splitting ratios.}
    \label{fig:ts_best}
\end{figure}
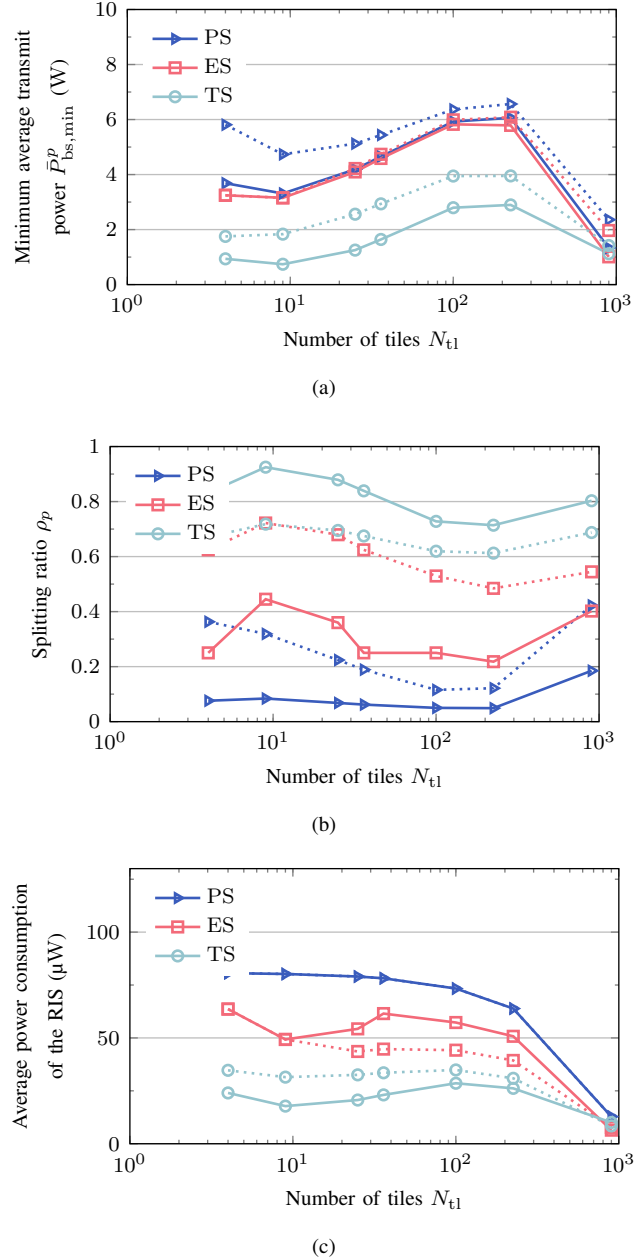

For our first scenario, we assume that the power consumption of the \gls{rf} phase shifters is larger than that of the \gls{ris} unit cells.
The optimal solutions (solid lines) and the benchmark performance (dotted lines) for all considered splitting schemes are shown in Fig.~\ref{fig:ts_best}, where we plot the minimum average transmit power at the \gls{bs} in Fig.~\ref{fig:ts_best-powers}, the splitting ratios in Fig.~\ref{fig:ts_best-split}, and the average power consumption of the \gls{ris} in Fig.~\ref{fig:ts_best-energy}.
Note that the benchmark curves in Fig.~\ref{fig:ts_best-split} show the splitting ratios averaged over their feasible domains, which are different subsets of \([0, 1]\) for each value of \(\gls{Number}_\tile\).
Thus, the average splitting ratio is different for each \(\gls{Number}_\tile\).
We also note that the \gls{ris} power consumption for the \gls{PowerSplit} scheme is independent of \(\rho_\policyPs\), leading to the same value for the optimized scheme and the benchmark scheme, cf. Fig.~\ref{fig:ts_best-energy}.

For \gls{PowerSplit} and \gls{TimeSplit}, Fig.~\ref{fig:ts_best-powers} shows that the optimized splitting ratio provides significant performance gains compared to the benchmark scheme using random \(\rho_\policy\).
Interestingly, this improvement is not observed for \gls{ElementSplit}, which is explained by the harvesting bound shown in Fig.~\ref{fig:tradeoff-harvesting-bounds} that is almost constant for all \(\rho_\policyEs\).

Fig.~\ref{fig:ts_best-powers} shows that the \gls{TimeSplit} scheme provides the best performance because the average transmit power at the \gls{bs} is significantly lower than that for the other schemes.
Moreover, all splitting schemes exhibit local minimum values for \(\gls{Number}_\tile = 9\) and \(\gls{Number}_\tile = \gls{Number}_\cell = 900\).
The highest powers are consumed for \(\gls{Number}_\tile \in \{100, 225\}\).
These characteristics of the curves in Fig.~\ref{fig:ts_best-powers} can be explained by the lower bound of the \gls{bs} transmit power that is required for energy harvesting.
A detailed explanation is provided in Appendix~\ref{sec:harvesting-bound}.

Moreover, the superior performance of the \gls{TimeSplit} scheme can be explained by the low power consumption of that scheme, cf. Fig.~\ref{fig:ts_best-energy}.
The reason is that the optimal splitting ratio for \gls{TimeSplit} is relatively large, see Fig.~\ref{fig:ts_best-split} and Fig.~\ref{fig:tradeoff}, which results in a long data transmission stage and a short energy harvesting stage.
Hence, the relatively large power consumption of the \gls{rf} phase shifters only has a minor impact on the average power consumption of the \gls{ris}, and thus on the minimum transmit power at the \gls{bs}.
In contrast, the \gls{ElementSplit} and \gls{PowerSplit} schemes harvest energy during the entire frame, which leads to a significantly larger power consumption compared to the \gls{TimeSplit} scheme.

In addition, the \gls{TimeSplit} scheme benefits from the deterministic transmit signal during energy harvesting, which leads to an efficient \gls{rf}-\gls{dc} conversion, cf. Section~\ref{sec:rectifier}, and is another reason for the short energy harvesting stage.
For the \gls{ElementSplit} and \gls{PowerSplit} schemes, however, most power is harvested during the data transmission stage, where the random transmit symbols reduce the efficiency of the \gls{rf}-\gls{dc} conversion.

\subsubsection{RIS UNIT CELLS DOMINATE POWER CONSUMPTION}\label{sec:evaluation-es_best}
\begin{figure}[!t]
    \centering
    \subfloat[]{
        \begin{tikzpicture}
            \begin{axis}[
                line plot style,
                xlabel=Number of tiles \(\gls{Number}_\tile\),
                ylabel={Minimum average transmit\\power \(\gls{PowerAverage}_{\bs,\minimum}^\policy\) (\unit{\watt})},
                ylabel style={align=center},
                xmode=log,
                legend pos=north east,
                xmin=1,
                xmax=1000,
                ymin=0,
                ymax=10,
                width=6.5cm,
                height=3.66cm,
                scale only axis,
            ]
                \addplot [color1,mark=triangle,mark options={rotate=270}] table [col sep=comma]
                {data/es_best-00-00.csv};
                \addlegendentry{\(\policyPs\)}

                \addplot [color1,dotted,mark=triangle,mark options={solid,rotate=270},forget plot] table [col sep=comma]
                {data/es_best-00-03.csv};
    
                \addplot [color2,mark=square] table [col sep=comma]
                {data/es_best-00-01.csv};
                \addlegendentry{\(\policyEs\)}

                \addplot [color2,dotted,mark=square,forget plot] table [col sep=comma]
                {data/es_best-00-04.csv};
    
                \addplot [color3,mark=o] table [col sep=comma]
                {data/es_best-00-02.csv};
                \addlegendentry{\(\policyTs\)}

                \addplot [color3,dotted,mark=o,forget plot] table [col sep=comma]
                {data/es_best-00-05.csv};
            \end{axis}
        \end{tikzpicture}%
        \label{fig:es_best-powers}
    }\\
    \subfloat[]{
        \begin{tikzpicture}
            \begin{axis}[
                line plot style,
                xlabel=Number of tiles \(\gls{Number}_\tile\),
                ylabel=Splitting ratio \(\rho_\policy\),
                xmode=log,
                legend pos=north east,
                xmin=1,
                xmax=1000,
                ymin=0,
                ymax=1,
                width=6.5cm,
                height=3.66cm,
                scale only axis,
            ]
                \addplot [color1,mark=triangle,mark options={rotate=270}] table [col sep=comma]
                {data/es_best-01-00.csv};
                \addlegendentry{\(\policyPs\)}

                \addplot [color1,dotted,mark=triangle,mark options={solid,rotate=270},forget plot] table [col sep=comma]
                {data/es_best-01-03.csv};
    
                \addplot [color2,mark=square] table [col sep=comma]
                {data/es_best-01-01.csv};
                \addlegendentry{\(\policyEs\)}

                \addplot [color2,dotted,mark=square,forget plot] table [col sep=comma]
                {data/es_best-01-04.csv};
    
                \addplot [color3,mark=o] table [col sep=comma]
                {data/es_best-01-02.csv};
                \addlegendentry{\(\policyTs\)}

                \addplot [color3,dotted,mark=o,forget plot] table [col sep=comma]
                {data/es_best-01-05.csv};
            \end{axis}
        \end{tikzpicture}%
        \label{fig:es_best-split}
    }\\
    \subfloat[]{
        \begin{tikzpicture}
            \begin{axis}[
                line plot style,
                xlabel=Number of tiles \(\gls{Number}_\tile\),
                ylabel={Average power consumption\\of the RIS (\unit{\micro\watt})},
                ylabel style={align=center},
                xmode=log,
                legend pos=north east,
                xmin=1,
                xmax=1000,
                ymin=0,
                ymax=130,
                width=6.5cm,
                height=3.66cm,
                scale only axis,
            ]
                \addplot [color1,mark=triangle,mark options={rotate=270},solid] table [col sep=comma,y expr=\thisrow{y}*1000000]
                {data/es_best-02-00.csv};
                \addlegendentry{\(\policyPs\)}

                \addplot [color1,mark=triangle,mark options={solid,rotate=270},dotted,forget plot] table [col sep=comma,y expr=\thisrow{y}*1000000]
                {data/es_best-02-03.csv};
    
                \addplot [color2,mark=square,solid] table [col sep=comma,y expr=\thisrow{y}*1000000]
                {data/es_best-02-01.csv};
                \addlegendentry{\(\policyEs\)}

                \addplot [color2,mark=square,dotted,forget plot] table [col sep=comma,y expr=\thisrow{y}*1000000]
                {data/es_best-02-04.csv};
    
                \addplot [color3,mark=o,solid] table [col sep=comma,y expr=\thisrow{y}*1000000]
                {data/es_best-02-02.csv};
                \addlegendentry{\(\policyTs\)}

                \addplot [color3,mark=o,dotted,forget plot] table [col sep=comma,y expr=\thisrow{y}*1000000]
                {data/es_best-02-05.csv};
            \end{axis}
        \end{tikzpicture}%
        \label{fig:es_best-energy}
    }
    \caption{Minimum average \gls{bs} transmit powers (a), splitting ratios (b), and average \gls{ris} power consumption (c) for \(\gls{PowerConsumed}_\static = \qty{1}{\micro\watt}\), \(\gls{PowerConsumed}_\cell = \qty{80}{\nano\watt}\), \(\gls{PowerConsumed}_\shifter = \qty{10}{\nano\watt}\). The solid lines represent the optimized solutions based on Algorithm~\ref{alg:grid-search}. The dotted lines indicate the average performance using random splitting ratios.}
    \label{fig:es_best}
\end{figure}
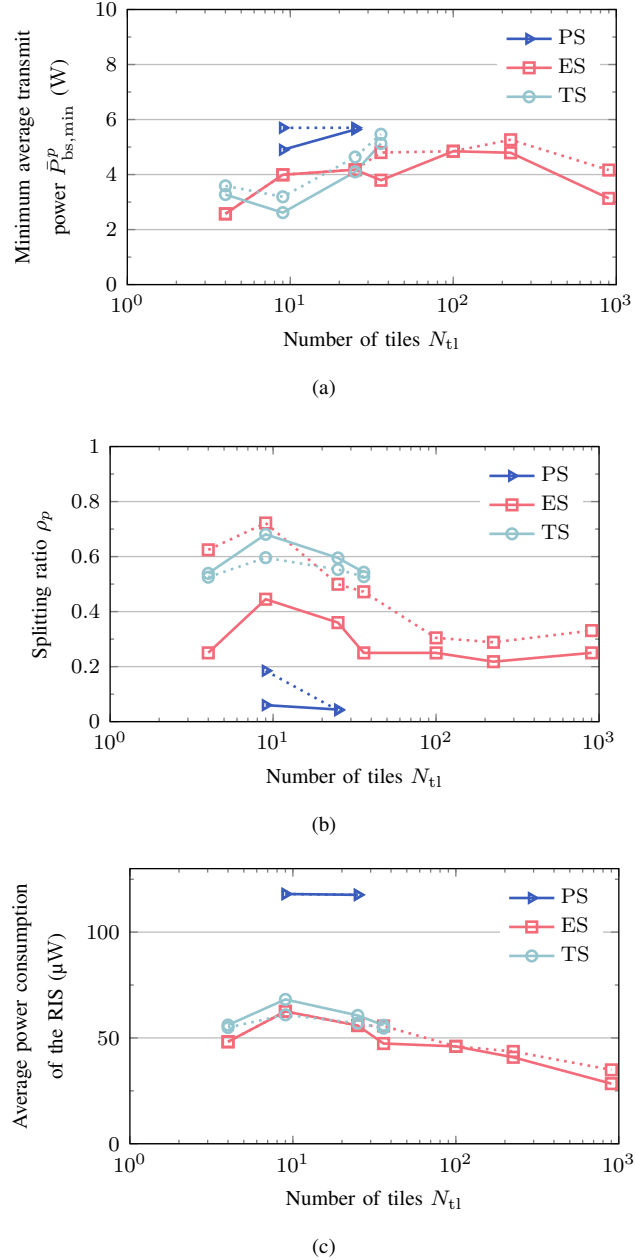

In contrast to the results above, Fig.~\ref{fig:es_best} shows the performance for the optimized and benchmark splitting schemes when the average power consumption is dominated by the \gls{ris} unit cells, i.e., we assume that the power consumption of a \gls{ris} unit cell is significantly larger than that of an \gls{rf} phase shifter.
In Fig.~\ref{fig:es_best-powers}, one can see that the proposed algorithm improves the performance compared to the benchmark scheme for most considered values of \(\gls{Number}_\tile\).
Moreover, Fig.~\ref{fig:es_best-powers} shows that the \gls{PowerSplit} and \gls{TimeSplit} schemes provide feasible solutions only for \(\gls{Number}_\tile \in \{9, 25\}\) and \(\gls{Number}_\tile \in \{4, 9, 25, 36\}\), respectively, while the \gls{ElementSplit} scheme achieves a similar performance as observed in Fig.~\ref{fig:ts_best-powers}.

The latter can be explained by the \gls{ris} power consumption for the \gls{ElementSplit} scheme, see Fig.~\ref{fig:es_best-energy}, which is similar to that in the previous case shown in Fig.~\ref{fig:ts_best-energy}.
Therefore, in both cases, the \gls{ElementSplit} schemes require approximately the same transmit power at the \gls{bs}, which leads to similar results in Fig.~\ref{fig:es_best-powers} and Fig.~\ref{fig:ts_best-powers}.

In contrast, assuming \(\gls{PowerConsumed}_\cell = \qty{80}{\nano\watt}\) for the power model significantly increases the \gls{ris} power consumption for the \gls{TimeSplit} scheme.
The reason is the relatively large \gls{TimeSplit} splitting ratio, which allocates more time of the transmission frame for signal reflection than for energy harvesting.
Nevertheless, due to the efficient \gls{rf}-\gls{dc} conversion, the \gls{TimeSplit} scheme still achieves the best performance for \(\gls{Number}_\tile = 9\).
For other values of \(\gls{Number}_\tile\), however, the efficiency is reduced and a larger transmit power at the \gls{bs} is required, which eventually exceeds the upper bound and makes the optimization problem infeasible.
For a visual explanation, one can refer to Fig.~\ref{fig:tradeoff} as follows.
Less efficient \gls{rf}-\gls{dc} conversion due to \(\gls{Number}_\tile \neq 9\) increases the lower bound for energy harvesting in Fig.~\ref{fig:tradeoff-harvesting-bounds}, which reduces the feasible set and the upper bound for \(\rho_\policy\).
Since \(\rho_\policy\) is also lower bounded by the minimum transmit power in Fig.~\ref{fig:tradeoff-data-bounds}, less efficient \gls{rf}-\gls{dc} conversion may lead to an empty feasible set.

Finally, Fig.~\ref{fig:es_best-energy} shows that assuming \(\gls{PowerConsumed}_\cell = \qty{80}{\nano\watt}\) has also a strong impact on the power consumption for the \gls{PowerSplit} scheme, caused by the fact that both energy harvesting and signal reflection is used at all tiles simultaneously.
Therefore, in Fig.~\ref{fig:es_best-powers}, the largest \gls{bs} transmit power is observed for the \gls{PowerSplit} scheme.
Moreover, only \(\gls{Number}_\tile \in \{9, 25\}\) leads to a feasible solution.
Similar to the \gls{TimeSplit} scheme, other values of \(\gls{Number}_\tile\) lead to a less efficient \gls{rf}-\gls{dc} conversion and require a larger transmit power at the \gls{bs}, which eventually exceeds the upper bound and makes the problem infeasible.

\subsubsection{CONSTANT POWER CONSUMPTION OF THE RIS}\label{sec:evaluation-ps_best}
\begin{figure}[!t]
    \centering
    \subfloat[]{
        \begin{tikzpicture}
            \begin{axis}[
                line plot style,
                xlabel=Number of tiles \(\gls{Number}_\tile\),
                ylabel={Minimum average transmit\\power \(\gls{PowerAverage}_{\bs,\minimum}^\policy\) (\unit{\watt})},
                ylabel style={align=center},
                xmode=log,
                legend pos=south east,
                xmin=1,
                xmax=1000,
                ymin=0,
                ymax=10,
                width=6.5cm,
                height=3.66cm,
                scale only axis,
            ]
                \addplot [color1,mark=triangle,mark options={rotate=270}] table [col sep=comma]
                {data/ps_best-00-00.csv};
                \addlegendentry{\(\policyPs\)}

                \addplot [color1,dotted,mark=triangle,mark options={solid,rotate=270},forget plot] table [col sep=comma]
                {data/ps_best-00-03.csv};
    
                \addplot [color2,mark=square] table [col sep=comma]
                {data/ps_best-00-01.csv};
                \addlegendentry{\(\policyEs\)}

                \addplot [color2,dotted,mark=square,forget plot] table [col sep=comma]
                {data/ps_best-00-04.csv};
    
                \addplot [color3,mark=o] table [col sep=comma]
                {data/ps_best-00-02.csv};
                \addlegendentry{\(\policyTs\)}

                \addplot [color3,dotted,mark=o,forget plot] table [col sep=comma]
                {data/ps_best-00-05.csv};
            \end{axis}
        \end{tikzpicture}%
        \label{fig:ps_best-powers}
    }\\
    \subfloat[]{
        \begin{tikzpicture}
            \begin{axis}[
                line plot style,
                xlabel=Number of tiles \(\gls{Number}_\tile\),
                ylabel=Splitting ratio \(\rho_\policy\),
                xmode=log,
                legend pos=north east,
                xmin=1,
                xmax=1000,
                ymin=0,
                ymax=1,
                width=6.5cm,
                height=3.66cm,
                scale only axis,
            ]
                \addplot [color1,mark=triangle,mark options={rotate=270}] table [col sep=comma]
                {data/ps_best-01-00.csv};
                \addlegendentry{\(\policyPs\)}

                \addplot [color1,dotted,mark=triangle,mark options={solid,rotate=270},forget plot] table [col sep=comma]
                {data/ps_best-01-03.csv};
    
                \addplot [color2,mark=square] table [col sep=comma]
                {data/ps_best-01-01.csv};
                \addlegendentry{\(\policyEs\)}

                \addplot [color2,dotted,mark=square,forget plot] table [col sep=comma]
                {data/ps_best-01-04.csv};
    
                \addplot [color3,mark=o] table [col sep=comma]
                {data/ps_best-01-02.csv};
                \addlegendentry{\(\policyTs\)}

                \addplot [color3,dotted,mark=o,forget plot] table [col sep=comma]
                {data/ps_best-01-05.csv};
            \end{axis}
        \end{tikzpicture}%
        \label{fig:ps_best-split}
    }\\
    \subfloat[]{
        \begin{tikzpicture}
            \begin{axis}[
                line plot style,
                xlabel=Number of tiles \(\gls{Number}_\tile\),
                ylabel={Average power consumption\\of the RIS (\unit{\micro\watt})},
                ylabel style={align=center},
                xmode=log,
                legend pos=south east,
                xmin=1,
                xmax=1000,
                ymin=0,
                ymax=130,
                width=6.5cm,
                height=3.66cm,
                scale only axis,
            ]
                \addplot [color1,solid,mark=triangle,mark options={rotate=270}] table [col sep=comma,y expr=\thisrow{y}*1000000]
                {data/ps_best-02-00.csv};
                \addlegendentry{\(\policyPs\)}

                \addplot [color1,dotted,mark=triangle,mark options={solid,rotate=270},forget plot] table [col sep=comma,y expr=\thisrow{y}*1000000]
                {data/ps_best-02-03.csv};

                \addplot [color2,solid,mark=square] table [col sep=comma,y expr=\thisrow{y}*1000000]
                {data/ps_best-02-01.csv};
                \addlegendentry{\(\policyEs\)}

                \addplot [color2,dotted,mark=square,forget plot] table [col sep=comma,y expr=\thisrow{y}*1000000]
                {data/ps_best-02-04.csv};

                \addplot [color3,solid,mark=o] table [col sep=comma,y expr=\thisrow{y}*1000000]
                {data/ps_best-02-02.csv};
                \addlegendentry{\(\policyTs\)}

                \addplot [color3,dotted,mark=o,forget plot] table [col sep=comma,y expr=\thisrow{y}*1000000]
                {data/ps_best-02-05.csv};
            \end{axis}
        \end{tikzpicture}%
        \label{fig:ps_best-energy}
    }
    \caption{Minimum average \gls{bs} transmit powers (a), splitting ratios (b), and average \gls{ris} power consumption (c) for \(\gls{PowerConsumed}_\static = \qty{80}{\micro\watt}\), \(\gls{PowerConsumed}_\cell = \qty{0}{\nano\watt}\), \(\gls{PowerConsumed}_\shifter = \qty{0}{\nano\watt}\). The solid lines represent the optimized solutions based on Algorithm~\ref{alg:grid-search}. The dotted lines indicate the average performance using random splitting ratios.}
    \label{fig:ps_best}
\end{figure}

Depending on the hardware design of the \gls{ris}, the power consumptions of the unit cells and the \gls{rf} phase shifters may be negligible compared to the static power consumption of the \gls{ris}.
For such a case, where \(\gls{PowerConsumed}_\static = \qty{80}{\micro\watt}\) and \(\gls{PowerConsumed}_\cell = \gls{PowerConsumed}_\shifter = 0\) is assumed, Fig.~\ref{fig:ps_best} shows the optimal and the benchmark solutions for the considered splitting schemes.
Again, the results in Fig.~\ref{fig:ps_best-powers} demonstrate that the optimized splitting ratios provide additional performance gains compared to the benchmark scheme using random splitting ratios.

Moreover, we observe that the performance of the \gls{TimeSplit} and \gls{ElementSplit} schemes is similar to that observed in Fig.~\ref{fig:es_best}.
For these schemes, the main difference is that the \gls{bs} transmit powers in Fig.~\ref{fig:ps_best-powers} are slightly larger than those in Fig.~\ref{fig:es_best-powers}, which is explained by the higher power consumption.
Interestingly, except for \(\gls{Number}_\tile = 9\), the \gls{PowerSplit} scheme achieves a lower \gls{bs} transmit power than the other schemes, which has not been observed in the previous cases.
The reason is that, for the considered power consumption model, the \gls{PowerSplit} scheme does not consume more power than the other schemes, although each tile performs energy harvesting and signal reflection simultaneously.
Therefore, the differences between the splitting schemes are mainly determined by the power lower bounds for signal reflection.
As shown in Fig.~\ref{fig:tradeoff-data-bounds}, the lowest bound is typically observed for the \gls{PowerSplit} scheme, which explains the superior performance in Fig.~\ref{fig:ps_best-powers}.

%% file: sections/conclusion.tex
\noindent
This paper studied a \gls{ris}-assisted communication system where the \gls{ris} is configured based on beam training and implements energy harvesting to enable self-sustainable operation.
For energy harvesting, we considered a tile-based hybrid combining scheme that takes the gains and losses for both \gls{rf} combining and \gls{dc} combining into account.
Moreover, we showed that different transmit strategies for beam training and data transmission lead to different rectifier characteristics, which can have a strong impact on the efficiency of the \gls{rf}-\gls{dc} conversion.
Furthermore, we formulated an optimization problem for \gls{PowerSplit}, \gls{ElementSplit}, and \gls{TimeSplit} to minimize the average transmit power at the \gls{bs}.
For a given minimum beam training \gls{snr} and a minimum data rate at the user, we showed that the optimization problem can be solved by an efficient grid search-based algorithm.
In addition, we evaluated the optimized splitting schemes for three different assumptions regarding the \gls{ris} power consumption, which unveiled the advantages and disadvantages of the considered schemes.
For example, due to the deterministic transmit signal for energy harvesting, the \gls{TimeSplit} scheme benefits from efficient \gls{rf}-\gls{dc} conversion leading to a low transmit power at the \gls{bs}.
Moreover, our analysis revealed that \gls{PowerSplit} can outperform the other splitting schemes when the \gls{ris} power consumption is dominated by the static component.
Furthermore, the \gls{ElementSplit} scheme is robust to changes in the power consumption model, but the restricted feasible set of the splitting ratio may be a limiting factor for its performance.
In addition, our results showed that the number of tiles that the \gls{ris} is divided into has a strong impact on the transmit power required at the \gls{bs}.

It is worth noting that the framework proposed in this work can be applied to multi-cell networks where the cell coverage of a specific \gls{bs} is extended by one \gls{ris}.
In addition, future extensions of this work may study a multi-\gls{ris} scenario where energy harvesting at multiple \glspl{ris} must be coordinated efficiently.
Moreover, efficient beam training and data transmission in a multi-user scenario is an interesting subject for future work because multiple users may share the same \gls{ris} beam.
Finally, prototypes and experimental studies of self-sustainable \glspl{ris} are required to validate the theoretical results reported in this paper.

%% file: sections/appendix.tex
\subsection{CODEBOOK SIZE}\label{sec:codebook-size}
\noindent
Partitioning the coverage area into subareas of equal size can be achieved by dividing \(\gls{SizeCoverageAreaY}\) and \(\gls{SizeCoverageAreaZ}\) into \(\gls{Number}_{y,\policy}\) and \(\gls{Number}_{z,\policy}\) intervals of equal size, respectively, which yields a subarea size of \(\gls{LengthSideSubarea}_{y,\policy} \gls{LengthSideSubarea}_{z,\policy}\), where
\begin{align}
    \gls{LengthSideSubarea}_{y,\policy} &= \frac{\gls{SizeCoverageAreaY}}{\gls{Number}_{y,\policy}}\\
    \gls{LengthSideSubarea}_{z,\policy} &= \frac{\gls{SizeCoverageAreaZ}}{\gls{Number}_{z,\policy}}.
\end{align}
Hence, the total number of subareas, which equals the number of codewords, is given by
\begin{equation}
    \abs*{\gls{Codebook}_\policy} = \gls{Number}_{y,\policy} \gls{Number}_{z,\policy} = \frac{\gls{SizeCoverageAreaY} \gls{SizeCoverageAreaZ}}{\gls{LengthSideSubarea}_{y,\policy} \gls{LengthSideSubarea}_{z,\policy}}.
\end{equation}
Since the beam training overhead increases with the number of codewords, it is desired to choose \(\abs*{\gls{Codebook}_\policy}\) as small as possible.
In other words, the subareas should be as large as possible while taking into account that each subarea must be fully covered by the footprint of a \gls{ris} reflection beam.
In the worst case, this is guaranteed when the footprint is wider than \(\sqrt{\gls{LengthSideSubarea}_{y,\policy}^2 + \gls{LengthSideSubarea}_{z,\policy}^2}\), i.e., the diagonal of a subarea.

The width of a footprint is given by \(L_l = d_{\reflected,l} 2 \tan\left(\frac{\Delta\theta}{2}\right)\), where \(\Delta\theta\) and \(d_{\reflected,l}\) denote the angular width of the \gls{ris} beam and the distance between the \gls{ris} and location \(l \in \gls{SetLocations}\), respectively.
Thus, the condition for full coverage is given by 
\begin{equation}\label{eq:codebook-size-condition}
    \sqrt{\gls{LengthSideSubarea}_{y,\policy}^2 + \gls{LengthSideSubarea}_{z,\policy}^2} \leq \min_{l \in \gls{SetLocations}} L_l = d_{\reflected,\minimum} 2 \tan\left(\frac{\Delta\theta}{2}\right),
\end{equation}
where \(d_{\reflected,\minimum} = \min_{l \in \gls{SetLocations}} d_{\reflected,l}\).
For \(\Delta\theta\), we adopt the \qty{3}{\deci\bel} beam width in broadside direction and assume \gls{ris} unit cells of size \(\frac{\gls{Wavelength}^2}{4}\), where \(\gls{Wavelength}\) denotes the wavelength.
Then, the beam width only depends on the number of unit cells used for signal reflection and is given by \(\Delta \theta = \abs*{\frac{\pi}{2} - \arccos\left(\frac{2.782}{\pi \sqrt{\gls{Number}\Ucr^\policy}}\right)} + \abs*{\frac{\pi}{2} - \arccos\left(\frac{-2.782}{\pi \sqrt{\gls{Number}\Ucr^\policy}}\right)}\)~\cite[Eq.~(27)]{han2021halfpowerbeamwidth}.
Using \(\tan(x) \approx x\) and \(\arccos(x) \approx \frac{\pi}{2} - x\) for \(\abs*{x} < 0.5\), the right-hand side of \eqref{eq:codebook-size-condition} simplifies to
\begin{equation}\label{eq:beam-width}
    d_{\reflected,\minimum} \Delta\theta = d_{\reflected,\minimum} 2 \frac{2.782}{\pi \sqrt{\gls{Number}\Ucr^\policy}}.
\end{equation}

Moreover, by solving \eqref{eq:codebook-size-condition} for \(\gls{LengthSideSubarea}_{z,\policy}\), the subarea size can be written as \(\gls{LengthSideSubarea}_{y,\policy} \gls{LengthSideSubarea}_{z,\policy} = \gls{LengthSideSubarea}_{y,\policy} \sqrt{d_{\reflected,\minimum}^2 \Delta\theta^2 - \gls{LengthSideSubarea}_{y,\policy}^2}\), which is maximized by \(\gls{LengthSideSubarea}_{y,\policy,\maximum} = \frac{1}{\sqrt{2}} d_{\reflected,\minimum} \Delta\theta\).
Similarly, we find \(\gls{LengthSideSubarea}_{z,\policy,\maximum} = \frac{1}{\sqrt{2}} d_{\reflected,\minimum} \Delta\theta\), i.e., the optimal subarea is a square.
Finally, considering that \(\gls{Number}_{y,\policy}\) and \(\gls{Number}_{z,\policy}\) are integers, the optimal codebook size is given by
\begin{equation}\label{eq:codebook-size-preliminary}
    \abs*{\gls{Codebook}_\policy}
    = \ceil*{\frac{\sqrt{2} \gls{SizeCoverageAreaY}}{d_{\reflected,\minimum} \Delta\theta}}
    \ceil*{\frac{\sqrt{2} \gls{SizeCoverageAreaZ}}{d_{\reflected,\minimum} \Delta\theta}}.
\end{equation}
Substituting \eqref{eq:beam-width} into \eqref{eq:codebook-size-preliminary} yields \eqref{eq:codebook-size}.

\subsection{IMPACT OF PHASE QUANTIZATION}\label{sec:impact-phase-quantization}
\noindent
In \eqref{eq:gain-tile-time-element} and \eqref{eq:gain-tile-power}, the optimal phase shift for coherent combining at the \(n\)th unit cell is given by \(\gls{ShiftCombining} = -\gls{PhaseIncident}\).
However, since we assume \gls{rf} phase shifters with a finite phase resolution, \(\gls{ShiftCombining} = -\gls{PhaseIncident}\) is quantized using \(\gls{Number}_\bits\) bits.
The resulting residual phase shift in \eqref{eq:gain-tile-time-element} and \eqref{eq:gain-tile-power} is given by \(e^{j\gls{ShiftCombining}} e^{j\gls{PhaseIncident}} = e^{j(-\gls{PhaseIncident} + \gls{ShiftErrorCombining})} e^{j\gls{PhaseIncident}} = e^{j\gls{ShiftErrorCombining}}\), where \(\gls{ShiftErrorCombining}\) denotes the phase error originating from quantization.
As a result, \(\gls{Gain}_{\tile,m}^\policy\) is smaller than its maximum value that is obtained with \(\gls{Number}_\bits \rightarrow \infty\) where \(\gls{ShiftErrorCombining} \rightarrow 0\).

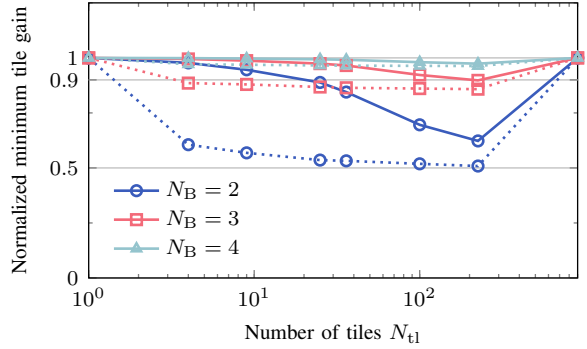
\begin{figure}[!t]
    \centerline{
        \begin{tikzpicture}
            \begin{axis}[
                line plot style,
                xlabel=Number of tiles \(\gls{Number}_\tile\),
                ylabel=Normalized minimum tile gain,
                xmode=log,
                legend pos=south west,
                ymax=1.25,
                ymin=0.0,
                extra y ticks={0.9},
                width=6.5cm,
                height=3.66cm,
                scale only axis,
            ]
                \addplot table [col sep=comma]
                {data/impact_quantization-00-00.csv};
                \addlegendentry{\(\gls{Number}_\bits = 2\)}

                \addplot table [col sep=comma]
                {data/impact_quantization-00-01.csv};
                \addlegendentry{\(\gls{Number}_\bits = 3\)}

                \addplot table [col sep=comma]
                {data/impact_quantization-00-02.csv};
                \addlegendentry{\(\gls{Number}_\bits = 4\)}

                \addplot [color1,dotted,mark=o,forget plot] table [col sep=comma]
                {data/impact_quantization-00-03.csv};

                \addplot [color2,dotted,mark=square,forget plot] table [col sep=comma]
                {data/impact_quantization-00-04.csv};

                \addplot [color3,dotted,mark=triangle,forget plot] table [col sep=comma]
                {data/impact_quantization-00-05.csv};
            \end{axis}
        \end{tikzpicture}
    }
    \caption{Normalized minimum tile gain as a function of number of tiles, plotted as the minimum (dashed lines) and the 0.1-quantiles (solid lines) with respect to all possible incident angles at the \gls{ris}. For \(\gls{Number}_\bits \geq 3\), the minimum tile gain for \qty{90}{\percent} of the possible incident angles is not less than \qty{90}{\percent} of the maximum tile gain. Thus, most incident angles lead to a negligible impact of phase quantization if \(\gls{Number}_\bits \geq 3\). \(\gls{Number}_\cell = 900\), \(\gls{LossInsertion}_0 = \qty{-0.5}{\deci\bel}\).\label{fig:impact-quantization}}
\end{figure}

Moreover, \(\gls{Gain}_{\tile,m}^\policy\) is different for each tile because \(\gls{ShiftErrorCombining}\) is different for each unit cell.
More specifically, the impact of \(\gls{ShiftErrorCombining}\) on \(\gls{Gain}_{\tile,m}^\policy\) depends on the number of bits used for quantization, the number of unit cells per tile, and the \gls{aoa} of the incident wave at the \gls{ris}.
For further analysis, we adopt the normalized minimum tile gain
\begin{equation}
    \bar{\gls{Gain}}_{\tile}^\policy = \frac{
        \min_{m \in \gls{SetTiles}_{\harvesting,\stage}^\policy} \gls{Gain}_{\tile,m}^\policy
    }{
        \max_{m \in \gls{SetTiles}_{\harvesting,\stage}^\policy} \gls{Gain}_{\tile,m}^\policy
    },
\end{equation}
which is a function of the \gls{aoa} and the number of bits for quantization.
In Fig.~\ref{fig:impact-quantization}, we show the minimum and the 0.1-quantile of \(\bar{\gls{Gain}}_{\tile}^\policy\) with respect to all \glspl{aoa} as dashed lines and solid lines, respectively.

One can see that the loss due to quantization increases with the number of tiles, except for \(\gls{Number}_\tile = \gls{Number}_\cell\), where each tile only comprises one unit cell and no phase alignment is required.
Moreover, the quantization loss decreases as the number of bits grows.
In addition, one can see that the loss significantly reduces when the 0.1-quantiles are considered.
This shows that a large quantization error is only observed for \qty{10}{\percent} of all possible \glspl{aoa}.
As an example, for \(\gls{Number}_\bits = 2\) and \(\gls{Number}_\tile = 36\), the worst-case \gls{aoa} can reduce the tile gain by a factor of 2, whereas \qty{90}{\percent} of all \glspl{aoa} only result in a loss of about \qty{10}{\percent}.
For \(\gls{Number}_\bits = 3\), the impact of the quantization error is further reduced.

Moreover, the larger quantization loss for larger \(\gls{Number}_\tile\) can be explained as follows.
For \(\gls{Number}_\tile = 1\), i.e., pure \gls{rf} combining, the entire \gls{ris} represents the only tile and the corresponding tile gain can be seen as the average over the quantization errors of all unit cells.
As \(\gls{Number}_\tile\) increases, there are multiple tiles and each tile gain \(\gls{Gain}_{\tile,m}^\policy\) corresponds to the quantization error averaged over a subset of unit cells.
Since these subsets are disjoint, \(\gls{Gain}_{\tile,m}^\policy\) is different for each tile.
These differences, i.e., the variance of \(\gls{Gain}_{\tile,m}^\policy\), increase as the tiles become smaller with larger \(\gls{Number}_\tile\) and constant \(\gls{Number}_\cell\), leading to lower values of \(\bar{\gls{Gain}}_{\tile}^\policy\).

\subsection{IMPACT OF NUMBER OF TILES}\label{sec:harvesting-bound}
\noindent
The shape of the curves in Fig.~\ref{fig:ts_best-powers}, in particular, the local minima and maxima, is typical for the problem at hand and can be explained as follows.

The minimum transmit power at the \gls{bs} is determined by the power required for energy harvesting, beam training, and data transmission, where energy harvesting typically is the limiting factor.
Thus, let us focus on the minimum powers required for energy harvesting, which are given by \eqref{eq:time-lower-bound-harvesting} and by the first arguments of the \(\max\) operators in \eqref{eq:element-lower-bound-training} and \eqref{eq:element-lower-bound-data}.

\begin{figure}[!t]
    \centering
    \begin{tikzpicture}
        \begin{axis}[
            line plot style,
            xlabel=Number of tiles \(\gls{Number}_\tile\),
            ylabel={Power (\unit{\milli\watt})},
            xmode=log,
            xmin=1,
            xmax=1000,
            ymin=0,
            ymax=1,
            width=6.5cm,
            height=3.66cm,
            scale only axis,
        ]
            \addplot+ [no marks] table [col sep=comma]
            {data/shape-00-02.csv};
            \addlegendentry{\(100 \frac{\gls{Number}_\tile}{B} \bar{\gls{FunctionModel}}_\stageEh^{-1}\left(\frac{A}{\gls{Number}_\tile}\right)\)}

            \addplot+ [no marks] table [col sep=comma]
            {data/shape-00-00.csv};
            \addlegendentry{\(\bar{\gls{FunctionModel}}_\stageEh^{-1}\left(\frac{A}{\gls{Number}_\tile}\right)\)}

            \addplot+ [no marks] table [col sep=comma]
            {data/shape-00-01.csv};
            \addlegendentry{\(\frac{\gls{Number}_\tile}{B}\)}
        \end{axis}
    \end{tikzpicture}
    \caption{Minimum transmit power required for energy harvesting \(\gls{PowerAverage}_{\stageEh,\minimum}^\policyTs\) and its two components defined in \eqref{eq:power-min-shape}. For visualization purpose, \(\gls{PowerAverage}_{\stageEh,\minimum}^\policyTs\) is scaled by a factor of \(100\).}
    \label{fig:shape}
\end{figure}
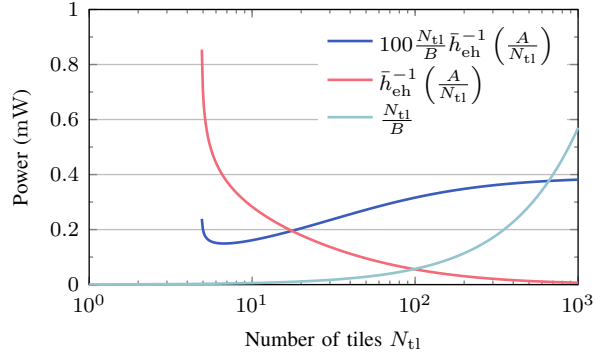

For the ease of presentation, we only consider \(\gls{PowerAverage}_{\stageEh,\minimum}^\policyTs\) in \eqref{eq:time-lower-bound-harvesting}.
For a simplified system configuration with \(\gls{PowerConsumed}_\cell = \gls{PowerConsumed}_\shifter = 0\), \(\gls{LossExponent} = 0\), and \(\TileGainMin^\policyTs = \frac{\gls{Number}_\cell}{\gls{Number}_\tile}\), one can write
\begin{equation}\label{eq:power-min-shape}
    \gls{PowerAverage}_{\stageEh,\minimum}^\policyTs = \frac{\gls{Number}_\tile}{B} \bar{\gls{FunctionModel}}_\stageEh^{-1}\left(\frac{A}{\gls{Number}_\tile}\right),
\end{equation}
where \(A = \frac{1}{\glsSubIncident{Gain} \glsSubTx{Gain} \gls{Number}_\cell}\) and \(B = \frac{\gls{PowerConsumed}_\static}{1-\rho_\policyTs}\).
In Fig.~\ref{fig:shape}, \(\gls{PowerAverage}_{\stageEh,\minimum}^\policyTs\) in \eqref{eq:power-min-shape} (scaled by \(100\) for better visualization) and the two factors \(\bar{\gls{FunctionModel}}_\stageEh^{-1}\left(\frac{A}{\gls{Number}_\tile}\right)\) and \(\frac{\gls{Number}_\tile}{B}\) are shown as functions of \(\gls{Number}_\tile\).
One can clearly see that \(\bar{\gls{FunctionModel}}_\stageEh^{-1}\left(\frac{A}{\gls{Number}_\tile}\right)\) is a scaled and horizontally flipped version of the inverse rectifier characteristic, which is a decreasing function of \(\gls{Number}_\tile\).
Moreover, the factor \(\frac{\gls{Number}_\tile}{B}\) is a linear function of \(\gls{Number}_\tile\).
Therefore, the product of both curves yields the typical characteristic of \(\gls{PowerAverage}_{\stageEh,\minimum}^\policyTs\), i.e., a fast decay for small \(\gls{Number}_\tile\), the minimum value near \(\gls{Number}_\tile = 9\), and the slow increase for larger \(\gls{Number}_\tile\).
This characteristic is observed for all results in Section~\ref{sec:evaluation}, except for \(\gls{Number}_\tile = \gls{Number}_\cell\).
In this particular case, the tile-based hybrid combining at the \gls{ris} reduces to pure \gls{dc} combining, where no \gls{rf} combining network is required that introduces additional power consumption and insertion loss.
Hence, the transmit power at the \gls{bs} can be reduced for \(\gls{Number}_\tile = \gls{Number}_\cell\), which is not captured by \eqref{eq:power-min-shape} but observed in Section~\ref{sec:evaluation}.

Finally, the minimum value of \(\gls{Number}_\tile\) that minimizes \(\gls{PowerAverage}_{\stageEh,\minimum}^\policyTs\) can be seen as the optimal operating point of the hybrid combining-based energy harvesting.
If the system parameters can be configured to achieve that operating point, the \gls{rf}-\gls{dc} conversion achieves the maximum efficiency, resulting in the lowest transmit power at the \gls{bs}.

\subsection{IMPACT OF INSERTION LOSS}\label{sec:imperfections}
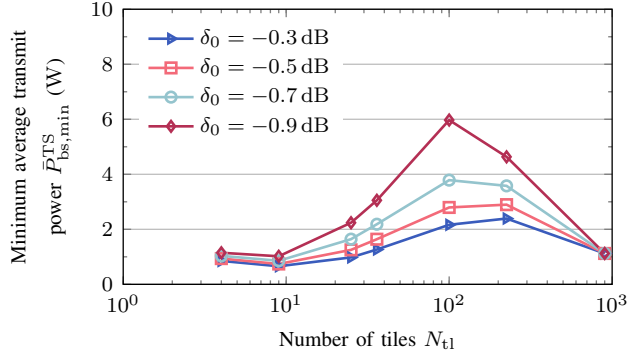
\begin{figure}[!t]
    \centering
    \begin{tikzpicture}
        \begin{axis}[
            line plot style,
            xlabel=Number of tiles \(\gls{Number}_\tile\),
            ylabel={Minimum average transmit\\power \(\gls{PowerAverage}_{\bs,\minimum}^\policyTs\) (\unit{\watt})},
            ylabel style={align=center},
            xmode=log,
            legend pos=north west,
            xmin=1,
            xmax=1000,
            ymin=0,
            ymax=10,
            width=6.5cm,
            height=3.66cm,
            scale only axis,
        ]
            \addplot [color1,mark=triangle,mark options={rotate=270}] table [col sep=comma]
            {data/ts_best_imperfect-00-00.csv};
            \addlegendentry{\(\gls{LossInsertion}_0 = \qty{-0.3}{\deci\bel}\)}

            \addplot [color2,mark=square] table [col sep=comma]
            {data/ts_best_imperfect-00-01.csv};
            \addlegendentry{\(\gls{LossInsertion}_0 = \qty{-0.5}{\deci\bel}\)}

            \addplot [color3,mark=o] table [col sep=comma]
            {data/ts_best_imperfect-00-02.csv};
            \addlegendentry{\(\gls{LossInsertion}_0 = \qty{-0.7}{\deci\bel}\)}

            \addplot [color4,mark=diamond] table [col sep=comma]
            {data/ts_best_imperfect-00-03.csv};
            \addlegendentry{\(\gls{LossInsertion}_0 = \qty{-0.9}{\deci\bel}\)}
        \end{axis}
    \end{tikzpicture}
    \caption{Minimum average \gls{bs} transmit power for \(\gls{PowerConsumed}_\static = \qty{1}{\micro\watt}\), \(\gls{PowerConsumed}_\cell = \qty{10}{\nano\watt}\), \(\gls{PowerConsumed}_\shifter = \qty{30}{\nano\watt}\), and different values of the insertion loss \(\gls{LossInsertion}_0\).}
    \label{fig:imperfections}
\end{figure}

\noindent
The impact of the insertion loss per bit, \(\gls{LossInsertion}_0\), on the system performance is depicted in Fig.~\ref{fig:imperfections}.
Note that \(\gls{LossInsertion}_0\) is modeled as a gain factor, i.e., negative values in dB scale correspond to a loss.
As expected, Fig.~\ref{fig:imperfections} shows that higher losses lead to a higher transmit power at the \gls{bs}, which has to compensate for the lower tile gain in \eqref{eq:gain-tile-time-element} and \eqref{eq:gain-tile-power}.

